\documentclass[review]{elsarticle}

\usepackage{lineno}
\usepackage{comment}
\usepackage{amsmath}
\usepackage{graphicx}
\usepackage{amsfonts}
\usepackage[colorinlistoftodos]{todonotes}
\usepackage[colorlinks=true, allcolors=blue]{hyperref}
\usepackage{subcaption}
\usepackage{booktabs}
\usepackage{tikz}
\usepackage{makecell}
\usepackage{multirow}
\usepackage{mathtools}
\usepackage[ruled,vlined]{algorithm2e}
\usetikzlibrary{shapes.geometric, arrows}
\usetikzlibrary{snakes,trees}
\usetikzlibrary{mindmap}
\usetikzlibrary{calc}
\usetikzlibrary{shapes,arrows,decorations.pathmorphing,positioning,fit}

\usetikzlibrary{calc,trees,positioning,arrows,chains,shapes.geometric,%
 decorations.pathreplacing,decorations.pathmorphing,shapes,%
    matrix,shapes.symbols,shapes.geometric,backgrounds,calc}
    
\tikzstyle{subprocess} = [rectangle, minimum width=1cm, minimum height=0.27cm, text centered, draw=black] %, fill=orange!30
\tikzstyle{state} = [rectangle, minimum width=1cm, minimum height=0.27cm, text centered, draw=black]
\tikzstyle{terminal} = [rectangle,rounded corners, minimum width=1cm, minimum height=0.27cm, text centered, draw=black]
\tikzstyle{arrow} = [thick,->,>=stealth]
\tikzstyle{dasharrow} = [thick,dashed,->,>=stealth]
\tikzstyle{dasharrownohead} = [thick,dashed,-,>=stealth]
\tikzstyle{textbf} = [draw,rectangle,text width=5cm,text centered]

\tikzset{
  basic box/.style = {
    shape = rectangle,
    align = center,
    draw  = #1,
    fill  = #1!25,
    rounded corners},
  header node/.style = {
    Minimum Width = header nodes,
    font          = \strut\Large\ttfamily,
    text depth    = +0pt,
    fill          = white,
    draw},
  header/.style = {%
    inner ysep = +1.5em,
    append after command = {
      \pgfextra{\let\TikZlastnode\tikzlastnode}
      node [header node] (header-\TikZlastnode) at (\TikZlastnode.north) {#1}
      node [span = (\TikZlastnode)(header-\TikZlastnode)]
        at (fit bounding box) (h-\TikZlastnode) {}
    }
  },
  hv/.style = {to path = {-|(\tikztotarget)\tikztonodes}},
  vh/.style = {to path = {|-(\tikztotarget)\tikztonodes}},
  fat blue line/.style = {ultra thick, blue}
}    

\journal{Elsevier}
\biboptions{authoryear}
\begin{document}

\begin{frontmatter}

\title{Market Design for Tradable Mobility Credits}
% \tnotetext[mytitlenote]{Fully documented templates are available in the elsarticle package on \href{http://www.ctan.org/tex-archive/macros/latex/contrib/elsarticle}{CTAN}.}


% %% or include affiliations in footnotes:
\author[mymainaddress]{Siyu Chen}
\author[mythirdaddress]{Ravi Seshadri \corref{mycorrespondingauthor}}
\cortext[mycorrespondingauthor]{Corresponding author}
\ead{ravse@dtu.dk}
\author[mythirdaddress]{Carlos Lima Azevedo}
\author[myfourthaddress]{Arun P. Akkinepally}
\author[mythirdaddress]{Renming Liu}
\author[myfifthaddress]{Andrea Araldo}
\author[mythirdaddress]{Yu Jiang}
\author[mymainaddress]{Moshe E. Ben-Akiva}

\address[mymainaddress]{Civil and Environmental Engineering Department, Massachusetts Institute of Technology, Cambridge, MA, US}
%\address[mysecondaryaddress]{Singapore-MIT Alliance for Research and Technology, Singapore}
\address[mythirdaddress]{Department of Technology, Management and Economics, Technical University of Denmark, Denmark}

\address[myfourthaddress]{Caliper Corporation, United States}

\address[myfifthaddress]{T\'el\'ecom SudParis - Institut Polytechnique de Paris, France}

%\fntext[corr]{Corresponding Author}

\begin{abstract} 
%\textcolor{red}{to be modified }
Tradable mobility credit (TMC) schemes are an approach to travel demand management that have received significant attention in recent years as a promising means to mitigate the adverse environmental, economic and social effects of urban traffic congestion. 
%In TMC schemes, a regulator provides an initial endowment of mobility credits (or tokens) to all potential travelers. In order to use the transportation system, travelers need to spend a certain amount of tokens (tariff) that could vary with their choice of mode, route, departure time etc. The tokens can be bought and sold in a market that is managed and operated by a regulator at a price dynamically determined by the demand and supply of tokens. 
This paper proposes and analyzes alternative market models for a TMC system -- focusing on market design aspects such as allocation/expiration of tokens, rules governing trading, transaction fees, and regulator intervention -- and develops a methodology to explicitly model the dis-aggregate behavior of individuals within the market. Extensive simulation experiments are conducted within a combined mode and departure time context for the morning commute problem to compare the performance of the alternative designs relative to congestion pricing and a no-control scenario. The simulation experiments employ a day-to-day assignment framework wherein transportation demand is modeled using a logit-mixture model with income effects and supply is modeled using a standard bottleneck model. 

The results indicate that small fixed transaction fees can effectively mitigate undesirable behavior in the market without a significant loss in efficiency (total welfare) whereas proportional transaction fees are less effective both in terms of efficiency and in avoiding undesirable market behavior. Further, an allocation of tokens in continuous time can be beneficial in dealing with non-recurrent events and avoiding concentrated trading activity. In the presence of income effects, despite small fixed transaction fees, the TMC system yields a marginally higher social welfare than congestion pricing while attaining revenue neutrality. Further, it is more robust in the presence of forecasting errors and non-recurrent events due to the adaptiveness of the market.  Finally, as expected, the TMC scheme is more equitable (when revenues from congestion pricing are not redistributed) although it is not guaranteed to be Pareto-improving when tokens are distributed equally.

\end{abstract}


\end{frontmatter}

%\linenumbers

\section{Introduction}
Historically, transportation network inefficiencies and externalities such as congestion and vehicular emissions have been addressed through road pricing, which although used in several cities worldwide, is plagued by issues of inequity and public acceptability \citep{tsekeris2009design,de2011traffic}. An alternative approach to travel demand management that has received increasing attention in the transportation domain in recent years is quantity control -- in particular, tradable mobility credit (TMC) schemes \citep{fan2013tradable,grant2014role,dogterom2017tradable}. 
%with information provision and pricing, despite their \cite{balakrishna2013information, tsekeris2009design}
%known deployment limitations. 
%Recently, quantity control, has been under the spotlight in transportation research, leveraging from successful applications in other economic sectors. 
%Limited supply is in the end a scarcity problem that can be dealt with a price instrument, a quantity instrument or a combination of both, such as tradable mobility credit schemes (TMC). 
Within a TMC system, a regulator provides an initial endowment of mobility credits or tokens to potential travelers. In order to use the road network or transportation system, users need to spend a certain number of tokens (i.e., tariff) that could vary with the attributes or performance of the specific mobility alternative used. The tokens can be bought and sold in a market at a price determined by token demand and supply. 

In principle, TMC schemes are appealing since they offer a means of directly controlling quantity, they are revenue neutral in that there is no transfer of money to the regulator, and they are viewed as being less vertically inequitable than congestion pricing \citep{de2020tradable}. Despite these promises, several important questions remain with regard to the design and functioning of the market within TMC schemes, an aspect critical to the effective operationalization of these schemes. For instance, how should the allocation and expiration of tokens be designed? What rules should govern trading behavior in the market so as to avoid undesirable speculation and trading (see \cite{brands2020tradable} for more on this), and yet ensure efficiency and revenue neutrality? How should the regulator intervene in the market in the presence of special or non-recurrent events? What is the role and impact of transaction fees? Despite the large body of literature on TMCs, issues of market design, market dynamics and market behavior have received relatively little attention although being critical to the successful real-world deployment of a TMC scheme. 

This paper aims to address these issues and contributes to the existing literature in several respects. First, we propose alternative market models (focusing on all aspects of market design including allocation/expiration of credits, rules governing trading, transaction fees, regulator intervention, and price dynamics) for a TMC system and develop a methodology that explicitly models the dis-aggregate behavior of individuals within the market. Second, we conduct extensive simulation experiments within a departure time and mode choice context for the morning commute problem to compare the performance of the alternative designs relative to congestion pricing and a no-control scenario. The simulation experiments employ a day-to-day assignment framework wherein transportation demand is modeled using a logit-mixture model (with income effects) and supply is modeled using a standard bottleneck model. The experiments yield insights into market design and the comparative performance of the TMC system relative to congestion pricing.

The results indicate that small fixed transaction fees can effectively mitigate undesirable behavior in the market without a significant loss in efficiency (total welfare) whereas proportional transaction fees are less effective both in terms of efficiency and in avoiding undesirable market behavior. Further, an allocation of tokens in continuous time provides the TMC system with additional flexibility (compared to a \textit{lump-sum} allocation) and can be beneficial in dealing with non-recurrent events. With regard to the relative performance vis-a-vis  
congestion pricing, the results indicate that the TMC scheme attains a marginally higher social welfare (under income effects and a small fixed transaction fee). Further, the TMC scheme is more robust in the presence of forecasting errors (during the optimization of 
the toll profiles) and by adjusting token allocation, can achieve a higher welfare than congestion pricing when actual demand and supply differs from
the anticipated demand and/or supply. Finally, as expected, the TMC scheme is more equitable (when revenues from congestion pricing are not redistributed) although it is not guaranteed to be Pareto-improving when tokens are distributed equally. 

The paper addresses a growing and imminent need to develop methodologies to realistically model TMCs that are suited for real-world deployments and can help us better understand the performance of these systems – and the impact in particular, of market dynamics. 

The rest of the paper is organized as follows. Section \ref{sec:Sec2} presents a review of the literature and identifies contributions of our paper. Section \ref{sec:Sec3} and \ref{sec:Sec4} propose a market design for TMCs and a framework for modeling market behavior, respectively. Section \ref{sec:Sec5} introduces the simulation model, including demand, supply and day-to-day learning. Section \ref{sec:Sec6} describes findings from extensive simulation experiments and Section \ref{sec:Sec7} provides concluding remarks and directions for further research.

\section{Review of Literature}\label{sec:Sec2}
Although early work on the use of tradable mobility credits (TMCs; also termed TCS or Tradable Credit Schemes in the literature) in transportation date back several years \citep{verhoef1997tradeable,raux2007tradable,goddard1997using}, formulations of the market and network equilibrium for TMCs is more recent, pioneered by the work of \cite{Yang2011} who proposed a user equilibrium variant for a TMC. Their work, along with advancements in technology and the widely recognized limitations of congestion pricing, has spurred interest in TMCs for transportation network management. Extensive reviews may be found in \cite{grant2014role,fan2013tradable,dogterom2017tradable}. We provide a brief summary of existing literature, limiting our attention to that of mobility management (in the context of both entire networks and single bottlenecks) although applications may also be found in parking. 

In the model of \cite{Yang2011}, the regulator distributes a pre-specified number of credits to travelers, charges a link-specific credit tariff and allows trading of credits within a market. They identify conditions under which the network and market equilibrium are unique. Extensions to their model have been proposed to incorporate heterogeneity in the value of time \citep{wang2012tradable} and multiple user classes \citep{zhu2015properties} using variational inequality formulations to establish existence and uniqueness of the equilibrium. \cite{he2013tradable} employed a similar equilibrium approach considering allocations of credits to not just individual travelers, but to transportation firms such as logistics companies and transit agencies; the effect of transaction costs in a TMC scheme with two types of markets (auction-based and negotiated) is considered by \cite{nie2012transaction}. In contrast with the aforementioned TMC schemes, \cite{kockelman2005credit,gulipalli2008credit} proposed a system of credit-based congestion pricing (termed CBCP) where credits are allowances used to pay tolls.

TMC schemes have also been studied in the context of managing congestion at a single bottleneck (or simple two route networks) by achieving peak spreading.   
\cite{nie2013managing} modeled a tradable credit scheme that manages commuters’ travel choices and attempts to persuade commuters to spread their departure times evenly within the rush hour and between alternative routes (see also \cite{nie2015new}) whereas \cite{tian2013tradable} investigated the efficiency of a tradable travel credit scheme for managing bottleneck congestion and modal split in a competitive highway/transit network with heterogeneity. Along related lines, \cite{xiao2013managing} studied a tradable credit system (consisting of a time-varying credit charge at the bottleneck wherein the credits can be traded and the price is determined by a competitive market)
to manage morning commute congestion with both homogeneous and heterogeneous users. More recently, \cite{bao2019regulating} examined the existence of equilibria under tradable credit schemes using different models of dynamic congestion and 
\cite{akamatsu2017tradable} proposed a tradable bottleneck credit scheme where the regulator issues link- and time-specific credits required for passing through a certain link or bottleneck in a pre-specified time period. 
%They develop a model to describe time-dependent flow patterns at equilibrium under a system of tradable bottleneck permits for general networks and show that the equilibrium obtained under this system is efficient in that it minimizes the social transportation cost. 
\cite{liu2022managing} considered distance-based token tariffs in a TMC scheme within a departure-time setting and examined its performance using a  trip-based MFD supply model.

In contrast with the previously described literature that largely focus on variants of the standard user equilibrium under TMC schemes, a related stream of research examines the design of the TMC schemes using bi-level optimization formulations in different contexts \citep{wu2012design,bao2017private,wang2014mpec}. On the other hand, the comparison of efficiency properties of tradable credits and congestion pricing has received relatively lesser attention. \cite{de2018congestion} performed a comparative analysis of the two instruments in a simple static transportation network (see also \cite{seshadri2021congestion} for a within-day dynamic setting) and showed that as long as there is no uncertainty, price and quantity regulation are equivalent as in the regular market case studied by \cite{weitzman1974prices}. In the presence of uncertainty and strongly convex congestion costs, the TMC instrument outperforms the pricing instrument in efficiency terms (see also \cite{de2020tradable} for comparisons in the case with one route and time period under elastic demand). \cite{akamatsu2017tradable} reached similar conclusions (see also \cite{shirmohammadi2013analysis}), demonstrating the equivalence of the tradable permit system and a congestion pricing system when the road manager has perfect information of transportation demands. 
On the behavior side, several stated preference studies have highlighted the importance of key factors from the perspective of behavioral economics and cognitive psychology towards tradable credits \citep{dogterom2017tradable}. 

To the best of our knowledge, \cite{brands2020tradable} is the only study thus far to examine issues of market design for tradable credits. They conducted a lab-in-the-field experiment in a parking context and examine performance of the credit system empirically in terms of several criteria including undesirable speculation, price stability and transaction costs.

%Studies that (1) propose and conceptualize the policy and economics of individual  TMC \cite{fan2013tradable}; (2) formulate a mathematical approach for understanding the user and market equilibrium under different theoretical assumptions\cite{grant2014role}; and (3) empirically investigate the individual behavior under such schemes\cite{dogterom2017tradable} have increasingly being put forward in the research field, generally focusing on road traffic inefficiencies. 

%While many proposals of TMC for environmental, congestion and even social surplus can be traced back to 1997 \cite{verhoef1997tradeable}, the mathematical formulations of market and network equilibrium for TMC operational design and assessment is much more recent, pioneered by the work of Yang and Wang\cite{Yang2011} who proposed a user equilibrium variant for the TMC. 

%Since then, research on the modelling of TMC has focused on the optimal design of TMC to achieve different system-level goals\cite{lahlou2017nash, miralinaghi2018multi}, on the relaxation of different assumptions of the trade market\cite{nie2012transaction}, on the impact of different traveler characteristics and behavior on TMC operations \cite{han2017stochastic, xu2016trip}, on the comparison with pricing mechanisms \cite{de2018congestion} and on identifying the equilibrium and stationary conditions of different theoretical models \cite{BAO2019225, Yang2011, YE2013}. 
 
 In summary, despite the large body of research on TMCs, several gaps remain. First, the modeling of the market has received little attention and almost all the studies employ an equilibrium approach to model the credit market (with the notable exception of \cite{YE2013} who model the price and flow dynamics of a tradable credit scheme). The literature has -- to the best of our knowledge -- thus far not attempted to realistically model the disaggregate behavior of individuals within the market. This would enable the consideration of empirically observed phenomena such as loss aversion, endowment effects, mental accounting, day-to-day learning \citep{dogterom2017tradable}. Second, despite being a critical step towards real-world deployment, design aspects of the credit market have received little attention. In particular, features such as token allocation/expiration, trading, intervention, and transaction fees, and their impact on efficiency and market behavior remain to be studied. Finally, income effects, which impact both efficiency and equity, have received relatively little attention (with the exception of \cite{wu2012design} who consider it in a route choice setting). This paper aims to address these gaps by proposing and analyzing alternative market designs of the TMC system and investigating their performance relative to congestion pricing using realistic models of traveler behavior (with heterogeneity and income effects) and congestion.  
%We first describe market design followed by a discussion of numerical experiments and findings.
%To overcome this shortcoming and to bring TMCs closer to practice, we are developing Trinity, a transportation demand management system of TMC based on the fundamental tenets of prediction, optimization and personalization, deployed via smartphones. 

%\section{Trinity}

%\subsection{Overall architecture}

\section{Market Design}\label{sec:Sec3}
%Traditional road pricing, which charges money for traveling (along certain routes or at certain departure times), can be thought of in a TMC perspective, as a TMC system in which only buying is allowed. Specifically, instead of charging a time-of-day toll (in units of dollars), the regulator charges the same toll but in electronic tokens and maintains a fixed token market price of \$1. Further, the regulator does not distribute any token endowment to travelers and is able to satisfy all buying demand. In order to avoid quantity buildup and market manipulation, travelers can only buy tokens for immediate travel use and any extra tokens after travel will expire immediately. Thus, the toll cost (in dollars) under road pricing is the same as the toll cost under TMC scheme, which is equal to the product of the toll in electronic tokens and the \$1 token price. Such a TMC scheme is clearly as efficient as road pricing in internalizing congestion externalities. However, similar to road pricing, this TMC scheme will be perceived as unfair or as just another flat tax. Under our proposed TMC scheme, revenue neutrality can potentially be achieved by providing a token endowment and allowing 1) token selling and 2) token price adjustment through a market. 
%Only allowing token selling without price adjustment cannot achieve revenue neutrality if travel demand fluctuates across days.

In this section, we focus on market design for a tradable credit scheme. Within the TMC scheme, the regulator provides a token endowment to all potential travelers. Although our design is generic, the application we explore involves a daily commute context where in order to use the network at a particular time-of-day (e.g., for a given departure time interval), travelers have to pay a pre-specified toll in tokens that does not vary from day to day. In other words, the toll in tokens is dynamic and varies by time-of-day, but is fixed across days. The rationale for this assumption is that modifying the toll in tokens from day to day would involve communicating the tariff or toll structure on a daily basis, which is complicated, particularly in large general networks (for instance, the electronic road pricing or ERP scheme in Singapore includes dynamic tolls, which are revised only every three months or longer). The market design we propose can be used for other applications  including parking management.

The regulator operates a market where tokens can be bought and sold at a prevailing market price and may also levy pre-specified transaction fees for buying and selling. The market price of the token varies across days and is adjusted by the regulator to achieve revenue neutrality, considering the demand and supply of tokens in the market. All transactions take place between an individual and the regulator directly, who guarantees all buying and selling requests. This central market with a regulator who acts as a price setting intermediary is similar to the \textit{virtual bank} in \cite{brands2020tradable}, who note that such a market can significantly reduce transaction costs (associated with information acquisition, negotiation, finding a potential buyer or seller etc.) compared to designs that involve consumer to consumer trading (and over existing designs such as Dutch and English auctions, sealed-bid auctions and Vickrey auction markets). The regulator may also intervene in the token market within the day by controlling token market price, token allocation, and transaction fees to manage non-recurrent events.

With regard to the token allocation or endowment, we adopt a ‘continuous time’ approach wherein tokens are acquired (provided by the regulator) at a certain rate over the entire day and each token has a lifetime (i.e., it expires after a certain period specified by the regulator). The expiration of tokens will avoid undesirable consequences of the TMC system that can compromise public acceptability such as speculative behavior and hedging in the market. 
%Although market price is only adjusted from day to day in this study, 
The `continuous' allocation avoids concentrated trading activities and excessive trading near a boundary (a time period when a large amount of tokens expire at the same time, such as for instance, in a \textit{lump-sum} allocation). It also provides more degrees of freedom for the regulator to intervene than that of a `lump sum' allocation which distributes tokens at the beginning of each day. A comparison between the two allocation approaches will be performed through numerical experiments presented in Section \ref{sec:robustness}.

As a result, each individual acquires tokens at a constant rate $r$ over the entire day (credited into a \textit{wallet}) and each token has a lifetime $L$ to avoid speculation and hoarding. Let $x_n^d(t)$ denote traveler $n$'s token account (or wallet) balance at time $t$ on day $d$. A full wallet state indicates that the number of tokens in the wallet has reached a maximum ($L r$), and in the absence of travelling or selling, does not change since the acquisition of new tokens is balanced by an expiry of old tokens. Thus, a full wallet implies that the oldest token in an individual's account has an age of $L$. In contrast, when the account is not in a full wallet state, it increases by an amount $r \Delta_t$ in a unit time interval $\Delta_t$.

Several additional assumptions regarding market design are noteworthy -- these serve to avoid quantity buildup and market manipulation. First, travelers can only buy tokens from the regulator at the time of traveling for immediate use, i.e., only if they wish to travel and are short of tokens. Second, when they sell tokens to the regulator, they have to sell all tokens in their wallet. Third, buying and selling cannot happen at the same time, i.e. travelers can sell all tokens anytime except at the time of buying. Note that the second assumption differs from the design of \cite{brands2020tradable}, who assume that tokens can be traded per piece, and implications of this assumption warrant more investigation, particularly when the market prices vary within-day. Since a large part of our experiments do not involve within-day dynamic prices and given that it considerably simplifies the modeling of selling behavior, we defer the relaxation of this assumption to future research.

\subsection{Account Evolution}
\label{sec:account-evolution}
Let $T(t)$ denote the toll in tokens to travel at time $t$, $\tilde{t}_n^d$ represent the departure time of traveler $n$ on day $d$ and $D$ represent the duration of one day. Note that in the simulation framework (Section \ref{sec:Sec5}), time will be discretized into intervals of a specified size; for now, we treat it as continuous. Let $r$ denote the allocation rate, $L$ denote token lifetime, and $x_n^d(t)$ denote traveler $n$'s token account balance at time $t$ on day $d$. At time $t$ on day $d$, traveler $n$ can perform one and only one of the following actions: 
%in order to avoid quantity buildup and market manipulation: 

  \begin{enumerate}
        \item Perform a trip if $t=\tilde{t}_n^d$. 
        \begin{itemize}
            \item  If $x_n^d (t)\ge T(t)$, she consumes $T(t)$ tokens. Her account balance at $t+ \Delta_t$,  $x_n^d(t+ \Delta_t)$, can be written as:
    	    \begin{align}
    	         x_n^d(t+ \Delta_t) = \textnormal{min}\left(x_n^d(t) - T (t) + r  \Delta_t,Lr\right),
    	    \end{align}
    	    where the cap $Lr$ ensures that tokens with life greater than $L$ expire.
    	    \item  If $x_n^d (t) < T(t)$, she needs to buy $T(t)-x_n^d(t)$ tokens. Her account balance $x_n^d(t+ \Delta_t)$ becomes:
    	    \begin{align}
    	         x_n^d(t+ \Delta_t) = r  \Delta_t
    	    \end{align}
    	    since all of $x_n^d(t)$ and the newly bought tokens are used to travel.
        \end{itemize}
       
	    \item Does nothing. Her account balance $x_n^d(t+ \Delta_t)$ becomes:
	    \begin{align}
	         x_n^d(t+ \Delta_t)= \textnormal{min}\left(x_n^d(t) + r  \Delta_t,Lr\right)
	    \end{align}
	     \item Sells all tokens $x_n^d(t)$. Her account balance becomes:
	     \begin{align}
	         x_n^d(t+ \Delta_t)=r  \Delta_t
	    \end{align}
    \end{enumerate}

\subsection{Buying and Selling}
\label{sec:buying-and-selling}

The token market price $p^d$ is fixed within day $d$ (in the absence of non-recurrent events) and is only adjusted day to day. Details of the price adjustment process are discussed in Section \ref{sec:price-adjustment}. We assume that the regulator levies a two-part (fixed and proportional) transaction fee for both buying and selling transactions. Let $F_S^P$, $F_B^P$ ($F_S^P,F_B^P\ge0$) denote the proportional part of selling and buying transaction fees (this component of the transaction fee is proportional to the amount of the trade), and $F_S^F$, $F_B^F$ ($F_S^F,F_B^F\ge0$) denote the fixed part of selling and buying transaction fees. The effect of transaction fees on market behavior and efficiency will be examined in Section \ref{sec:Overall_market}. 

% We also consider an additional feature of the market design wherein the price of a token (during selling) depends on both the prevailing market price and the age of the token. Specifically, we assume that the selling price of tokens decays linearly with the age of the token. 
% Let $t_a$ denote acquisition time of the oldest token in the wallet, $t_a^i$ denote acquisition time of token $i$, and $t$ denote current time. The selling price of token $i$ on day $d$ at a time $t$, denoted as $\dot{p}(t,t_a^i)$ is defined as, 
% \begin{align}
%     \dot{p}(t,t_a^i) = p^d(1-c_s)(1- \nu \frac{t-t_a^i}{D}) = p^d_s(1-\nu\frac{t-t_a^i}{D})
% \end{align}

% where $p^d$ represents token market price on day $d$, $c_s$ represents proportional part of the selling transaction cost, $\nu$ represents a decay factor and $p_s^d=p^d(1-c_s)$. If $\nu=0$, the selling price of all tokens is equal to the market price $p^d$; if $\nu=1$, the selling price of token $i$ at an age of one day ($t-t_a^i=D$) is equal to 0 (its value decays linearly to 0 at expiration). This may offer a potential means to further mitigate speculation and market manipulation although it may lead to losses in efficiency.  
% %as tokens expire in order to account time values of tokens and avoid quantity buildup but it may lead to efficiency loss. 
% The effect of decaying selling prices will be investigated through numerical experiments presented in Section \ref{sec:results}. 

The revenue obtained from selling $y$ tokens ($y \le Lr$) with transaction fees on day $d$ at time $t$ can be written as,
\begin{align} \label{sellingRev}
   S(y) = y p_s^d-F_S^F,
\end{align}

where $p_s^d=p^d(1-F_S^P)$ is the token market price adjusted for the proportional selling transaction fee. Transaction fees and price are not expressed in function inputs for conciseness. 
 
%  In the case when we have the linearly decaying selling price, the revenue obtained from selling $y$ tokens is given by (see Appendix B for a proof),
 
%  \begin{align}
% S^d(y,t) = y p^d_s-\nu \frac{1}{2}\frac{p^d_s y^2}{Dr} -TC_s 
% \end{align}

The cost of buying $y$ tokens ($y \le Lr$) with transaction fees at time $t$ on day $d$ can be written as,
\begin{align} \label{buyingCost}
    B(y)=yp_b^d+F_B^F,
\end{align}

where $p_b^d=p^d(1+F_B^P)$ is the token market price adjusted for the proportional buying transaction fee.

\subsection{Price Adjustment}
\label{sec:price-adjustment}
The marketplace dictates the token price $p^d$ on day $d$, which is adjusted according to an apriori rule established by the regulator to achieve revenue neutrality. The price $p^d$ is modified daily with a deterministic rule considering the regulator revenue $K^{d-1}$ (net revenue from all buying and selling transactions of users) from the previous day as follows 

\begin{align}
p^d =
        \begin{cases}
                p^{d-1} & \text{$K^{d-1} \in [-K_t,K_t]$} \\
                p^{d-1}+\Delta p & \text{$K^{d-1}<-K_t$} \\
                p^{d-1}-\Delta p & \text{$K^{d-1}>K_t$}, 
        \end{cases}
\end{align}

where $\Delta p$ currently is a constant parameter representing the price change. $K_t$ is a constant parameter representing a regulator revenue threshold to adjust the price and ensures that price will not fluctuate for small regulator revenues close to zero. Price is ensured to be positive and below a certain cap $p_{m}$ as follows:
\begin{align}
    p^d = \max\left(0,\min\left(p^d,p_{m}\right)\right)
\end{align}

Although token price is typically constant within a day, the regulator may intervene in the market to adjust the market price during a day in the presence of unusual events. For example, if road capacity drops because of an accident, or if demand increases due to a concert, the regulator can intervene, increasing token price in a certain period to discourage travel and reduce congestion. Numerical experiments are conducted to study this in Section \ref{sec:robustness}.

Market elements discussed in this section are summarized in Table \ref{tab:marketele}.
\begin{table}[h!]
   
  \begin{center}
    \caption{Market elements for the tradable mobility credits system}
     \vspace{2ex}
    \label{tab:marketele}
    \begin{tabular}{|l|l|l|} % <-- Alignments: 1st column left, 2nd middle and 3rd right, with vertical lines in between
    \hline
      \textbf{Elements} & \textbf{Design} & \textbf{Motivation} \\
      \hline
      \multirow{2}{4em}{Allocation} 
        & Lump-sum & Simple; automated trading\\\cline{2-3}
        & Continuous & \makecell[l]{Avoid concentrated trading; \\additional control}\\
        \hline
      Expiration & Lifetime & Avoid quantity buildup \\
      \hline
       \multirow{2}{7em}{Transaction fee} & Proportional &\multirow{2}{15em}{Avoid undesirable market behavior (e.g. frequent selling)}  \\\cline{2-2}
       &Fixed &  \\
       \hline
       Price adjustment & \makecell[l]{Day to day \\ constant adjustment} & Balance demand and supply  \\
       \hline
       \multicolumn{3}{|l|}{Market rules governing trading} \\
       \hline
    \end{tabular}
  \end{center}
\end{table}

\section{Market Behavior}\label{sec:Sec4}

As buying behavior is governed by the previously specified buying rule, this section primarily discusses individual selling behavior. It is assumed that the individual selling decision and mobility decision (departure time and mode) are inter-dependent. In other words, selling decisions are made conditional on a departure time/mode chosen at the beginning of the day, which in turn is based on a forecast of the account balance over the entire day. This forecast is based on historical travel and selling decisions of the user and his/her past experience (described in more detail in Section \ref{sec:d2dl}). 
We note that one could think of this selling behavior as a strategy or an automated operation performed through (or programmed into) for example a smartphone application, since in practice, it may be onerous to expect users to constantly make these selling decisions `manually'. In this respect, one may also view it as an element in the overall design of the tradable credit scheme. However, note that our modeling framework does not preclude the use of an actual behavioral model of selling in the market (in place of the selling strategy we formulate next), which would require the collection of empirical data; we defer this to future research.  

From the perspective of simply maximizing profit (which is a reasonable selling strategy), the decision to sell can be formulated as a dynamic programming or optimal control problem, where the optimal selling strategy is characterized by Bellman's equation \citep{kirk2004optimal}. However, this is complicated, both from the standpoint of computational complexity and system design, and instead, we derive a simpler heuristic approach to characterize an individual's selling strategy.

At time $t$ on day $d$, assume traveler $n$ has an upcoming planned trip at a time denoted by $\tilde{t}_n$, where $\tilde{t}_n=\tilde{t}_n^d$ if $t\le \tilde{t}_n^d$, and $\tilde{t}_n=\tilde{t}_n^{(d+1)}$, if $t> \tilde{t}_n^d$. Given the next trip, a conditional profit function $ \Pi_n^d(t)$, which represents the profit obtained by selling all tokens at time $t$ (with no further selling until the next departure $\tilde{t}_n$) can be written as follows,
 \begin{align} \label{eq:Pi}
 \Pi_n^d(t) &= S\left(x^d_n(t)\right)-\mathbb{I}\left(T(\tilde{t}_n)\ge \hat{x}_n(\tilde{t}_n)\right) \cdot B\left(T(\tilde{t}_n)-\hat{x}_n(\tilde{t}_n)\right) \\\notag
 & = x^d_n(t) p_s^d-F_S^F-\mathbb{I}\left(T(\tilde{t}_n)\ge\hat{x}_n(\tilde{t}_n)\right) \cdot
 \left(\left(T(\tilde{t}_n)-\hat{x}_n(\tilde{t}_n)\right)p_b^d+F_B^F\right)
\end{align}

where $\hat{x}_n(\tilde{t}_n)$ represents the expected account balance at the time of the next trip $\tilde{t}_n$. Since it is assumed there will be no further selling until the next departure $\tilde{t}_n$, it can be written as,
\begin{align}
   \hat{x}_n(\tilde{t}_n)= \min\left[(\tilde{t}_n-t)r,Lr\right]
\end{align}

For other notation in the conditional profit function $\Pi_n^d(t)$, $T(\tilde{t}_n)$ represents the toll in tokens of traveling at departure time $\tilde{t}_n$. A buying cost is incurred only if the toll at $\tilde{t}_n$ is greater than or equal to traveler $n$'s expected account balance (i.e. $T(\tilde{t}_n)\ge \hat{x}_n(\tilde{t}_n)$), which is represented by the indicator function. Note that in defining the profit function above, we have made the critical assumption that if a decision to sell at the current time is made, no further selling will occur until the next trip. This simplification allows us to derive an optimal selling strategy analytically and is partly justifiable given that we also assume that during selling, an individual needs to sell all tokens in her wallet, and that prices do not vary within-day. However, observe that the selling strategy we derive, when applied, involves a decision made at every time point $t$, implying that it does not preclude the possibility of an individual making multiple selling decisions in the time period until the next trip if this is beneficial.

Under our assumptions, at time $t$ on day $d$, traveler $n$ will consider selling tokens only if the profit value is positive, i.e., $\Pi_n^d(t)>0$. If the profit value is positive, she may still decide to wait if the derivative of the profit function is positive (meaning that the profit is expected to increase if she defers the decision to sell). Therefore, the selling strategy depends on both the profit function and its derivative, which can be analyzed from the following three cases:

% Without considering transaction fees, the conditional profit function can be simplified and written as
%  \begin{align} \label{eq:PiwoTF}
%  \Pi_n^d(t) &= S(x^d_n(t))-\mathbb{I}(T(\tilde{t}_n)\ge\hat{x}_n(\tilde{t}_n)) \cdot B(T(\tilde{t}_n)-\hat{x}_n(\tilde{t}_n)) \\\notag
%  & = x^d_n(t) p^d-\mathbb{I}(T(\tilde{t}_n)\ge\hat{x}_n(\tilde{t}_n)) \cdot
%  ((T(\tilde{t}_n)-\hat{x}_n(\tilde{t}_n))p^d)
% \end{align}

%where its value and its derivative can be analyzed from two cases:

\begin{enumerate}
    \item $T(\tilde{t}_n)<   \hat{x}_n(\tilde{t}_n)$ (no tokens need to be bought for the next trip)
    
    The profit function $\Pi_n^d(t) $ can be written as
    \begin{align}
        \Pi_n^d(t) &= x^d_n(t) p_s^d-F_S^F
    \end{align}
    
    and the derivative can be written as 
      \begin{align}
            \frac{d \Pi_n^d(t)}{d t} =
                \begin{cases}
                    0 &\text{$x_n^d(t)=Lr$}\\
                    rp_s^d &\text{otherwise},
                \end{cases}
        \end{align}
    which implies that profit will continue to increase until a full wallet is reached. It does not make sense to wait longer at a full wallet because newly acquired tokens simply replace expired tokens. Hence, selling should be at a full wallet. 
    
    However, it is worth noting that, without fixed transaction fees, the selling revenue at full wallet is the same as that obtained from selling every time when one receives new tokens. In fact, as long as one avoids token expiration, any selling strategy is equivalent in the absence of fixed transaction fees. It is fixed transaction fees that prevent frequent selling.
    
    \item $T(\tilde{t}_n)> \hat{x}_n(\tilde{t}_n)$ (tokens need to be bought for the next trip)
    
    The profit function $\Pi_n^d(t)$ can be written as 
    \begin{align} 
 \Pi_n^d(t)
 & = x^d_n(t) p_s^d-F_S^F-
 \left(\left(T(\tilde{t}_n)-\hat{x}_n(\tilde{t}_n)\right)p_b^d+F_B^F\right)
\end{align}
    
    and its derivative can be written as
    \begin{align}
            \frac{d \Pi_n^d(t)}{d t} =
                \begin{cases}
                    -rp_b^d &  \text{$x_n^d(t)=Lr$ }\\
                    rp_s^d-rp_b^d & \text{otherwise},
                \end{cases}
        \end{align}
    which is always negative since $p_s^d < p_b^d$ given $F_B^P$ or $F_S^P$ is greater than 0. This implies that profit obtained from waiting and selling at any time in the future (until the next trip) is guaranteed to be less than the profit from selling now. Hence, she should sell now if the profit is positive.
    
    Without transaction fees, the profit function $\Pi_n^d(t)$ can be written as 
    \begin{align} \label{eq:pibuy}
 \Pi_n^d(t)
 & = x^d_n(t) p^d-
 \left(T(\tilde{t}_n)-\hat{x}_n(\tilde{t}_n)\right)p^d
\end{align}
    
    and its derivative can be written as 
    \begin{align}
            \frac{d \Pi_n^d(t)}{d t} =
                \begin{cases}
                    -rp^d &  \text{$x_n^d(t)=Lr$ }\\
                    0 & \text{otherwise},
                \end{cases}
        \end{align}
    which means that as long as account balance is not full, it does not matter whether one sells now or later. However, once we introduce fixed transaction fees, it is better to sell at a full wallet to minimize the number of transactions. With additional proportional transaction fees, it is better to sell immediately and not worth waiting anymore as the derivative is always negative. 
    
    \item $T(\tilde{t}_n) = \hat{x}_n(\tilde{t}_n)$ (the expected account balance is just enough to cover the toll of the next trip)
    
    The profit function $\Pi_n^d(t) $ can be written as
    \begin{align}
        \Pi_n^d(t) &= x^d_n(t) p_s^d-F_S^F
    \end{align}
    
    but its derivative does not exist because the conditional profit function is discontinuous at $t$ due to the transaction fees of buying. To avoid any buying transaction fees (either fixed or proportional), it is optimal to sell immediately if profit $\Pi_n^d(t)$ is positive. Without transaction fees, similarly, it does not matter whether one sells now or later as long as token expiration is avoided.
\end{enumerate}

%  In other words, she will sell now only if the derivative of the profit function, computed as follows, is non-positive (otherwise, she waits):

Based on the analysis in this section, the effect of fixed transaction fees is to prevent multiple transactions while the effect of proportional transaction fees is to make one sell as soon as possible when the conditional profit is positive (if tokens need to be bought for the next trip). The proportional transaction fee is not preferable because it does not prevent frequent selling but instead prevents selling at a full wallet. Numerical experiments in Section \ref{sec:Sec6} will provide further justification for the use of only a fixed transaction fee from an efficiency perspective.

The selling strategy for an individual $n$ at any time $t$ on day $d$ considering positive transaction fees is summarized in Algorithm \ref{Selling}. 

\begin{algorithm}[!ht]
\SetAlgoLined
% \KwResult{Write here the result }
\SetKwInOut{Input}{input}
\Input{$d,t,n,p^d,\tilde{t}_n,x_n^d(t),L,r, F_S^P, F_B^P,F_S^F, F_B^F$}
%  initialization\;
At time $t$ on day $d$, calculate $\Pi_n^d(t)$\;
and expected account balance  $\hat{x}_n(\tilde{t}_n)= \min\left[(\tilde{t}_n-t)r,Lr\right]$\;
\eIf{$ \Pi_n^d(t)>0$}{
    \uIf{$T(\tilde{t}_n)\ge \hat{x}_n(\tilde{t}_n)$}{
        Sell now\;
    }
    \Else{
        \eIf{$x(t)=Lr$}{
            Sell now;
        }{
            Do nothing;
        }
    }
}{
    Do nothing;
}
 \caption{Selling Rule}\label{Selling}
\end{algorithm}

\section{Simulation Framework}\label{sec:Sec5}
\begin{figure}[!h]
    \centering
    \includegraphics[width=0.9\linewidth]{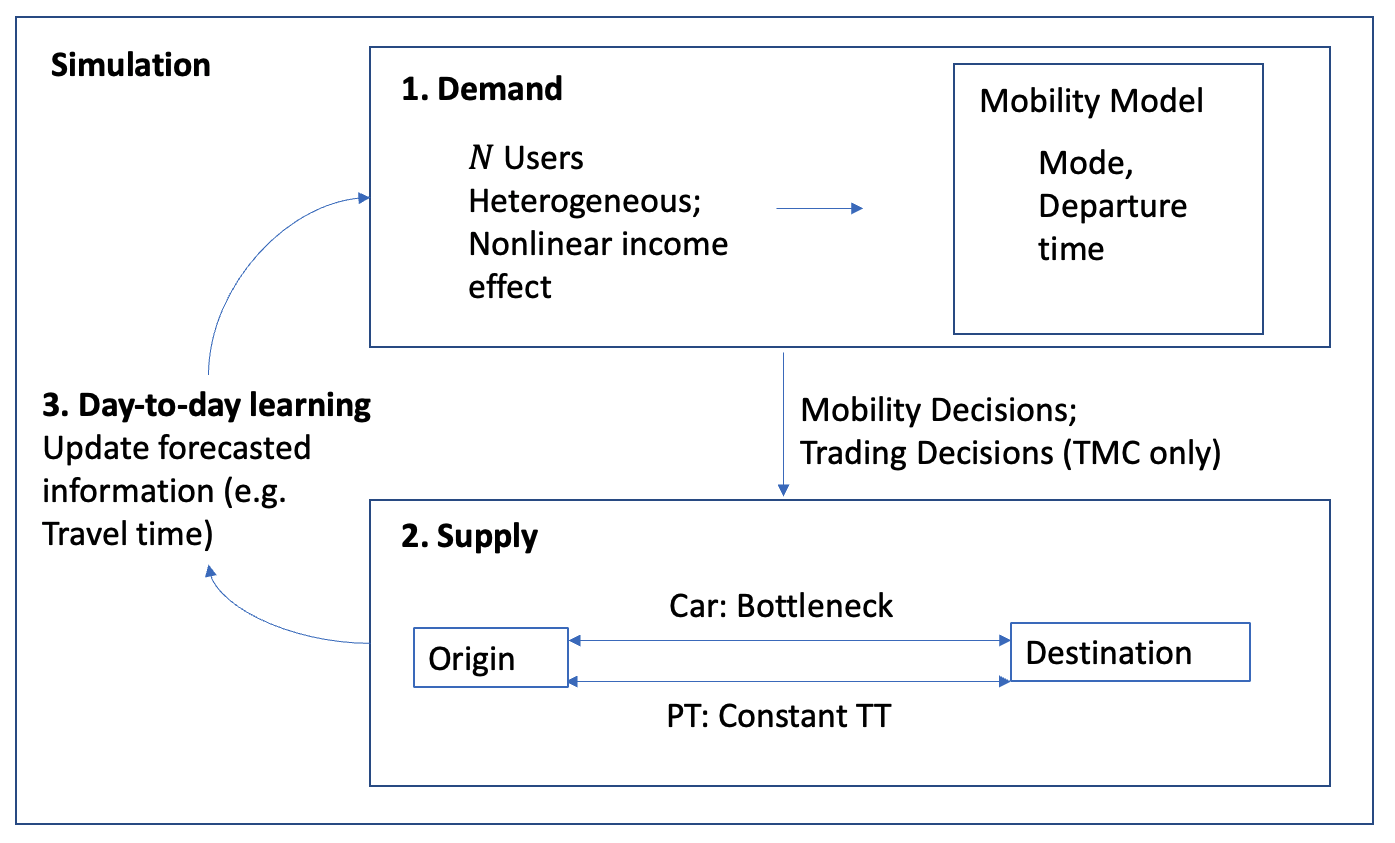}
    \caption{Simulation Framework}
\label{fig:sim}
\end{figure}

This section describes the modeling and simulation framework for evaluating the performance of the designed instruments including a no-toll or no-control benchmark, referred to as NT, congestion pricing, referred to as CP, and the tradable mobility credit scheme termed TMC. The overall simulation framework is shown in Figure \ref{fig:sim}. 

$N$ travelers perform a daily commute between a single origin-destination pair. For the sake of simplicity, each traveler performs a single morning trip and a single evening trip. Only their morning commute trip will be explicitly simulated and their evening trip is assumed to be a mirror of the morning trip. 

At the beginning of each day, every traveler uses forecasted information of travel times, schedule delays and their account balance over the entire day to make a \textit{pre-day mobility decision}, which is the combination of a choice of mode (between car and public transit, hereafter PT) and departure time (over a individual set of departure time choices) for their morning commute trip. Travelers who choose to drive may be subject to a time-of-day toll. For TMC, the time-of-day toll profile is in units of tokens. Note that mobility credits can only be used for toll road payment. The individual mobility decision is modeled using a logit mixture model allowing for heterogeneity and non-linear income effects.

Next, the mobility decisions along with trading decisions -- which occur over the entire day (i.e., are within-day) -- are simulated on a simple network connected by a single driving path and an alternative public transit (PT) line. Congestion (for driving) is modeled by a point queue model (bottleneck of finite capacity), in which a queue develops once flow exceeds capacity. Travel time of PT is assumed to be constant.

Travelers' day-to-day learning is modeled through an exponential smoothing filter to update forecasts of travel time and account balance over the day. The day-to-day framework in Figure \ref{fig:sim} is used to simulate the evolution of the system state (departure flows, travel times) until a measure of convergence has been reached. The performance measures (overall welfare, distribution of user benefits, congestion, and mode shares) at convergence are used to evaluate the different instruments.

In the following sections, we first describe the models of demand, supply and day-to-day learning, termed the \textit{system model}, in more detail. Next, we discuss social welfare computation and the simulation-based toll optimization problem (to determine optimal tolls) for the different instruments.  Relevant notation is summarized in Table \ref{tab:notation}.

\begin{table}[!]
  \begin{center}
    \caption{Notation}
     %\vspace{1ex}
    \label{tab:notation}
    \renewcommand{\arraystretch}{0.9}
    \begin{tabular}{l|l} % <-- Alignments: 1st column left, 2nd middle and 3rd right, with vertical lines in between
     \hline
     \textbf{Variables} & \textbf{Description}\\
      \hline
      $h$ & Departure time interval \\
      $\bar{t}$ & Simulation time step\\
      $d$ & Day $d$ \\
      $t_h$ & Start time of interval $h$\\
      $\Delta_h$ & Duration of departure time interval\\
      $\Delta_t$ & Duration of simulation time step \\
      $\Delta_a$ & Size of desired arrival window \\
      $n$ & Individual $n$ \\
      $\alpha_n $ &  Value of time of individual $n$ \\ 
      $\beta_{En}/\beta_{Ln} $ & Value of schedule delay early/late of individual $n$\\ 
      $\lambda$ & Coefficient of nonlinear income effect\\
      $\gamma$ &  Nonlinear income effect adjustment parameter\\
      $\mu_n$ & Random component scale parameter of individual $n$ \\
      $\epsilon_{in}$ & Random utility component for mobility decision $i$ of individual $n$ \\
      $I_n$ & Disposable income of individual $n$ \\
      $H_n$ & Departure time choice set of individual $n$ \\
      $M_n$ & Mode choice set of individual $n$ \\
      $\hat{t}_n$ & Desired arrival time of individual $n$\\
      $\eta$ & Departure time window size parameter\\
      $p$ &  Market price\\
      $\mathrm{T}^{j}(h)$  & Toll of instrument $j$ in $h$\\
      $\tilde{\tau}_i$ & Forecasted travel time of choice $i$ \\
      $\tilde{c}_{in}$ &  Expected cost for mobility decision $i$\\
      $x_n^d(t)$ & Account balance of individual $n$ at time $t$ \\
      $L$ & Token lifetime\\
      $r$ & Token allocation rate\\
      $t_f$ & Free flow travel time \\
      $t_v(t)$ & Delay in queue at $t$\\
      $Q(t)$ & Number of drivers in queue at $t$ \\ 
      $\theta_\tau/\theta_{t}$ & Weights on previous day's forecasts \\
      $(\cdot)^{j}$ & Variable associated with instrument $j$ \\
    \hline
    \end{tabular}
  \end{center}
\end{table}

\subsection{System Model}\label{sec:SysModel}
As noted previously, the setting we consider involves $N$ users traveling between a single origin-destination pair connected by a path containing a bottleneck of finite capacity and a PT line. Users wish to arrive at the destination within a certain ``preferred arrival time window'' in the morning, and can choose between PT and car. If they decide to drive, they can adjust their departure times to avoid congestion (similar to the model in \cite{ben1984dynamic}, which is a dynamic extension of \cite{de1983individual}). The system is modeled using a stochastic process approach that can be viewed as a simplification of the model in \cite{cascetta1991day}, who consider the stochastic assignment problem in general networks. Day to day adjustment is modeled using suitable learning and forecasting filters, within-day departure time decisions and mode choices are modeled using a logit-mixture model, and supply is modeled using a point queue model. We refer to \cite{cantarella1995dynamic} and  \cite{watling1999stability} for a nuanced discussion of terminology and a detailed description of deterministic and stochastic process models. % (with probabilistic assignment or a probabilistic model for users’ choice behavior). %They propose conditions for existence and uniqueness of fixed-point attractors of the deterministic process, which extend results for the traditional user and stochastic user equilibrium.  In case of the stochastic process, conditions for regularity are proposed which ensure existence and uniqueness of a stationary distribution of network states. 
%It is noted that the model of \cite{ben1984dynamic} may be viewed as a deterministic process model with probabilistic assignment. 

% The deterministic process approach (with a deterministic assignment model of user choices) is discussed at length in \cite{friesz1994day} (see also \cite{huang2002modeling}, \cite{friesz2011approximate} and \cite{xiao2016day} for the dynamic user equilibrium problem in different settings).  Finally, day-to-day (or inter-periodic) traffic assignment models are also discussed in \cite{watling2003dynamics}. Further, \cite{watling1999stability} provides a taxonomy of day-to-day traffic assignment models, which are classified into four types based on whether the dynamic process is deterministic or stochastic, and continuous or discrete. 

The mobility demand model, network model, and demand-supply interactions are discussed in detail next.

\subsubsection{Demand Model}
The demand model (preday mobility decision) is a combined model of departure time and mode choice. Unless otherwise specified, the discussion in this subsection pertains to a specific day $d$ (we omit $d$ in all quantities for notational convenience). The day is discretized into  $h = 1 \ldots H$ time intervals of size $\Delta_h$ (let the set of all time intervals in the day be denoted by $\mathcal{H} = \{ 1, \ldots, h, \ldots, H \})$, and it is assumed that each individual $n$ has a preferred/desired arrival time $\hat{t}_n$ (more specifically, users are assumed to wish to arrive within a time window of size $2\Delta_a$ centered around $\hat{t}_n$; this is discussed in more detail later). The day is also discretized into smaller time intervals of size $\bar{t}=1 \ldots \bar{T}$ of size $\Delta_t$, which is the resolution of the supply model and trading (selling) decisions.  

The choice set of mode for individual $n$ is defined as $M_n = \{C,PT\}$, where $C$ represents car and $PT$ represents transit. The choice set of feasible departure time intervals $H_n \subset \mathcal{H}$ is individual-specific and defined as $H_n = \{\tilde{t}_{0n}-\eta\Delta_h,\tilde{t}_{0n}-(\eta-1)\Delta_h ,\dots,\tilde{t}_{0n}+\eta\Delta_h\}$, where $\eta$ is a parameter, and $\tilde{t}_{0n}$ represents the initial departure time interval on day 0, which is computed based on the preferred arrival time $\hat{t}_n$ and the free flow travel time. Thus, the departure time choice set $H_n$ consists of $2\eta$ time intervals of size $\Delta_h$ centered around the preferred departure time interval on day 0, $\tilde{t}_{0n}$. 
Note that because we model income effects, the individual departure time choice set is also subject to a budget constraint (i.e., an individual cannot choose a departure time that is not affordable). Thus, we define the set of feasible departure time intervals under instrument $j$ ($j = NT, P, M$ for the No Toll scenario, congestion pricing, and the tradable mobility credit scheme respectively), 
as $H_n^j \subseteq H_n$. Under the No Toll scenario, $H_n^{NT} = H_n$. We will revisit the budget constraint later when discussing income effects. 
Let $i=(m,h)$ represent an individual's mobility decision as a combination of mode and departure time choice ($i\in\{m,h|m\in M_n, h\in H_n^{j}\}$).

Each individual is assumed to be rational and wishes to maximize her money-metric utility from the choice situation. The utility of the mobility decision $i$ for individual $n$ is denoted by $U_{in}$, which consists of two parts: a systematic utility $V_{in}$ which is a function of observable variables and a random utility component $\epsilon_{in}$ that represents the analyst’s imperfect knowledge. $\epsilon_{in}$ is assumed to follow an i.i.d. extreme value distribution with zero mean and individual specific scale parameter $\mu_n$. It is also assumed that the individual random error component is perfectly correlated across days and across instruments (i.e. remains the same before the `change' and after the `change') assuming before and after periods are not too far apart (e.g., \cite{mcfadden2001economic,de2005switching}). This assumption can be relaxed in future work (see for example \cite{delle2013transition,zhao2008welfare}).

The systematic money-metric utility for individual $n$ departing in time interval $h$ by car under instrument $j$ is denoted by  $V_{in} (\boldsymbol{\boldsymbol{\tilde{\phi}}}_i^j)$, where $i\in\{m=C,h| h\in H_n^{j}\}$. $\boldsymbol{\tilde{\phi}}_i^j$ is a vector of forecasted information in the systematic utility that affects the choice of departure time interval for driving and consists of five components. The first is forecasted/expected travel time $\tilde{\tau}^{j}_i$, which determines the expected schedule delay early (second component) and schedule delay late (third component). The fourth component is expected cost $\tilde{c}^j_{in}$ which is explained in more detail next. The last component is remaining income, which is equal to the disposable income for transportation $I_n$ minus expected cost $\tilde{c}^j_{in}$.

The marginal utility of an additional unit of travel time for individual $n$ is denoted by $\alpha_n$. For simplicity, we assume travelers have common knowledge of forecasted travel times (more on this in section \ref{sec:d2dl}). The desired arrival time window for individual $n$ is defined as $[\hat{t}_n-\Delta_a, \hat{t}_n+\Delta_a]$, where $\hat{t}_n$ represents the center of the period and $\Delta_a$ represents arrival flexibility. If she arrives outside of the desired time period, she incurs a schedule delay. The marginal utility of an additional unit of schedule delay early is $\beta_{En}$ and an additional unit of schedule delay late is $\beta_{Ln}$, where $\beta_{En}\le \alpha_n \le \beta_{Ln}$ from empirical evidence (e.g., \cite{small1982scheduling}).

The expected cost $\tilde{c}^j_{in}$ warrants additional discussion. Under the No Toll (NT) scenario, it is equal to the operational cost $c_f$ (fuel cost). Under pricing ($j=P$), it is equal to the toll in dollars charged for departing in time interval $h$, $T^{P}(h)$, plus the operational cost $c_f$, which can be written as
\begin{align}
    \tilde{c}^{P}_{in} = T^{P}(h) + c_f
\end{align}

Under the TMC ($j=M$) scheme, it depends on an individual's expected opportunity cost of tokens $\tilde{R}_{in}$ (which can be negative if one has a net revenue from selling tokens) plus the operation cost $c_f$ as follows:
\begin{align}
    \tilde{c}^{M}_{in} = \tilde{R}_{in} + c_f
\end{align}

Recall that the selling revenue of $y$ tokens with transaction fees ($F_S^F,F_S^P$) and token price ($p$) can be written as (selling revenue function), 
\begin{align} 
   S(y) = y p\left(1-F_S^P\right)-F_S^F
\end{align}

and similarly, the buying cost of $y$ tokens can be written as (buying cost function),
\begin{align} 
   B(y)=y p\left(1+F_B^P\right)+F_B^F
\end{align}

Let $t_h$ be the start time of interval $h$, $\tilde{x}_n(t_h)$ be the expected account balance at time $t_h$, the beginning time of the time interval $h$. If a traveler does not need to pay any toll, she can sell the entire day's token allocation completely. Hence, the opportunity cost (or negative opportunity benefit) is equal to the negative of selling revenue of the entire day's allocation, $-S(Lr)$.

If a traveler needs to pay a toll $T^{M}(h)$ in $h$ but the expected account balance $\tilde{x}_n(t_h) $ is greater or equal to $T^{M}(h)$ (no buying), her opportunity cost is equal to the negative of selling revenue of the one-day allocation $Lr$ minus the toll in tokens $T^{M}(h)$, which can be written as
\begin{align}
       \tilde{R}_{in} = -S\left(Lr-T^{M}(h)\right)
\end{align}

However, if she does not have enough account balance to cover the toll $T^{M}(h)$, she has to buy additional tokens equal to $T^{M}(h)-\tilde{x}_n(t_h)$ in order to travel in $h$. The amount of tokens she can sell for profit is equal to the one-day allocation $Lr$ minus her expected account balance $\tilde{x}_n(t_h)$ since all of her tokens will be used for toll payment if she departs in $h$. The opportunity cost can be written as
\begin{align}
            \tilde{R}_{in} = -S \left( Lr-\tilde{x}_n(t_h) \right) + B\left(T^{M}(h)-\tilde{x}_n(t_h)\right) 
\end{align}

In summary, the expected opportunity cost $\tilde{R}_{in}$ of departing by car in interval $h$ depends on an individual's forecasted account balance $\tilde{x}_n(t_h)$, market price $p$, the toll in tokens $T^{M}(h)$ and transaction fees as follows:
\begin{align}\label{eqn:tollcost}
   \tilde{R}_{in} =
                \begin{cases}
                    -S\left(Lr-T^{M}(h)\right)  & \text{$\tilde{x}_n(t_h) \ge \mathrm{T}^{M}(h)$} \\
                    -S\left(Lr-\tilde{x}_n(t_h)\right) + B\left(T^{M}(h)-\tilde{x}_n(t_h)\right) & \text{otherwise}
                \end{cases}
\end{align}

Note that if transaction fees are zero, the opportunity cost in Equation \ref{eqn:tollcost} reduces to the one-day allocation minus the toll in tokens times token price, i.e.,  $\tilde{R}_{in}= -\left(Lr-\mathrm{T}^{M}(h)\right)p$. In the absence of non-linear income effects, $Lrp$ can be ignored because it is a constant (appearing in all alternatives) that does not affect the choice and the expression reduces to $\mathrm{T}^{M}(h)p$, which is intuitive.

Regarding the income effect, the diminishing marginal utility of income suggests that as an individual's income increases, the extra benefit to that individual decreases. It is thus natural to model this nonlinear effect of remaining income by a quasiconcave function (as per \cite{mcfadden2017foundations}). Hence, we add the remaining income plus a natural log of the remaining income to the systematic money-metric utility.

The utility of an individual $n$ driving and departing in time interval $h$ (choosing a mobility decision $i\in\{m=C,h| h\in H_n^{j}\}$) under instrument $j$ can thus be written as,
\begin{align}\label{eqn:utility}
    U_{in} \left(\boldsymbol{\tilde{\phi}}_i^j \right) =& V_{in}\left(\boldsymbol{\tilde{\phi}}_i^j\right)+ {\epsilon_{in}} \\ \notag
         = & -2\alpha_n \tilde{\tau}^j_i-\beta_{En}SDE\left(h,\hat{t}_n,\tilde{\tau}^j_i\right)-\beta_{Ln}SDL\left(h,\hat{t}_n,\tilde{\tau}^j_i\right) \\ \notag & +I_n - 2\tilde{c}^j_{in}+\lambda ln\left(\gamma+I_n-2\tilde{c}^j_{in}\right)+{\epsilon_{in}},
\end{align}

where 
\begin{align}
    SDE\left(h,\hat{t}_n,\tilde{\tau}^j_i\right)=\max\left(0,\hat{t}_n-\Delta_a-(t_h+\tilde{\tau}^j_i)\right)
\end{align}
\begin{align}
    SDL\left(h,\hat{t}_n,\tilde{\tau}^j_i\right)=\max\left(0,(t_h+\tilde{\tau}^j_i)-\hat{t}_n-\Delta_a\right)
\end{align}

Schedule delay of the evening trip is ignored because it is assumed to be more flexible. 

% Individual choice $h\in H_n$ has to satisfy income constraint, such that $I_n-\tilde{c}_n^{C,h}>0$. If all choices in choice set $H_n^{C}$ are infeasible, $H_n^{C}$ shifts early to have $I_n-\tilde{c}_n^{C,h}>0$ for all $h\in H_n^{C}$.   

% The choice set of feasible departure time intervals for using PT $H_n^{P}\subset\mathcal{H}$ only contains one time interval $h$ with corresponding arrival time the closest to the desired arrival time $\hat{t}$ as both PT travel time and headway are assumed to be constant. 

\DeclarePairedDelimiter\floor{\lfloor}{\rfloor}

The systematic money-metric utility function of user $n$ who departs in time interval $h$ by PT is denoted as $V_{in}(\boldsymbol{\tilde{\phi}}_i^j)$, where $i\in\{m=PT,h| h=\floor{\hat{t}_n-\tau_{PT} }\}$. Since the travel time and headway of PT are constant, we only need to consider one departure time interval $h$, which has a corresponding arrival time closest to the desired arrival time $\hat{t}_n$. For PT, the input vector $\boldsymbol{\tilde{\phi}}_i^j$ for the systematic utility consists of four components: PT travel time $\tau_{PT}$, expected waiting time $W_{PT}$, expected PT cost $\tilde{c}^j_{in}$ and remaining income $I_n-\tilde{c}^j_{in}$.

The marginal utility of an additional unit of PT travel time of individual $n$ is assumed to be the same as that of car travel time, $\alpha_n$. The marginal utility of an additional unit of waiting time is $\beta_{Wn}$. 

The expected PT cost $\tilde{c}^j_{in}$ is equal to the PT fare $c_{PT}$ under the No Toll (NT) scenario and pricing. Under the TMC scheme, it depends on an individual's expected opportunity cost of tokens $\tilde{R}_{in}$ and the PT fare $c_{PT}$, where $\tilde{R}_{in}$ is equal to the negative of selling revenue of a full wallet $Lr$ since travelers who choose PT can sell all of their tokens acquired in one day for maximum return. It can be written as $\tilde{R}_{in}=-S(Lr)$.

Hence, the expected PT cost $\tilde{c}^j_{in}$ under the TMC scheme can be written as
\begin{align}
    \tilde{c}^M_{in} = \tilde{R}_{in} + c_{PT}
\end{align}

The utility of an individual $n$ using PT who departs in interval $h$ (choosing a mobility decision $i\in\{m=PT,h| h=\floor{\hat{t}_n-\tau_{PT} }\}$) can be thus written as,
\begin{align}
    U_{in}\left(\boldsymbol{\tilde{\phi}}_i^j\right) =& V_{in}\left(\boldsymbol{\tilde{\phi}}_i^j\right)+ \epsilon_{in} \\ \notag = & -2\alpha_n\tau_{PT}-2\beta_{Wn}W_{PT}+I_n-2\tilde{c}^j_{in}+\lambda ln\left(\gamma+I_n-2\tilde{c}^j_{in}\right) + \epsilon_{in}
\end{align}

% The probability of driving and departing in a feasible time interval $h$ can be written as,

% \begin{align}
%     P_n^h= \frac{\exp(V_n^h(\tilde{\tau}(h),I_n-\tilde{c}_n(h))\mu_n)}{\sum_{h\in H_n}\exp(V_n^h(\tilde{\tau}(h),I_n-\tilde{c}_n^ h))\mu_n)+\exp(V_n^P(\tau^P,W^P, I_n-\tilde{c}_n^P))\mu_n}, \forall h \in H_n
% \end{align}

% while the probability of using PT can be written as,

% \begin{align}
%     P_n^P = \frac{\exp(V_n^P(\tau^P,W^P, I_n-\tilde{c}_n^P)\mu_n)}{\sum_{h\in H_n}\exp(V_n^h(\tilde{\tau}(h),I_n-\tilde{c}_n^ h))\mu_n)+\exp(V_n^P(\tau^P,W^P, I_n-\tilde{c}_n^P)\mu_n)}
% \end{align}

\subsubsection{Supply Model}
The network is assumed to be a single origin-destination pair connected by a single path containing a bottleneck of fixed capacity $s$ \citep{arnott1990economics}. A first-in-first-out (FIFO) queue develops once the flow of travelers exceeds $s$. The free flow travel time is $t_f$ and the extra delay time for a traveler departing from home at time $t$ is $t_v (t)$. Thus, the total travel time for a traveler departing from home at time $t$ is:
\begin{align}
    \tau(t)=t_v(t)+t_f
\end{align}

Let $Q(t)$ be the number of travelers in the queue at time $t$. The delay at time $t$ is derived from the deterministic queuing model as follows:
\begin{align}
    t_v(t)=\frac{Q(t)}{s},
\end{align}

where $Q(t)=0$ and $t_v(t) =0$ when there is no congestion.	

Note that within our simulation, the capacity $s$ is defined for time intervals $\bar{t}$ of size $\Delta_t$. The travel time for a given departure time interval $h$ is obtained by averaging the travel times of all travelers departing in $h$. Further, the exact time of departure of a traveler within the supply model is randomly (uniformly) drawn within the chosen departure time interval $h$.

The alternative PT line has a constant travel time $\tau_{PT}$. Its headway is also constant, which is equal to twice the expected waiting time $W_{PT}$. 

%In order, to   a time interval contains several time steps and travel time of a time interval is equal to averaged travel time of all travelers departing in that time interval.

\subsubsection{Day-to-Day Learning} \label{sec:d2dl}
Let $\tau^{d-1,j}_i$ denote the actual or experienced car travel time on day $d-1$ of choice $i$ under instrument $j$, where $i\in\{m=C,h| h\in H_n^{j}\}$. As we specified in the demand model, travelers are assumed to make their choices of departure time according to forecasted car travel times $\Tilde{\tau}^{d,j}_i, \; \forall h \in \mathcal{H}$ from their memory and learning. We use an exponential smoothing filter, a type of homogeneous filter \citep{cantarella1995dynamic}, to model the learning and forecasting process by weighting actual and forecasted costs of the previous day as follows:
\begin{align}
 \Tilde{\tau}^{d,j}_i = (1-\theta_\tau) \Tilde{\tau}^{d-1,j}_i + \theta_\tau{\tau}^{d-1,j}_i
 \end{align}

where $\theta_\tau \in [0,1]$ is a learning weight given to the previous day's realized travel time.

Under the TMC scheme, in order to obtain the individual forecasted account balance on day $d$, denoted by $\tilde{x}^d_n(\bar{t})$, the individual forecasted departure time $\tilde{t}^d_{n}$ is first computed by applying a similar filter as follows:
\begin{align}
 \tilde{t}^{d}_{n} = (1-\theta_{{t}}) \tilde{t}^{d-1}_{n} +\theta_{{t}} {{t}}^{d-1}_{n} 
\end{align}
where $\theta_{\tilde{t}} \in [0,1]$. 

Next, the trading model presented in Section \ref{sec:buying-and-selling} is applied using the individual forecasted departure time to determine their forecasted account balance over the entire day, which is used to compute the expected toll costs under the TMC scheme through Equation \ref{eqn:tollcost}.

\subsection{Simulation-based optimization}

The problem of determining the optimal toll in dollars for congestion pricing, $T^{P}(h), \; \forall h \in \mathcal{H}$ and the optimal toll in tokens for the TMC scheme, $T^{M}(h), \; \forall h \in \mathcal{H}$ can be formulated as a simulation-based optimization problem with the objective of maximizing total social welfare ($SW$). 
%With social welfare of NT scheme normalized to 0, 
The social welfare of the CP and TMC instruments is calculated relative to the NT scenario and is equal to the sum of user benefits ($UB$) and regulator revenue ($K$).  
%For the regulator, it is assumed that the operation cost is constant across instruments. Therefore, its surplus is only represented by regulator revenue ($K$). 
Under congestion pricing (P), the regulator revenue is given by,  
\begin{align}\label{eq:kp}
    K^{P}=\sum_{n=1}^N \sum_{i\in M_n \times H^{P}_n} \hat{c}^{P}_{in} \mathbb{I}_n\left(i|\boldsymbol{T}^{P}\right), 
\end{align}

where $i$ is the mobility decision, which is a combination of mode and departure time choice; $\mathbb{I}_n(i|\boldsymbol{T}^{P})$ is an indicator if traveler $n$ chooses mobility choice $i$ given toll vector $\boldsymbol{T}^{P}=\{T^{P}(h)|h\in\mathcal{H}\}$; $\hat{c}^{P}_{in}$ is equal to the toll payment for driving ($T^{P}(h)$) or the PT fare payment for PT ($c_{PT}$); and $H^{P}_n$ is the set of feasible departure time intervals taking into account budget constraints.

Under the TMC scheme, the regulator revenue $K^{M}$ consists of two parts. The first part is the sum of PT fare payments and the second part is the sum of tokens bought (by individuals) minus tokens sold over one day. $K^{M}$ can be written as,

\begin{align}\label{eq:km}
      K^{M}=\sum_{n=1}^N \left(c_{PT} \mathbb{I}_n\left(PT|T^{M}(h)\right)+
      \sum_{\bar{t}\in \{1...\bar{T}\}} \left( B\left(T^M\left(\bar{t}\right)-x_n(\bar{t})\right)\mathbb{I}^B_n\left(\bar{t}|\boldsymbol{T}^{M}\right)-
      S\left(x_n(\bar{t})\right)\mathbb{I}^S_n(\bar{t}|\boldsymbol{T}^{M})\right)\right),
\end{align}

where $B(\cdot)$ is cost of buying function and $S(\cdot)$ is revenue of selling function; $\mathbb{I}^B_n\left(\bar{t}|\boldsymbol{T}^{M}\right)$ and $\mathbb{I}^S_n\left(\bar{t}|\boldsymbol{T}^{M}\right)$ are indicators of buying or selling at $\bar{t}$ given the toll in tokens $\boldsymbol{T}^{M}$. Note that the price adjustment mechanism described in Section \ref{sec:price-adjustment} is designed to ensure that the regulator revenue of the TMC scheme (second part in Equation \ref{eq:km}) at equilibrium is close to zero, thereby achieving revenue neutrality. 
%Regulator revenue is the sum of users' out-of-pocket costs, which is equal to total toll revenue under the CP scheme and the total user buying amount minus total user selling amount under the TMC scheme. 

%Note that the price adjustment mechanism described in Section \ref{sec:price-adjustment}, in designed to ensure that the regulator revenue of the TMC scheme at equilibrium is close to zero, thereby achieving revenue neutrality. 

The user benefits, which are a measure of consumer well being can be computed in three ways in the presence of non-linear income effects (we refer the reader to \cite{mcfadden2017foundations} for a detailed discussion): Market Compensating Equivalent (MCE), Hicksian Compensating Variation (HCV), and Hicksian Equivalent Variation (HEV). MCE is equal to the difference in indirect utilities between a ``but for" scenario and ``as is" scenario, scaled to money metric units by dividing by the marginal utility of income (MUI) at the ``as is" scenario. It differs from the commonly used Marshallian consumer surplus (MCS) only in the MUI scaling factor. It can be easily computed when the indirect utility function and its derivatives are known. HCV is equal to the net decrease in the ``but for" scenario income that equates utility in the two scenarios while HEV is equal to the net increase in the ``as is"" scenario income that equates utility in the two scenarios. 

\begin{comment}
In this paper, we use MCE to measure indvidual $n$ benefit as follows:
\begin{align}
    UB_n = \frac{\max_{h\in H_n}( U_n (h,{\tau}^d(h),I_n-{c}_n^d(h)))-\max_{h\in H_n^{NT}}(U_n^{NT} (h,{\tau}^d(h),I_n))}{MUI^{NT}(I_n)}
\end{align}

where $U_n (h,{\tau}^d(h),I_n-{c}_n^d(h)))$ is the experienced utility from departing in time interval $h$ of individual $n$ at equilibrium under the CP or the TMC scheme; $U_n^{NT} (h,{\tau}^d(h),I_n))$ is the experienced utility of departing in time interval $h$ for individual $n$ at equilibrium under the NT scheme. $MUI^{NT}(I_n)$ is marginal utility of income of individual $n$, which, from equation \ref{eqn:utility} is given by:

\begin{align}
    MUI^{NT}(I_n) = 1+\frac{\lambda}{\gamma+I_n}
\end{align}
\end{comment}

A potential drawback of these three measures is that their ethical implications are not defensible as pointed out by \cite{blackorby1990review}. Well-being measured in units of income treat increases in income as equally socially valuable no matter who receives them. This is not the case with net utility improvement since the nonlinear effect of income improvement is captured by the income effect term in the utility specification (lower income users have a higher marginal utility of income). Hence, we measure user benefits ($Z^j$) under instrument $j$ as the sum of all users' net experienced utilities relative to NT denoted as $z^j_n$ (along the lines of \cite{de2004congestion}).
%used except they summed up the log transformations of individual utilities. 
Since the utilities adopted in this study are money-metric, the net utility amount serves as a meaningful measurement of improvement directly. An individual $n$'s net experienced utility is the difference between maximum utility under instrument $j$ and under NT, which can be written as,

\begin{align}
    z^j_n = \max_{i\in M_n \times H^{j}_n}\left(U_{in}\left(\boldsymbol{\phi}_i^j\right)\right) - \max_{i\in M_n \times H^{NT}_n}\left(U_{in}\left(\boldsymbol{\phi}_i^{NT}\right)\right),
\end{align}

where $\boldsymbol{\phi}_i^j$ is a vector of experienced variables under instrument $j$ and $\boldsymbol{\phi}_i^{NT}$ is a vector of experienced variables under $NT$.

Hence, the user benefits $Z^j$ can be written as 

\begin{align} \label{eq:zj}
    Z^j = \sum_{n=1}^N z^j_n
\end{align}

As noted before, in the case of the CP scheme, we determine the toll in dollars which maximizes social welfare $SW$, computed at the equilibrium (or after convergence of the day-to-day model). For the TMC system, in addition to the toll in tokens, which is optimized,  other parameters like the allocation rate $r$ and transaction fees are set exogenously (explained in more detail in Section \ref{sec:Opt_TMC}). This is formulated as,

\begin{equation} \label{eq:tollop}
\begin{aligned}
\max_{\boldsymbol{T}^j} \quad & Z^j +K^j\\
\textrm{s.t.} \quad & Z^j, K^j= SM\left(\boldsymbol{T}^j,\boldsymbol{\xi},\boldsymbol{\psi}\right)\\
  &\boldsymbol{T}^{j}=\{T^{j}(h)|h\in\mathcal{H}\}    \\
  &\boldsymbol{T}^j \geq 0    \\
\end{aligned}
\end{equation}

where $j$ can be either $P$ or $M$; the toll profile $\boldsymbol{T}^{j}$ is a set of toll values over the entire day. $\boldsymbol{\xi}$ represents all input data for simulation, such as individual income, preferred arrival time, and choice attributes. $\boldsymbol{\psi}$ represents all model parameters, such as demand model coefficients, bottleneck capacity, user learning weights, and market parameters for the TMC scheme. The $SM(\cdot)$ function is the system model discussed in Section \ref{sec:SysModel}. The toll function that we consider is a step toll profile (of the kind implemented in Singapore and Stockholm), which consists of five step toll values and six break points.

%We adopt two different time of day toll profile specifications to characterize $T^{CP}(h)$ and $T^{TMC}(h)$. The first one expresses the tolling function  (or profile) by a mixture of $M$ Gaussian functions, which are specified by $3M$ parameters, mean $\xi_m$, variance $\sigma_m$ and amplitude $A_m$, $m=1,2,\dots, M$. However, in practice, implementing a smooth and continuous toll profile of this nature is complex. Hence, 

Clearly, our optimization problem has no closed-form since the objective function for a given toll profile is the outcome of a simulation of the stochastic process (a simulation-based optimization problem), or more specifically, the system model presented in \ref{sec:SysModel}, which includes traveler behavior, regulator states and actions, and the resulting network and market conditions. In order to solve this simulation-based optimization problem, a differential evolution (DE) algorithm is adopted as it is derivative-free and performs well for global optimization problems of this kind \citep{storn1997differential}.
%Bayesian optimization (BO) approach is adopted as it can approximate the simulation-based objective function using a few evaluations.

\subsubsection{Differential evolution algorithm}\label{sec:Optimization_Algo}

In this section, we illustrate the application of the DE algorithm. Let $\mathbf{X}$ represent the decision variables of the simulation-based optimization problem (i.e., the parameters of the step toll profile).
%, $\mathbf{X}=(\xi_m,\sigma_m,A_m),~m=1,2,\dots,M$. 
The DE algorithm essentially has three iterative operators, mutation, crossover and selection, to iteratively improve candidate solutions.

The mutation operator uses individuals from the current solution population to generate variant vectors. The $g$th variable of vector $i$ at generation $k$, $\mathbf{Y}_{g,i}^k$, is given by
\begin{align}
 \mathbf{Y}_{g,i}^k=\mathbf{X}_{g,r1}^k+F\cdot(\mathbf{X}_{g,r2}^k-\mathbf{X}_{g,r3}^k)
\end{align}
where $r1,~r2,~r3\in [1,NP]$, $i\neq r1 \neq r2 \neq r3$, $F$ is a scale factor and $NP$ is the solution population size.

Next, the crossover operator creates a trial vector $\mathbf{U}_{i}^k$ by combining the variant vector and original vector as follows,
\begin{align}
\mathbf{U}_{g,i}^k =
\begin{cases}
    \mathbf{Y}_{g,i}^k & \text{if}~rand(0,1)<CR~\text{or}~g=rg \\
    \mathbf{X}_{g,i}^k & \text{otherwise}
\end{cases}
\end{align}
where $CR\in [0,1]$ is the crossover rate, $rand(0,1)$ represents a random uniformly distributed variable within $(0,1)$, and $rg$ is a random integer in $[1,3M]$ ensuring at least one variable of the trial vector $\mathbf{U}_{i}^k$ is from the variant vector $\mathbf{Y}_{i}^k$.

Finally, the selection operator produces the next generation of vectors by comparing the original vector $\mathbf{X}_{i}^k$ and the trial vector $\mathbf{U}_{i}^k$ in terms of social welfare as follows,
\begin{align}
\mathbf{X}_{i}^{k+1} =
\begin{cases}
    \mathbf{U}_{i}^k & \text{if}~SW(\mathbf{U}_{i}^k)>SW(\mathbf{X}_{i}^k) \\
    \mathbf{X}_{i}^k & \text{otherwise}
\end{cases}
\end{align}

\section{Numerical Experiments}\label{sec:Sec6}

%\subsection{Experiment Setup}

%In order to perform credible comparisons of the various tolling instruments, it is important to have realistic parameters and input data. 
This section describes numerical experiments that examine the performance of the TMC scheme using the simulation framework described in Section \ref{sec:Sec5}. Data and parameters of the demand model, supply model, and the day-to-day learning model are introduced in Section \ref{sec:Data_parameters}. The parameters are calibrated based on empirical evidence to represent a realistic base case. Selected parameters will be varied in later experiments.
The objectives of the experiments are to: 1) examine the overall performance of the TMC system under alternative market designs -- more specifically, the effect of transactions fees in avoiding undesirable market behavior whilst retaining efficiency, 2) compare the efficiency and equity of the TMC scheme (market design based on experiments under 1) and CP under varying levels of congestion, heterogeneity, income effects, and 3) examine the robustness (adaptiveness) of the TMC scheme with a lump sum allocation versus a continuous allocation in the presence of unusual events. 

\subsection{Data and Model Parameters}\label{sec:Data_parameters}
\begin{figure}[!]
    \centering
    \includegraphics[width=0.9\linewidth]{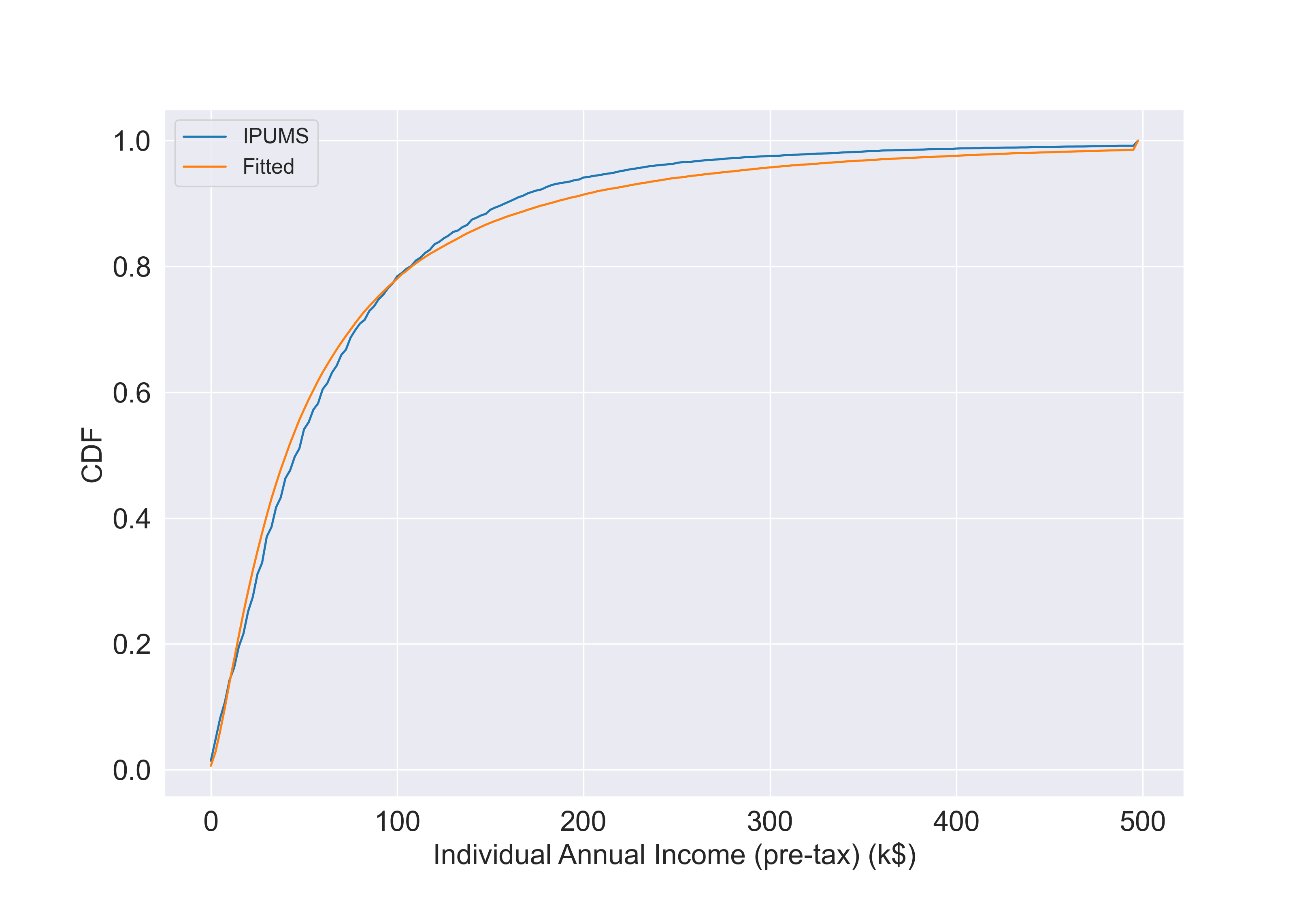}
    \caption{Individual pre-tax annual income distribution}
\label{fig:IPUMS}
\end{figure}

The demand model requires both choice attributes and individual characteristics as input data. The individual characteristics include disposable income $I_n$ and preferred arrival time $\hat{t}_n$. Recall that disposable income $I_n$ in our context is defined as personal net income after taxes and subtracting necessary living expenses (e.g. housing, health, food). 
%In other words, it represents the individual's available income for traveling. 
The individual pre-tax annual income is assumed to follow a lognormal distribution and is fitted using the Integrated Public Use Microdata Series (IPUMS) 2019 census data \citep{IPUMS2019}. The cumulative distribution function (CDF) of IPUMS data and fitted data are shown in Figure \ref{fig:IPUMS}. 
%The x-axis represents individual pre-tax annual income in units of one thousand dollars and the y-axis represents cumulative probability values. 
Note that all annual incomes greater than 500 thousand dollars are grouped together. As we can see, the lognormal distribution fits the income distribution reasonably well.

Individual daily income is computed as the annual income divided by 260 working days per year and the individual hourly wage rate is computed by dividing daily income by 8 working hours per day. The minimum wage rate is set to \$7.25 per hour as per the U.S. Department of Labor. It is less straightforward to
obtain disposable income after taxes since necessary living expenses could vary significantly based on income and disaggregate data on this is sparse. Based on average data from the Bureau of Labor Statistics  \citep{BLS2019}, we assume that each traveler's daily disposable income after taxes and necessary living expenses is equal to 60\% of their pre-tax daily income. 
%However, it is not straightforward to obtain disposable income after taxes because it depends on individual income and other attributes. Further, individual necessary living expenses could vary based on income and there is limited disaggregate data on this. 
%According to data from the Bureau of Labor Statistics, the average pretax household income in the United States in 2019 was \$82,852, while average household expenditures except transportation added up to \$52,294, which means that the average American has 63\% of their pretax or market income available for transportation and other expenditures \citep{BLS2019}. Hence, it is assumed that each traveler's daily disposable income after taxes and necessary living expenses is equal to 60\% of their pre-tax daily income. 

\begin{figure}[!]
    \centering
    \includegraphics[width=0.9\linewidth]{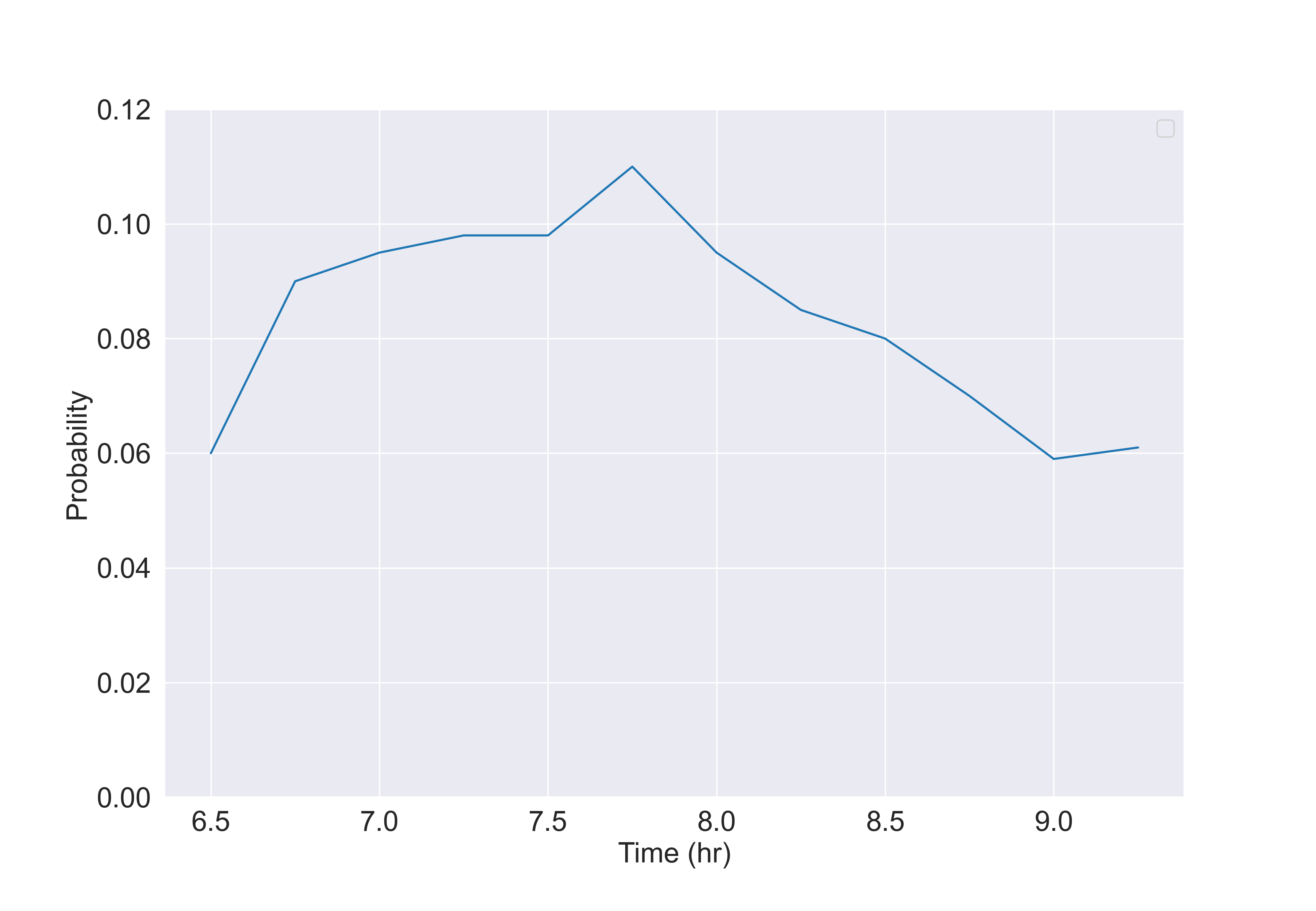}
    \caption{Individual preferred departure times distribution}
\label{fig:PDT}
\end{figure}

The preferred departure time distribution (preferred arrival time $\hat{t}_n$ minus the free flow travel time) is based on a recent empirical study of road users in Stockholm \citep{kristoffersson2018estimating}, and is shown in Figure \ref{fig:PDT}. For simplicity, the size of the preferred arrival window $\Delta_a$ is set to 0, which implies that individuals have a single preferred arrival time as in the standard Vickrey model. The departure time window size parameter $\eta$ is set to 30, which means the individual departure time choice set $H_n$ ranges over a 60-minute interval. 

%The preferred arrival time $\hat{t}_n$ represents the optimal or desired arrival time. The preferred departure time is just a shift of preferred arrival time by free flow travel time. Many researchers have assumed it follows a uniform distribution \citep{de2002comparison}. A more recent study estimated preferred departure of morning road users in Stockholm \citep{kristoffersson2018estimating}. The distribution of preferred departure times estimated from their study is adopted in this study and is shown in Figure \ref{fig:PDT}. The x-axis represents departure time in hours and the y-axis represents probability values. As we can see, it does not follow the uniform distribution exactly.  

% done
% value of time, sde, sdl, vow, 
% lambda, gamma based on income effect
% scale parameter based on price elasticity
% TODO
% preferred departure time window
% departure time choice set
% PT departure time choice
% PT share calibration

Empirical evidence (e.g. \cite{small2005uncovering}) indicates that there is significant heterogeneity in the value of time across drivers. We assume that the individual value of time $\alpha_n$ is perfectly correlated with income and is one third of the wage rate \citep{white2016revised}. %Therefore, they . 
Note that this assumption will be relaxed in some experimental scenarios to consider different levels of heterogeneity. 

%For bottleneck models of congestion, schedule delay costs are another important part of congestion costs. 
Values of schedule delay early $\beta_{En}$ and late $\beta_{Ln}$ are also likely to be distributed across individuals. Due to the lack of empirical data on this, the literature on bottleneck models incorporate heterogeneity by making assumptions on the ratios between values of schedule early/late to values of time. Proportional heterogeneity (first considered by \cite{vickrey1973pricing}) assumes that values of time and schedule delays vary proportionally or in other words, the ratio of the parameters is identical for all individuals \citep{van2011congestion}. Ratio heterogeneity assumes that the values of schedule delays are fixed and only values of time are distributed \citep{de2002congestion}. As a result, ratios of parameters are distributions. \cite{van2011winning} considers a more general heterogeneity, assuming the ratio of values of time to values of schedule delay early follows a symmetric triangular distribution from 1 to 3 based on intuition and ratio of values of schedule delay late to values of schedule delay early is a constant 3.9 based on \cite{arnott1990economics}.
%As summarized in \citep{van2011congestion,van2011winning}, there are mainly two types of heterogeneity for bottleneck models. The first type of heterogeneity is termed proportional heterogeneity (first considered by \cite{vickrey1973pricing}), which assumes that values of time and schedule delays vary proportionally. In other words, the ratio of the parameters is identical for all individuals. A different case studied by \cite{de2002congestion} assumes values of schedule delays are fixed and only values of time are distributed. As a result, ratios of parameters are distributions. This type of heterogeneity is known as ratio heterogeneity. Behavioral interpretations of these two types can be found in \cite{van2011winning}. Further, \cite{van2011winning} considers a more general heterogeneity, assuming the ratio of values of time to values of schedule delay early follows a symmetric triangular distribution from 1 to 3 based on intuition and ratio of values of schedule delay late to values of schedule delay early is a constant 3.9 based on \cite{arnott1990economics}.
Along similar lines, we assume that the ratio of values of schedule delay early $\beta_{En}$ to values of time $\alpha_n$ follows a triangular distribution from 0.1 to 1 with a mode at 0.5. The ratio of values of schedule delay late $\beta_{Ln}$ to $\alpha_n$ is assumed to follow a triangular distribution from 1 to 3 with a mode at 2 (the selection of the modes as 0.5 and 2 are based on \cite{small2012valuation}). The bounds are set based the empirical relationship  $\beta_{En}\le \alpha_{n} \le \beta_{Ln}$ \citep{small2012valuation}.

Further, as pointed out by \cite{small2012valuation}, waiting times are onerous compared to in--vehicle times by multiples of two to three by most assessments. For simplicity, the ratio of values of time $\alpha_n$ to values of waiting time $\beta_{Wn}$ is assumed to be a constant equal to 3.
%that $\beta_{Ln}$ is twice large as $\alpha_n$ and $\alpha_n$ is twice large as $\beta_{En}$ \cite{small2012valuation}. The bounds are set based on empirical relationships that $\beta_{En}\le \alpha_{n} \le \beta_{Ln}$ \citep{small2012valuation}. As pointed out by \cite{small2012valuation}, waiting times are onerous compared to in vehicle times by multiples of two to three by most assessments. For simplicity, the ratio of values of time $\alpha_n$ to values of waiting time $\beta_{Wn}$ is assumed to be a constant equal to 3.
With regard to the income effect, $\gamma$ is set to 2 and $\lambda$ is calibrated to be 3 to have the highest marginal utility of income to be less than 1.34 \citep{layard2008marginal}.
 
%Regarding utility coefficients of income effect $\lambda$ and $\gamma$, $\lambda$ measures the strength of nonlinear income effect and $\gamma$ ensures the logarithmic function is defined when remaining income is equal to 0. In this study, $\gamma$ is set to 2 and $\lambda$ is calibrated to be 3 to have the highest marginal utility of income to be less than 1.34 \cite{layard2008marginal}.

% As such, the larger (smaller) the error variance, the smaller (larger) the parameters of the deterministic compo- nent of utility will be. Any observed differences in estimated parameters could be the result of different marginal utilities, different error variances, or both
The scale parameter $\mu_n$ is known to be both confounded with the systematic utility and inversely related to error variance within the choice data \citep{ben1985discrete}. As pointed out in the literature (e.g. \cite{louviere2006confound}), the modeled heterogeneity can come from heterogeneity in individual coefficients and scale heterogeneity that is shared across coefficients. We assume the scale parameter follows a lognormal distribution. The mean value is calibrated based on price elasticity (discussed subsequently) and the coefficient of variation is set to 0.5 based on judgement. %The mean of scale parameter .

Next, we calibrate the mean value of the scale parameter to ensure that the aggregate price elasticities of the mobility model (for departure time choice) are reasonable and accord with empirical evidence. The price elasticity of peak hour demand (7 AM - 8AM) is computed assuming there exists a flat toll profile during the period from 6:30-9:30 AM. 
%Since coefficients of mobility demand models are distributed, it requires simulation to obtain price elasticity. Each iteration consists of two simulation runs. Run 0 is the base case and run 1 is the case with operation costs and peak hour toll (7 to 8AM) increased by 5\% to calculate peak hour price elasticity. Several iterations are done for different initial toll levels. For simplicity, flat toll profiles are used from 6:30 to 9:30AM.
From the literature, the aggregate elasticities of peak hour travel vary greatly from case to case as they are dependent on the model structure, physical environment, activity type, initial toll levels and many other factors. \cite{ding2015cross} estimated the elasticity of departing during the peak in Washington D.C. to be -0.0906 for driving alone. \cite{sasic2013modelling} estimated that the elasticity of departing by car in the AM peak for work trips in Toronto is between -0.067 to -0.12. \cite{holguin2005evaluation} found price elasticity of using crossings (tunnels and bridges) in NYC ranges from -0.31 to -1.97 for weekdays depending on the time-of-day. When there is no initial toll, the price elasticity simply represents fuel price elasticity. \cite{lipow2008price} estimated fuel price elasticity as -0.17 and \cite{gillingham2014identifying} estimated fuel price elasticity in California as -0.15.

\begin{table}[!]
    \centering
       \caption{Price elasticities across income groups by toll levels}
    \label{tab:priceelas}
    \renewcommand{\arraystretch}{0.75}
    \begin{tabular}{ |c||c|c|c|c|c|c|  }

 \hline
 Toll & $\le 25\%$ &$25\%$ to $50\%$ & $50\%$ to $75\%$& $75\%$ to $90\%$&$90\%\le  $ &Total\\
 \hline
 0& -0.34& -0.29& -0.12& 0.00& 0.00& -0.19\\
2.5& -1.14& -0.59& -0.10& -0.04& -0.03& -0.38\\
5&-1.57&-1.07&-0.20&-0.09&-0.06&-0.53\\
 \hline
\end{tabular}
    
\end{table}

From calibration, the mean of scale parameter is determined to be 0.5. The corresponding price elasticities across different income groups and initial toll levels are presented in Table \ref{tab:priceelas}. As we can see, low income users are more sensitive to price than high income users, and when there is no toll, the aggregate price elasticity is similar to the empirical fuel elasticity. As the toll level increases, the aggregate price elasticity also increases and is similar to empirical values found in \cite{holguin2005evaluation}.

Regarding the supply model, 
%attributes of car and public transit (PT) have to be specified. 
the free flow speed of car is set to be 45mph \citep{ali2007prediction} and the one way driving distance is assumed to be 18 miles (free flow travel time of 24 minutes). The operational cost of driving is assumed to be composed of only fuel cost, which is equal to \$3.13 (driving distance times 1/23 gallon per mile times 4 dollars per gallon). For public transit, based on the New York City MTA data, the fare is set to \$ 2, average speed is 25mph, and headway is 10 minutes. The PT distance is also assumed to be 18 miles, and the resulting PT travel time is 43 minutes since both headway and travel time of PT are constant. The expected waiting time is assumed to be 5 minutes. 

\begin{table}[h!]
  \begin{center}
    \caption{Model and simulation parameters}
     %\vspace{2ex}
    \label{tab:simvar}
    \renewcommand{\arraystretch}{0.75}
    \begin{tabular}{l|l|l} % <-- Alignments: 1st column left, 2nd middle and 3rd right, with vertical lines in between
      \hline 
      \textbf{Variables} & \textbf{Description} & \textbf{Values}\\
      \hline
      $N$ & Population & $7,500$\\ 
      $\Delta_t$ & Duration of a simulation time step & 1 min  \\
      $\Delta_h$ & Duration of a departure time interval & 5 min  \\
      $\Delta_a$ & Size of desired arrival window & 0 min  \\
      $\eta$ & Departure time window size parameter & 30\\
      $\lambda$ & Coefficient of nonlinear income effect & 3 \\
      $\gamma$ & Nonlinear income effect adjustment parameter & 2\\
      $s$ & Bottleneck capacity (per min) & 39 \\
      $t_0$ & Free flow travel time & $24$ mins \\
      $c_f$ & Operation cost of car & \$3.13 \\
      $\tau_{PT}$ & PT travel time & $43$ mins \\
      $W_{PT}$ & Expected waiting time & $5$ mins \\
      $c_{PT}$ & Operation cost of PT & \$2 \\
      $\theta_{\tau} /\theta_{\bar{t}} $ & Learning weights & 0.1 \\
      \hline
    \end{tabular}
  \end{center}
  
\end{table}

The bottleneck capacity $s$ is determined based on a calibration of the travel time index (TTI), which represents the ratio between actual travel time and free flow travel time. 
%For instance, a TTI value of 1.2 means actual travel time is 20\% longer than free flow travel time. In this study, 
The base case capacity is determined to be 2340 vehicles per hour to have a reasonable level of congestion with a TTI of 1.68 \citep{chen2010travel} under the no toll (NT) scenario.

As described in Section \ref{sec:d2dl}, an exponential smoothing filter is adopted to update travel time information and individual departure time. The greater the learning weights are, the more unstable the system becomes \citep{cantarella1995dynamic}. The learning weights $\theta_{\tau}$ and $\theta_{\Tilde{t}}$ are assumed to be 0.1.

Recall that we focus on the morning commute and hence, we simulate half a day (12 hours) with a simulation time interval $\Delta_t$ of 1-minute, yielding 720 time intervals, $\bar{t} = 0 \ldots 719$. The market elements (allocation, expiration, and price adjustment) and trading behavior are also simulated for the first half. The second half of a day is assumed to be a mirror of the first half. The departure time interval ($\Delta_h$) is assumed to be 5 minutes and the population size $N$ is 7500. Descriptions and values of key parameters are summarized in Table \ref{tab:simvar}.

% supply: capacity, travel time, operation costs

% day to day learning weights

\subsection{Existence and Uniqueness of Equilibrium}
%\textcolor{red}{[Move material to appendix, summarize here]} In this section, we examine existence and uniqueness of the equilibrium in the day-to-day dynamic model. 
%It is important to check whether an equilibrium exists since the analysis and comparisons are meaningless when system states are still changing. Once equilibrium existence is confirmed, it is important to confirm uniqueness of equilibrium to avoid issues with multiple equilibria. As pointed out by \cite{lindsey2004existence}, some problems include the questionable validity of comparative statistics under multiple equilibria, the possible instability of equilibrium, the difficulty of determining which equilibrium will prevail, and sensitivity to starting values.
Existence and uniqueness of equilibrium in the bottleneck model have been widely studied (e.g. \cite{hendrickson1981schedule,hurdle1983effects,smith1984existence,daganzo1985uniqueness}), although heterogeneity is typically limited to the desired arrival time. \cite{lindsey2004existence} considers a more general heterogeneity in values of time, desired arrival times, and values of schedule delay and establishes conditions for existence and uniqueness of a deterministic departure time user equilibrium in the bottleneck model. \cite{de1983stochastic} examine a stochastic user equilibrium where travelers are assumed to have identical systematic travel cost functions. They show that the equilibrium departure rate is unique but their findings cannot be directly applied to our model as we consider heterogeneity. \cite{miyao1981discrete} establish conditions for the existence, uniqueness and stability of equilibrium for discrete choice models. Although their framework is relatively general, it assumes individuals have the same choice set whereas in our case, choice sets (for departure time choices) are individual specific. To the best of our knowledge, there are no analytical results on uniqueness of the equilibrium for our specific model. 
%literature has been found to support the problem considered in this study has unique equilibrium 
However, simulations with different initial conditions suggests that the day-to-day model converges to the same equilibrium solution. The tests on equilibrium and stationarity are summarized in \ref{sec:App_uniqueness}.

Note that the travel time of driving is used as a representative measure of the system state partly because it is central to the day-to-day learning process of travelers (alternatively, departure flows could also be used). The infinity norm (supremum norm) of day $d-1$'s travel time vector of driving and day $d$'s travel time vector of driving is calculated and used as a measure of convergence across days. %The expression of this norm can be written as follows:
This is given by,
\begin{align}\label{eqn:Infinity_norm}
    ||\boldsymbol{\tau}^{d-1}-\boldsymbol{\tau}^{d}||_{\infty} = \sup{\{|\tau^{d-1}_i-\tau^{d}_i|: i\in\{m=C,h| h\in \mathcal{H}\} \}}
\end{align}

where $\boldsymbol{\tau}^{d-1}$ is the vector of travel times of day $d-1$ defined over a set of time intervals $\mathcal{H}$.

\begin{figure}[!]
  \begin{subfigure}[b]{0.48\textwidth}
    \centering\includegraphics[width=\textwidth]{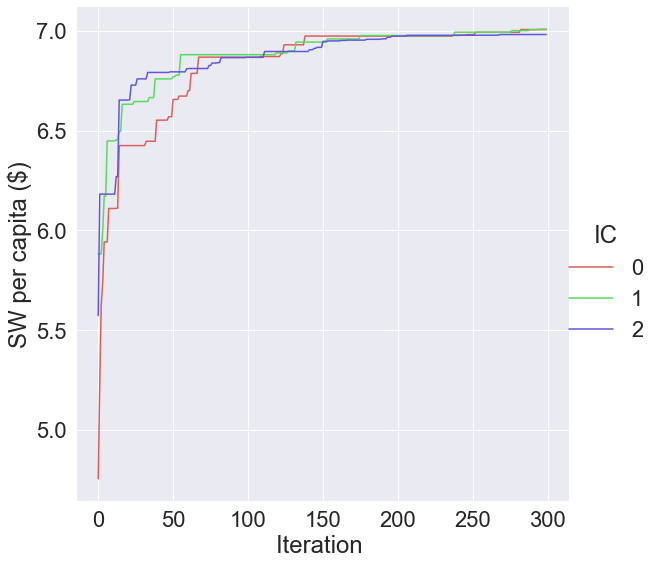}
    \caption[]{\small{Convergence of the social welfare }}
    \label{fig:PToll}
  \end{subfigure}
  \begin{subfigure}[b]{0.48\textwidth}
    \centering\includegraphics[width=\textwidth]{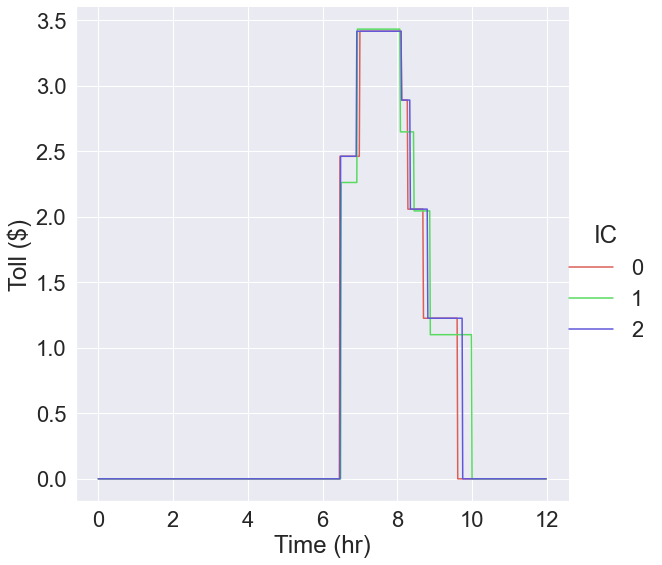}
    \caption[]{\small{Optimal toll profile}}
  \end{subfigure}
  
  \caption{Convergence of the social welfare and optimal time-of-day step toll profile of Congestion Pricing for different initial values }
  \label{fig:OPTPER}
\end{figure}

\subsection{Optimization of tolls in TMC and Congestion Pricing}\label{sec:Opt_TMC}
 %\textcolor{red}{[Move material to appendix, summarize here]} 
 As discussed in Section \ref{sec:Optimization_Algo}, the time-of-day step toll profile is optimized using a type of metaheuristic algorithm known as Differential Evolution (DE). Metaheuristic algorithms have been shown to work well for nonconvex and nonlinear toll design problems (e.g. \cite{shepherd2004genetic,zhang2004optimal}). 
%They are fairly robust to initial values because of mutation, crossover and selection steps in the algorithm.
The population size of the DE algorithm $NP$ is set to 15. %which means 15 candidates of the time-of-day step toll profile are evaluated in one iteration. Taking advantage of parallelization, they can be simulated and evaluated concurrently. 
Using data and parameters discussed in Section \ref{sec:Data_parameters}, we first examine the 
performance of the optimization algorithm for the congestion pricing (CP) instrument using three different initial populations (with 15 candidates). Next, the performance of the optimized TMC instrument and convergence properties are examined.   
%comparative performance of TMC and CP a pricing without distribution ($P-$) case is used to test the performance of the optimization algorithm. 
%are used for the algorithm to optimize the time-of-day step toll profile. 

For CP, the convergence of social welfare along with the optimal time-of-day step toll profile with three different initial populations are shown in Figure \ref{fig:OPTPER}. The algorithm converges to the same optimal welfare  (within a tolerance of \$0.01) for three different initial populations. The optimal toll profiles are near identical although there are minor differences at the locations when the toll changes (`steps').
%The total number of iterations is 300 implying a total of 4500 function evaluations. Recall that social welfare is equal to the sum of user benefits relative to NT and regulator revenue. As we can see, although the starting social welfare is very different because of different initial values, they converge to the same social welfare (within a tolerance of \$0.01) after about 150 iterations. The optimal toll profiles are also very similar with small differences.

For the TMC instrument, we set the total daily allocation of tokens for each individual as the per capita regulator revenue from congestion pricing (this should in theory yield a 1 dollar equilibrium market price and a toll in tokens that is identical to the optimal toll in dollars from CP, assuming there are no transaction costs and income effects). Note that the daily allocation can be set arbitrarily. Next, we optimize the toll profile in tokens given the specified token allocation. %For completeness, the optimization performance under different allocation rates is examined in the next section.
We assume that tokens are distributed uniformly to everyone and are allocated in continuous time. 
%Also, continuous allocation is the default method for token allocation because of its advantages discussed in Chapter \ref{chap3}, which is also demonstrated later in the case of sub-optimal toll profiles. 
Every traveler is assumed to have a random account balance at the beginning of the simulation (to reflect different times of entry into the system; more on this in Section \ref{sec:Overall_market}). %The effect of the initial account balance is examined in the following section \ref{sec:trading} 
Recall also that we assume the evening period is a mirror of the morning period, and hence we only simulate half a day, and assume the lifetime $L$ of tokens is 720 minutes. For simplicity, in these experiments, transaction fees are set to zero. Other parameters of the TMC are summarized in Table \ref{tab:marketvar}.

%Since the base case $P-$ has revenue as $4.1040$ dollars per capita over the entire day, the individual allocation rate $r$ is $0.00285$ tokens per minute. The token lifetime $L$ should be one day since the context we consider is a daily commute problem and the frequency of trip making is one day. However, since we only simulate half day, the lifetime $L$ is 720 minutes.

%The effect of both proportional and fixed transaction fees are investigated later in this section. In all prior experiments they are assumed to be 0. The token price is adjusted daily by a constant based on the demand and supply of tokens, which is represented by regulator revenue of MU directly $K^{MU}$. The greater the revenue is, the more demand for buying tokens and price increases and vice versa. The initial price ($p^0$) is by default set at 1 dollar but is varied in later experiments. The constant price change $\Delta p$ is 5 cents (0.05 dollars). To avoid unnecessary price fluctuations for small amounts of regulator revenue, the price does not change when revenue is in the interval $[-K_t,K_t]$, where $K_t$ is equal to 300 dollars in this study (based on preliminary experiments). In addition, the price cannot fall below 0 but there is no maximum cap for it.

%Descriptions and values of key parameters are summarized in Table \ref{tab:marketvar}.

\begin{table}[h!]
  \begin{center}
    \caption{Market parameters for tradable mobility credits (base case)}
     \vspace{2ex}
    \label{tab:marketvar}
    \renewcommand{\arraystretch}{0.75}
    \begin{tabular}{l|l|l} % <-- Alignments: 1st column left, 2nd middle and 3rd right, with vertical lines in between
    \hline
      \textbf{Variables} & \textbf{Description} & \textbf{Values}\\
      \hline
      $r$ & Token allocation rate & 0.00285 per min\\
      $L$ & Token lifetime & 720 mins\\
      $F^F_B/F^F_S$ & Proportional transaction fee of buying/selling & \$0  \\
      $F^P_B/F^P_S$ & Fixed transaction fee of buying/selling & \$0  \\
      $p^0$ & Initial token price & \$1  \\
      $\Delta p$ & Price change & \$0.05  \\
      $K_t$ & Regulator revenue threshold & \$300  \\
    \hline
    \end{tabular}
    
  \end{center}
\end{table}

\begin{figure}[!ht]
  \begin{subfigure}[b]{0.48\textwidth}
    \centering\includegraphics[width=\textwidth]{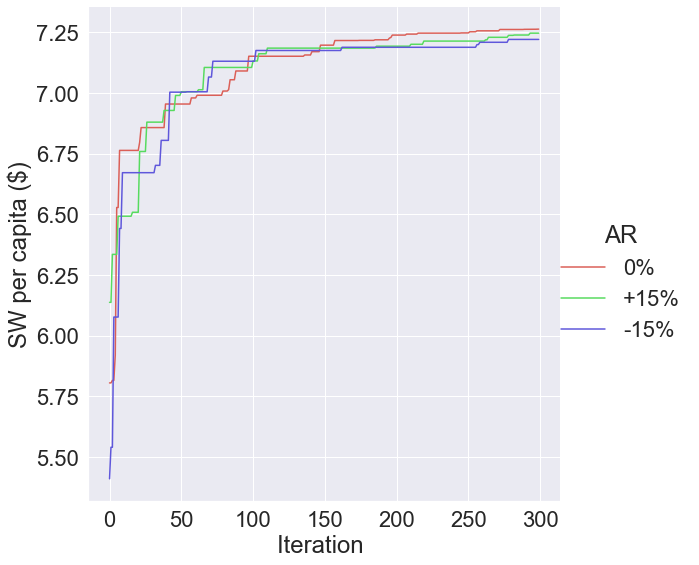}
    \caption[]{\small{Convergence of the social welfare }}
  \end{subfigure}
  \begin{subfigure}[b]{0.48\textwidth}
    \centering\includegraphics[width=\textwidth]{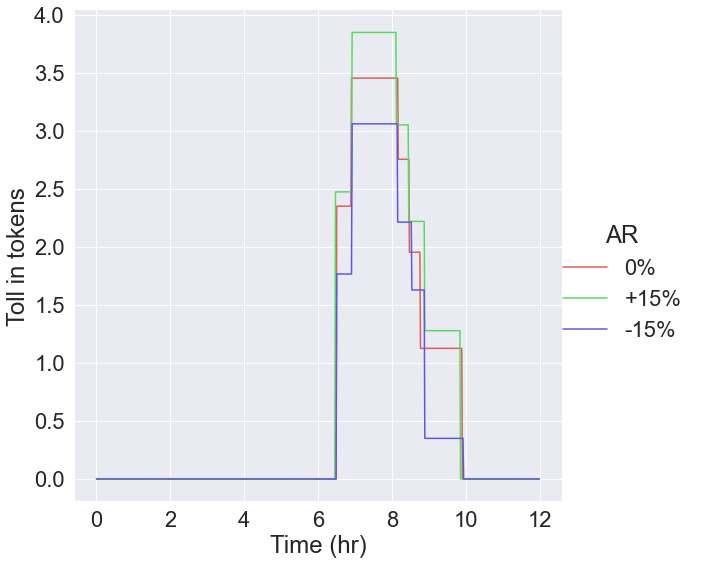}
    \caption[]{\small{Optimal toll profile in tokens}}
      \label{fig:TriTollAR}
  \end{subfigure}
    %\begin{subfigure}[b]{0.48\textwidth}
    %\centering\includegraphics[width=\textwidth]{Plots/Optimization/TriOPTTOLLP.png}
    %\caption[]{\small{Optimal toll profile in dollars}}
    %\label{fig:TriTollPAR}
  %\end{subfigure}
  \caption{Convergence of the social welfare, optimal time-of-day step toll profile in tokens and in dollars of TMC for three allocation rates}
  \label{fig:TriOPTPER}
\end{figure}

The convergence of the optimized objective values (social welfare) along with the optimal time-of-day step toll profile in tokens 
%and in dollars (tokens times token price) 
with three different allocation rates ranging from 15\% less than the baseline to 15\% more than the baseline are shown in Figure \ref{fig:TriOPTPER}. 
As we can see, 
%although the starting social welfare are very different because of different allocation rates, 
the different allocation rates converge to the same social welfare (within a tolerance of \$0.02) at the end of 300 iterations. The token price of the baseline allocation rate is close to \$1 (as expected) while the lower allocation rate has a higher token price of \$1.1 and the higher allocation rate has a lower token price equal to \$0.9. This is consistent with our expectation that the lower allocation rate leads to a higher token price due to less supply and vice versa. The optimal toll profiles in tokens in Figure \ref{fig:TriTollAR} show that the lower allocation rate leads to the overall higher tolls in tokens and vice versa, again, as expected. 
%This is because as the token price increases (decreases) the toll in tokens has to decrease (increase) to maintain the product of the price and toll in tokens similar to the optimal toll in dollars.
Interestingly, due to the non-linear income effects, the social welfare of the TMC scheme is slightly greater than that of congestion pricing (we return to this later). 
%The product of toll profiles in tokens and token prices are shown in Figure \ref{fig:TriTollPAR}. The black line represents the optimal toll profile in dollars for $P-$, which is also the red line shown in Figure \ref{fig:PToll}. In the peak hour (from 7AM to 8:30AM) when the most people are traveling, the toll profiles in dollars are very similar with small differences. The differences among the toll profiles in the off peak indicate that the objective function (social welfare function) might have the relatively flat landscape near the optimal values. 

\begin{figure}[!ht]
  \begin{subfigure}[b]{0.48\textwidth}
    \centering\includegraphics[width=\textwidth]{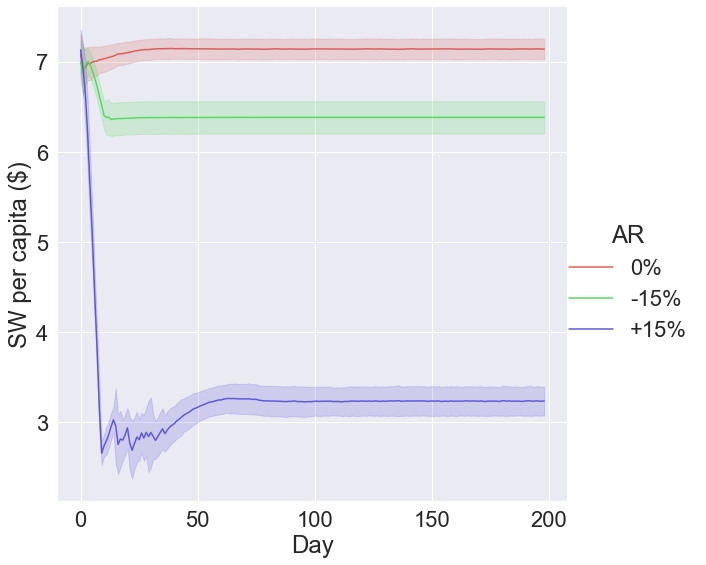}
    \caption[]{\small{Convergence of social welfare }}
  \end{subfigure}
  \begin{subfigure}[b]{0.48\textwidth}
    \centering\includegraphics[width=\textwidth]{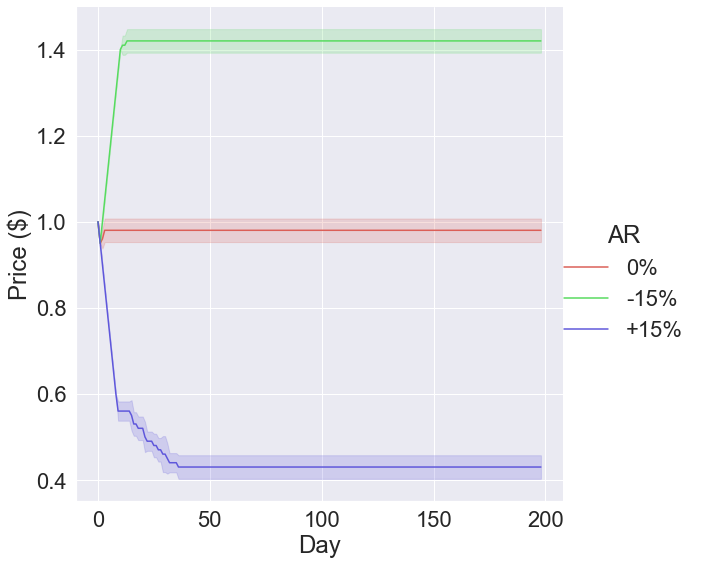}
    \caption[]{\small{Convergence of token price}}
  \end{subfigure}
  \begin{subfigure}[b]{0.48\textwidth}
    \centering\includegraphics[width=\textwidth]{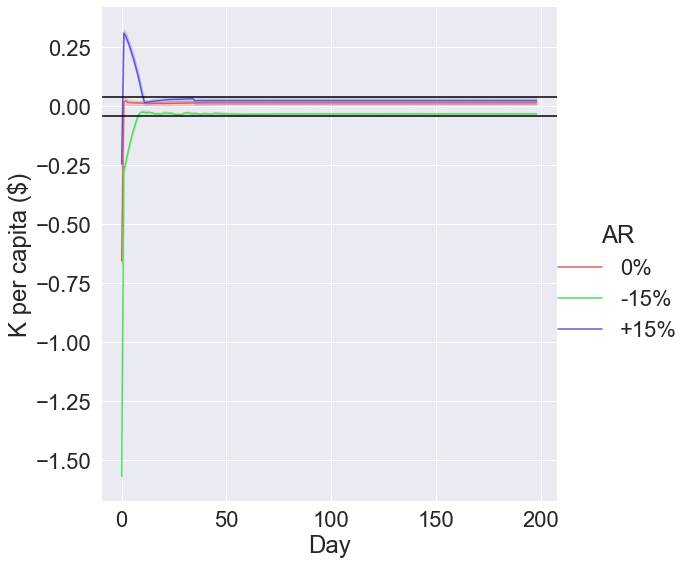}
    \caption[]{\small{Convergence of regulator revenue}}
    \label{fig:ARICRR}
  \end{subfigure}
    \begin{subfigure}[b]{0.48\textwidth}
    \centering\includegraphics[width=\textwidth]{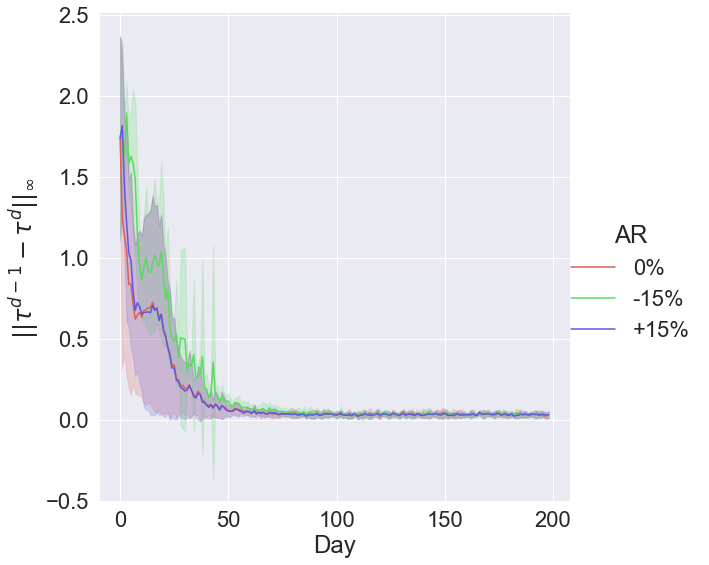}
    \caption[]{\small{Convergence of travel time of driving}}
    \label{fig:ARICTT}
  \end{subfigure}
  
  \caption{Convergence of social welfare, token price, regulator revenue, and travel time of driving by allocation rate $r$ }
  \label{fig:ARIC}
\end{figure}

Given that disaggregate market behavior is modeled, it is also worthwhile to examine stability and convergence of the market prices. We examine this under different allocation rates $r$ (where the toll in tokens is not varied and is set to the optimal profile under the baseline allocation rate) and convergence of social welfare, market price, regulator revenue, and travel time of driving are shown in Figure \ref{fig:ARIC}. As we can see, since tolls are not re-optimized, different allocation rates lead to different social welfare and price values. The results also indicate stability of the credit market, and that the variation of equilibrium price and welfare with allocation rate is intuitive. The effect of various initial market prices on convergence of token price and social welfare are also examined in \ref{sec:App_convg_market_price}. %At the baseline allocation rate %(recall that this is computed based on an optimized toll in dollars from pricing without distribution), token price of MU (TMC with uniform allocation) converges to \$1. With smaller allocation rate, token price converges to be higher than \$1 due to more demand and less supply; with greater allocation rate, token price converges to be less than \$1 due to less demand and more supply. The social welfare of greater or smaller allocation rates are less than that of the baseline allocation rate. The social welfare of the baseline allocation rate is greater than that of pricing without distribution. The regulator revenues under the three allocation rates converge to be within the regulator revenue threshold band (the black lines) as shown in Figure \ref{fig:ARICRR}. Travel times of driving under the three allocation rates converge too as shown in Figure \ref{fig:ARICTT}. 

\subsection{TMC design and market behavior}\label{sec:Overall_market}
In the literature, a transaction fee has been used to prevent undesirable market behavior like frequent selling. For example, \cite{brands2020tradable} apply a small transaction fee of 0.01 euro to prevent frequent selling in their experiment. However, it has also been shown that transaction fees could reduce system efficiency \citep{nie2012transaction}. Our analysis in Section \ref{sec:Sec4} shows that a fixed transaction fee has the effect of preventing multiple transactions while the proportional transaction fees has the effect of making one sell as soon as possible when the conditional profit is positive (if buying is required at the time of the next trip). As an alternative to transaction fees, the regulator may also impose a minimum threshold (in dollar amounts) below which transactions are not permitted. 

Numerical experiments in this section examine the effect of proportional and fixed transaction fees on social welfare and undesirable behavior. Specifically, undesirable behavior is defined as buying back tokens sold previously. We deem this undesirable because we would like to have users strictly being either sellers or buyers (not both). Sellers are the ones who travel in the off peak (or by transit) and sell their tokens while buyers are the those with a high willingness to pay to travel by car during the peak period. %Suitable transaction fees are determined to eliminate buyback behavior while yielding the least efficiency loss. 

\begin{figure}[!ht]
  \begin{subfigure}[b]{0.48\textwidth}
    \centering\includegraphics[width=\textwidth]{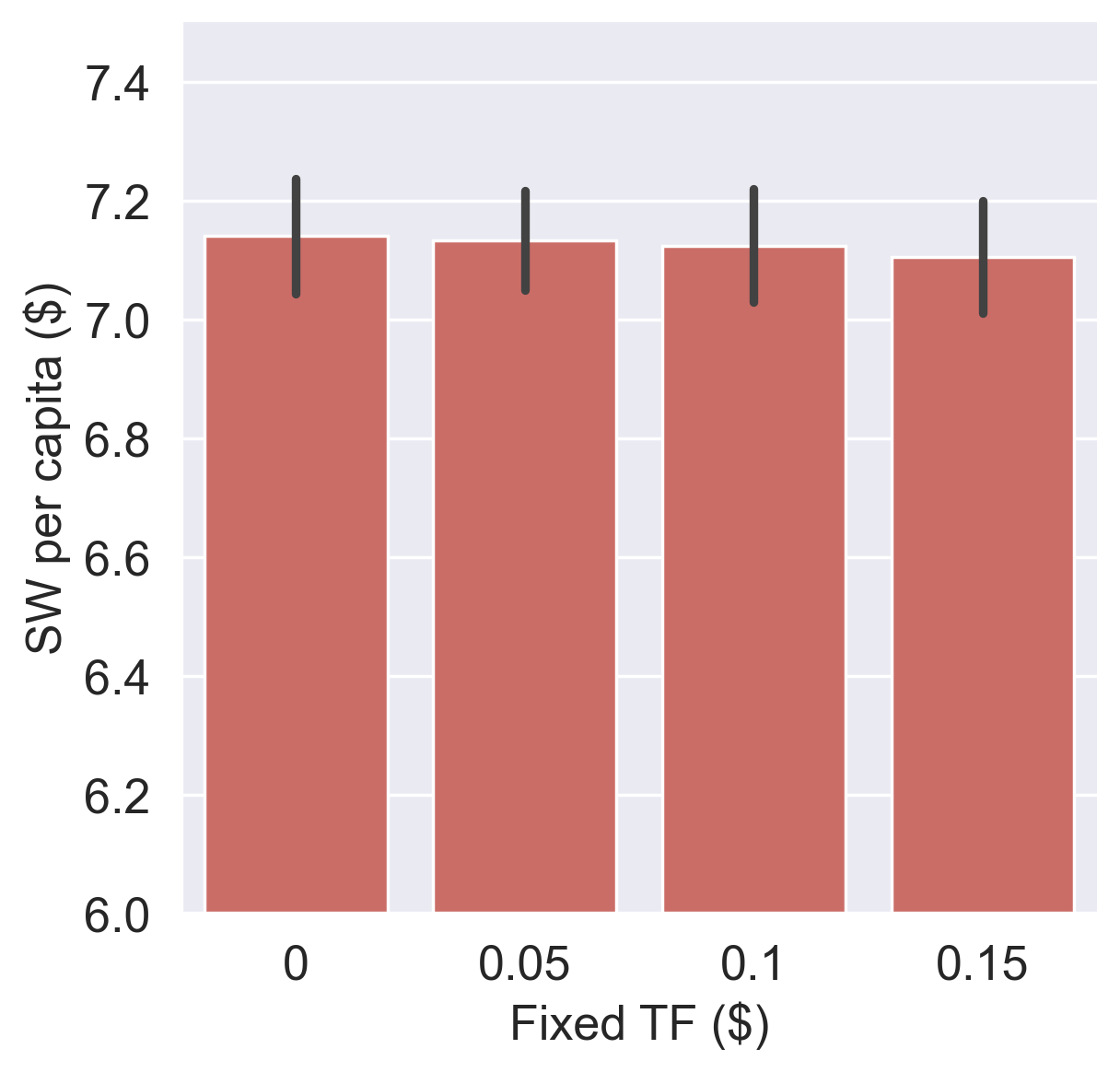}
    \caption[]{\small{Effect of Fixed TF on SW}}
        \label{fig:FTFSW}
  \end{subfigure}
    \begin{subfigure}[b]{0.48\textwidth}
    \centering\includegraphics[width=\textwidth]{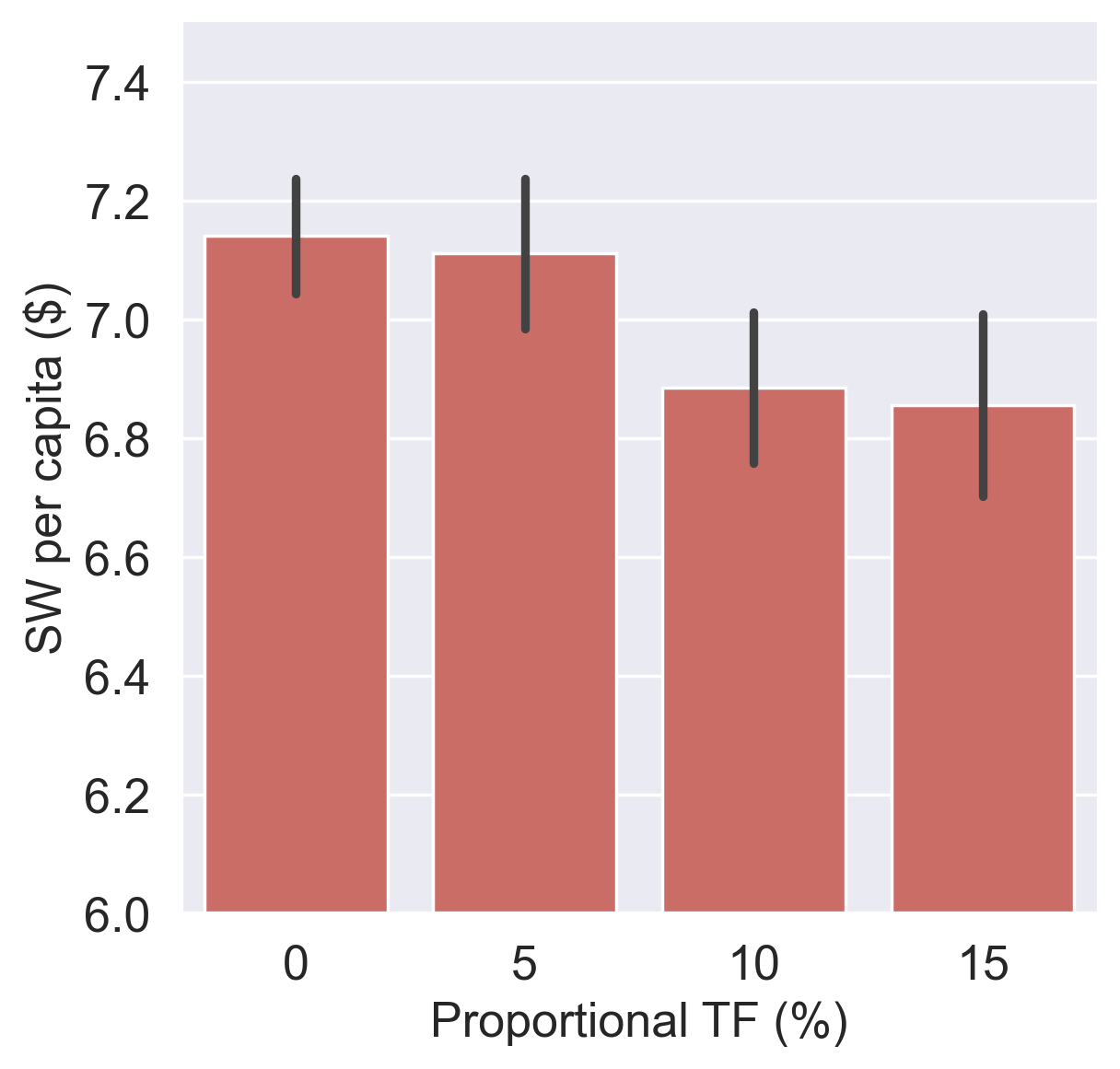}
    \caption[]{\small{Effect of Proportional TF on SW}}
     \label{fig:PTFSW}
  \end{subfigure}
  \begin{subfigure}[b]{0.48\textwidth}
    \centering\includegraphics[width=\textwidth]{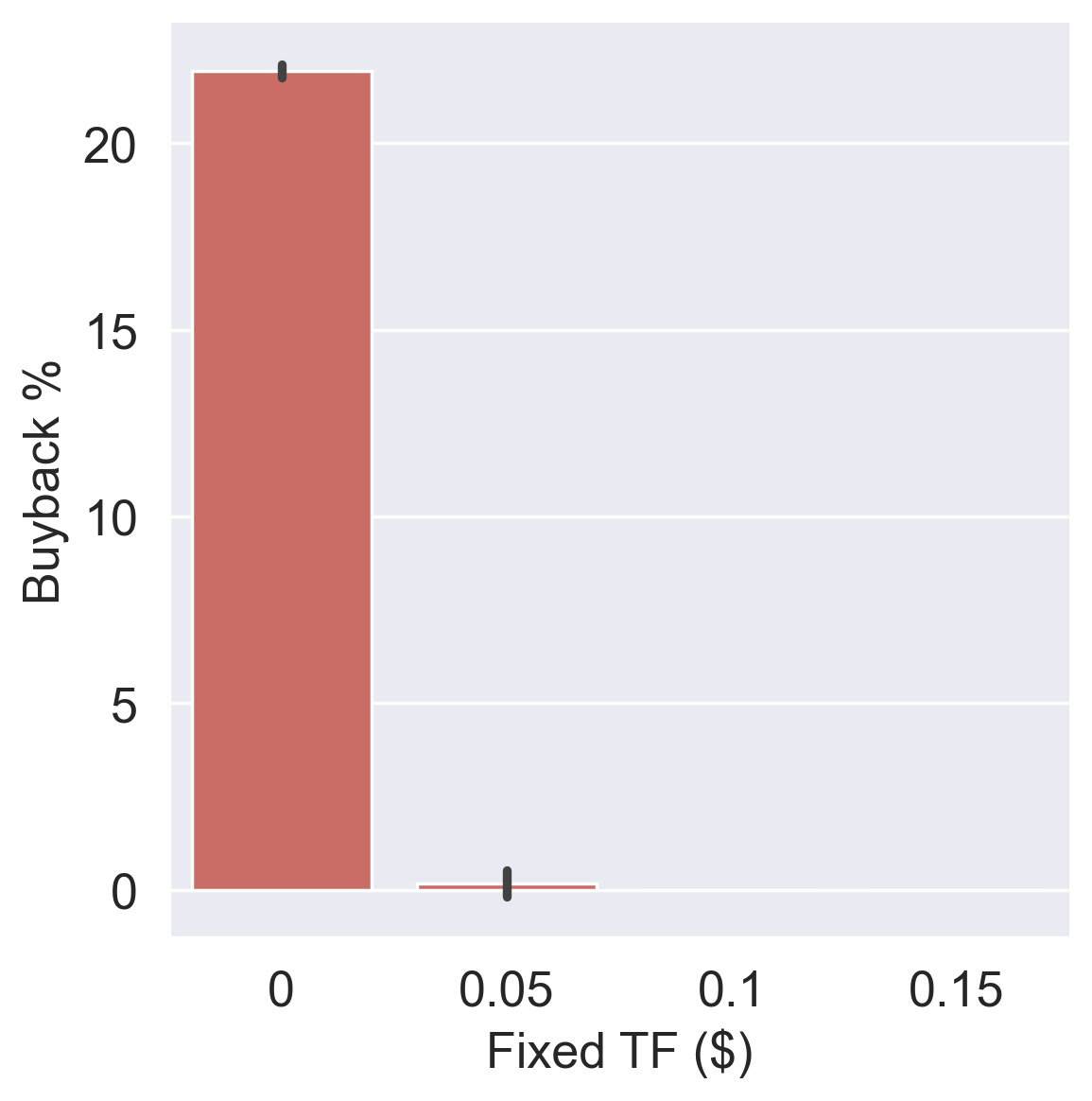}
    \caption[]{\small{Effect of Fixed TF on Buyback}}
     \label{fig:FTFBuy}
  \end{subfigure}
   \begin{subfigure}[b]{0.48\textwidth}
    \centering\includegraphics[width=\textwidth]{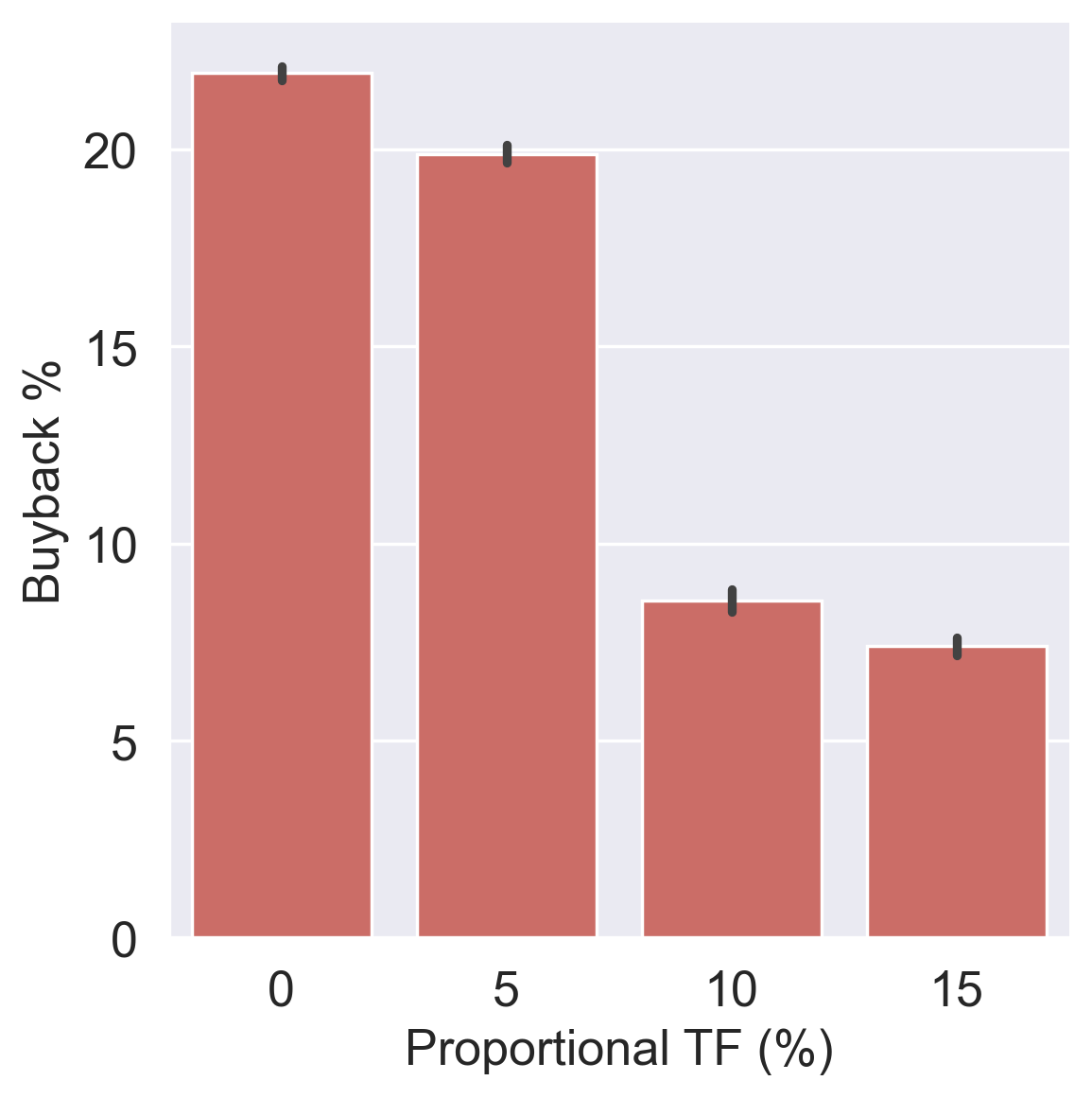}
    \caption[]{\small{Effect of Proportional TF on Buyback}}
     \label{fig:PTFBuy}
  \end{subfigure}

  \caption{The effect of fixed and proportional transaction fees on social welfare and buyback behavior }
  \label{fig:TF}
\end{figure}

For simplicity, the fixed transaction fees of buying and selling are varied together with the proportional transaction fees set to zero and vice versa. The effects of fixed and proportional transaction fees on social welfare and buyback behavior are shown in Figure \ref{fig:TF}. From the simulation experiments, a small fixed transaction fee (5 cents in this study) is seen to be able to eliminate buyback behavior in Figure \ref{fig:FTFBuy} and reduce welfare only slightly in Figure \ref{fig:FTFSW}. On the other hand, there are higher social welfare losses in the case of the proportional transactions fees in Figure \ref{fig:PTFSW}, which is also less effective in reducing buyback behavior in Figure \ref{fig:PTFBuy}. This is consistent with the findings from the literature on efficiency losses from proportional transaction costs \citep{nie2012transaction}. 

Next, we examine the effect of full initial account balances and random initial account balances on the transaction numbers and amount by time-of-day at equilibrium. The plots are based on simulations with a particular random seed because stochasticity can make the visualizations hard to interpret.% but findings and insights from the plots are general.

\begin{figure}[!ht]
  \begin{subfigure}[b]{0.52\textwidth}
    \centering\includegraphics[width=\textwidth]{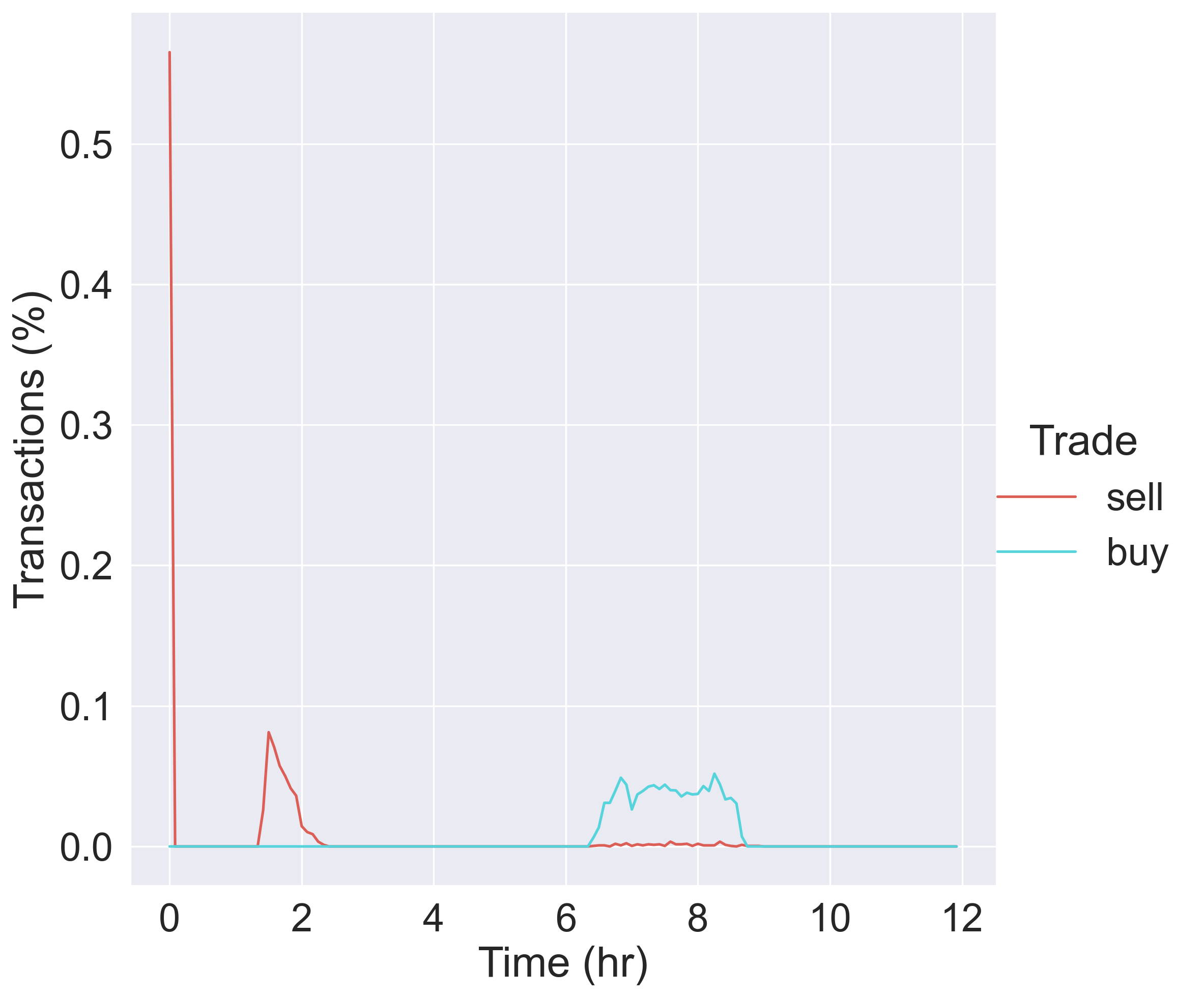}
    \caption[]{\small{Number of transactions: full wallets}}
        \label{fig:FWTN}
  \end{subfigure}
    \begin{subfigure}[b]{0.48\textwidth}
    \centering\includegraphics[width=\textwidth]{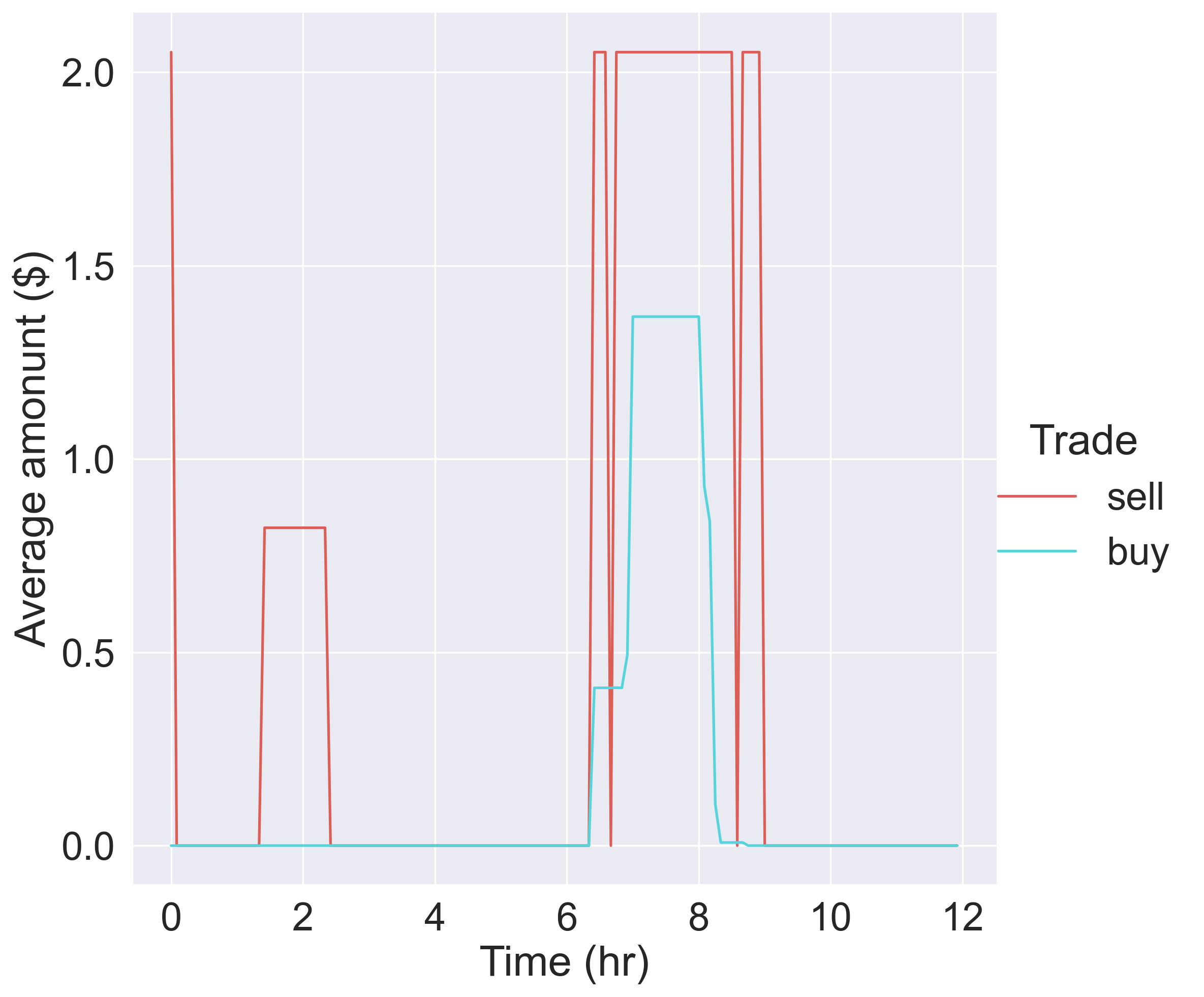}
    \caption[]{\small{Avg. trading amount: full wallets}}
     \label{fig:FWAVGA}
  \end{subfigure}
  \begin{subfigure}[b]{0.52\textwidth}
    \centering\includegraphics[width=\textwidth]{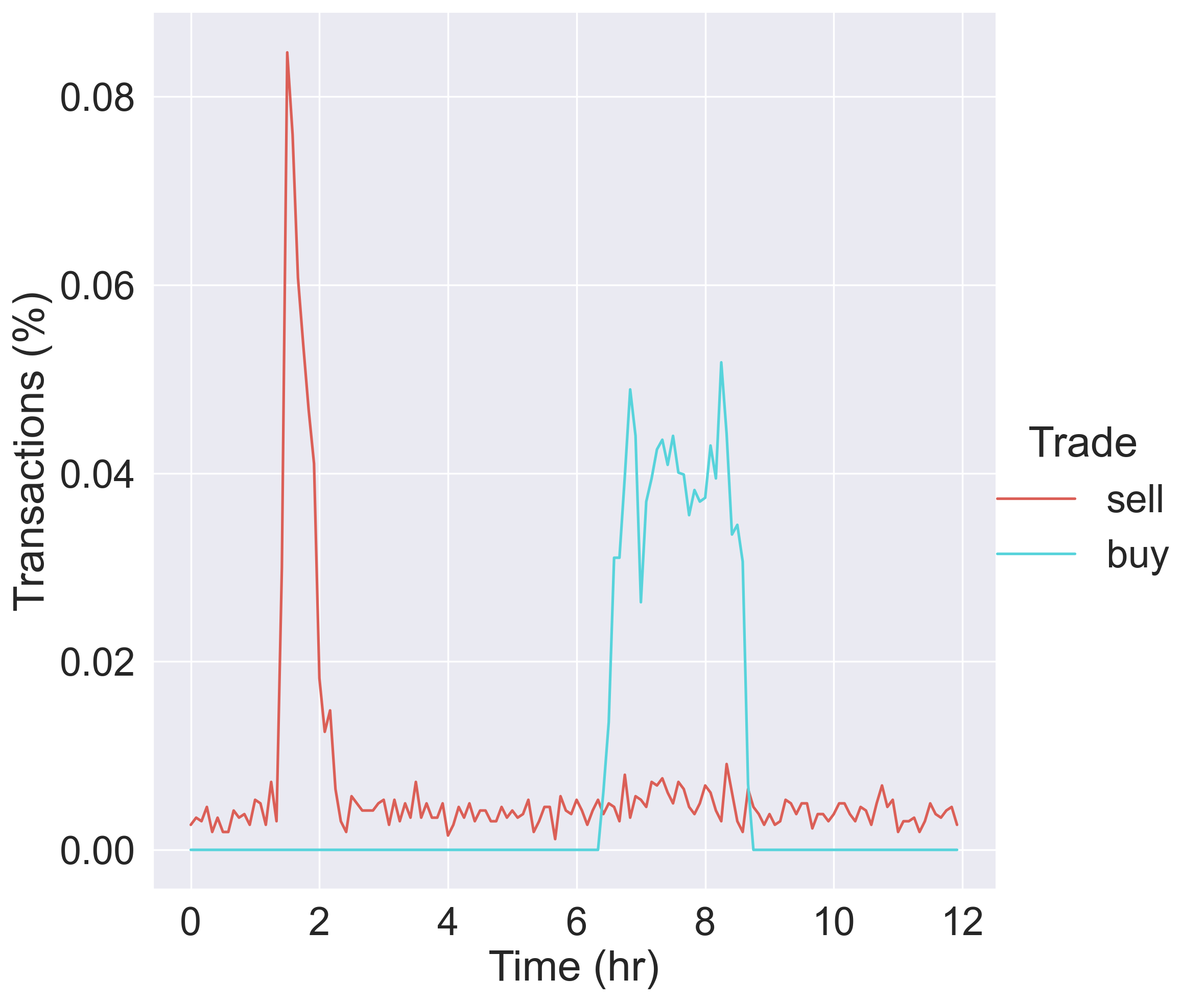}
    \caption[]{\small{Number of transactions: random wallets}}
     \label{fig:RWTN}
  \end{subfigure}
   \begin{subfigure}[b]{0.48\textwidth}
    \centering\includegraphics[width=\textwidth]{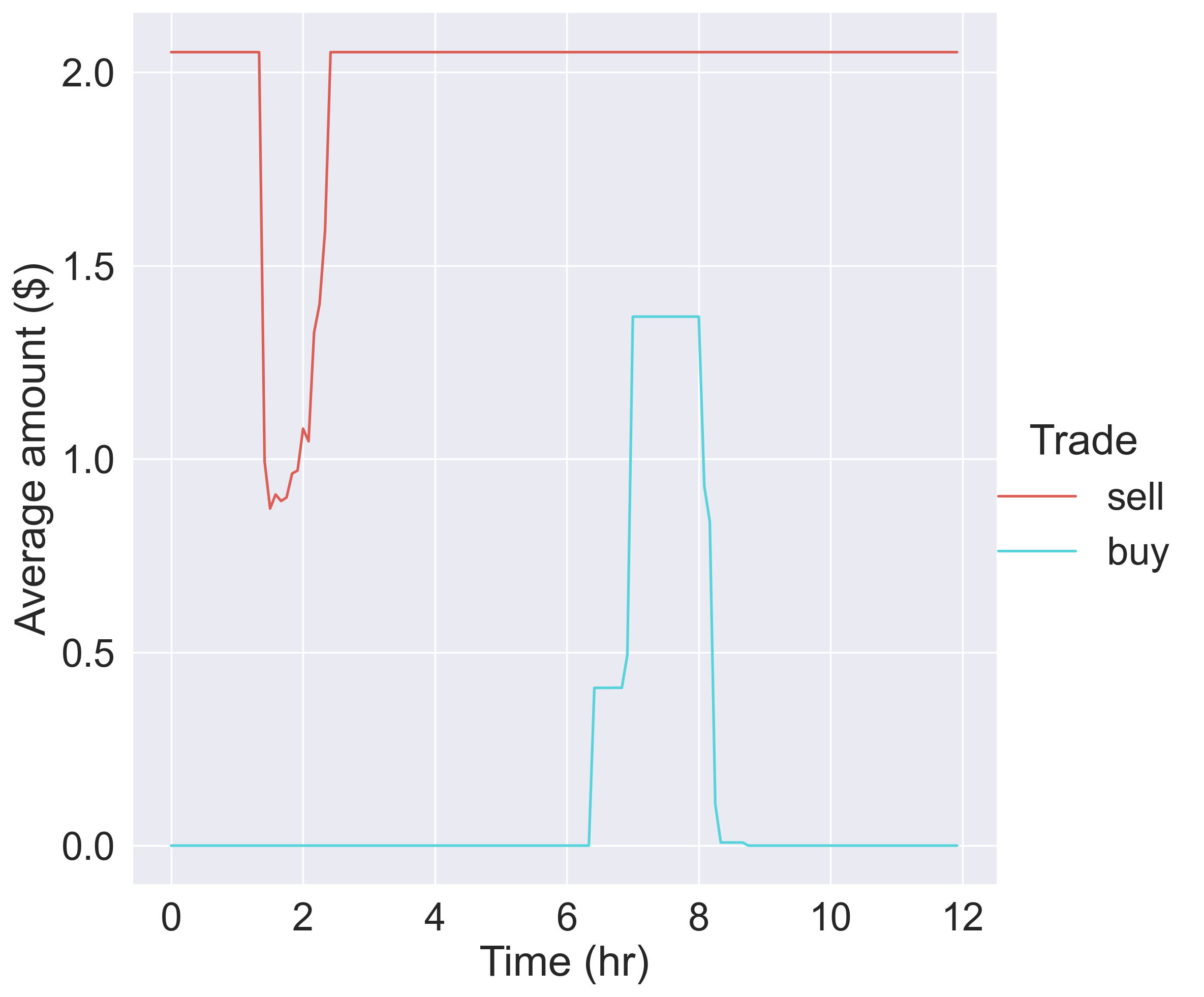}
    \caption[]{\small{Avg. trading amount: random wallets}}
     \label{fig:RWAVGA}
  \end{subfigure}

  \caption{The effects of full and random initial account balances on the transaction numbers and amount by time-of-day at equilibrium}
  \label{fig:Trading}
\end{figure}

In Figure \ref{fig:FWTN}, the numbers of buying and selling transactions as a percentage of the corresponding total number by time-of-day at equilibrium for full initial account balances are plotted. As we can see, buying transactions only happen in the peak hour because travelers can only buy tokens at time of traveling if they are short of tokens. In contrast, selling transactions happen at the beginning of the day, in the early morning, and peak hour, which can be explained by the plot of the average transaction amount by time-of-day for full initial account balances in Figure \ref{fig:FWAVGA}. For travelers selling at the beginning of the day, all of them sell at full wallets as shown in Figure \ref{fig:FWAVGA} (2.052 tokens as equilibrium token price is \$1) because they travel in the off peak and do not need to use tokens; for travelers that sell in the early morning (around 2AM), their account balances at time of selling are not full because their future token allocations until their departure times can cover their toll and it is optimal for them to sell now; for travelers that sell in the peak period, they sell at full wallets because their account balances reach the full wallet after paying small toll charges. The selling behavior is consistent with the derived selling strategy but the excessive trading at the beginning of the day may be undesirable and avoiding this was in fact one of the motivations of the continuous allocation. 
%Part of the reason is the fact that we assume error terms are perfectly correlated across days, and assumption that should be relaxed in future research. This will likely lead to more variability in day to day trading behavior and result in less concentrated trading activity at the beginning of the day. 

In practice, it is plausible that travelers will register for the program at different times in the day (one may think of the system as being implemented via a smartphone app). As a result, their account balances at the beginning of the day will be different, and we now assume the initial account balances are distributed uniformly between 0 and the maximum account balance (2.052 tokens). As shown in Figure \ref{fig:RWTN}, the selling transactions are now spread across the day with a relatively mild peak in the early morning (around 2AM). Apart from these travelers who sell in the early morning not at full wallets, other travelers sell only at full wallets as shown in in Figure \ref{fig:RWAVGA}. Under this assumption, we see a much more desirable pattern of transactions over the day, and the usefulness of the continuous allocation. However, the fact that different initial account states lead to different patterns of selling behavior at equilibrium is a problematic property of the system (despite the fact that the optimal welfare and associated market prices and flows are unique), and one that deserves further investigation.

\subsection{Performance of the TMC Scheme under varying levels of congestion, heterogeneity and income effects}\label{sec:Performance_TMC}

\begin{table}[h]
    \centering
       \caption{Factor levels for experiments}
    \label{tab:expdes}

    \begin{tabular}{ |c||c|c|c|  }

 \hline
 Factor & Level 1 &  Level 2 &  Level 3\\
 \hline
 Capacity ($s$) & -15\%& \textcolor{red}{0\%}& 15\%\\
Income Effect ($\lambda$)& 0 & \textcolor{red}{3} & 6\\
Heterogeneity (c.o.v)& 0.2 & 0.9 & \textcolor{red}{1.6}\\
 \hline
\end{tabular}
    
\end{table}

We next examine the performance of the TMC scheme relative to congestion pricing at varying levels of three important experimental factors: capacity, income effect and heterogeneity. For the TMC scheme, fixed transaction fees are set to \$0.05 and proportional transaction fees are set to 0 based on the experiments in the previous section. The factors are varied one at a time across three levels as presented in Table \ref{tab:expdes}. Values used in the base case are highlighted in red (when varying a given factor, other factors are fixed at the base level). With regard to capacity, bottleneck capacity $s$ is varied from 15\% less capacity than the baseline to 15\% more capacity than the baseline; for the income effect, the nonlinear income effect coefficient in the utility specification $\lambda$ is varied from 0 to 6; for heterogeneity, the coefficient of variation of value of time $\alpha_n$ is varied from 0.2 to 1.6. In the following discussion, the TMC system is denoted by MU (U denotes the uniform allocation of tokens) and congestion pricing is denoted by P$-$ to indicate that we do not assume a redistribution of toll revenues in any form. The No-toll scenario is denoted by NT. For each scenario in the experimental design, the two instruments and NT are simulated with five different random seeds until convergence.

\begin{figure}[h!]
  \begin{subfigure}[b]{0.55\textwidth}
    \centering\includegraphics[width=\textwidth]{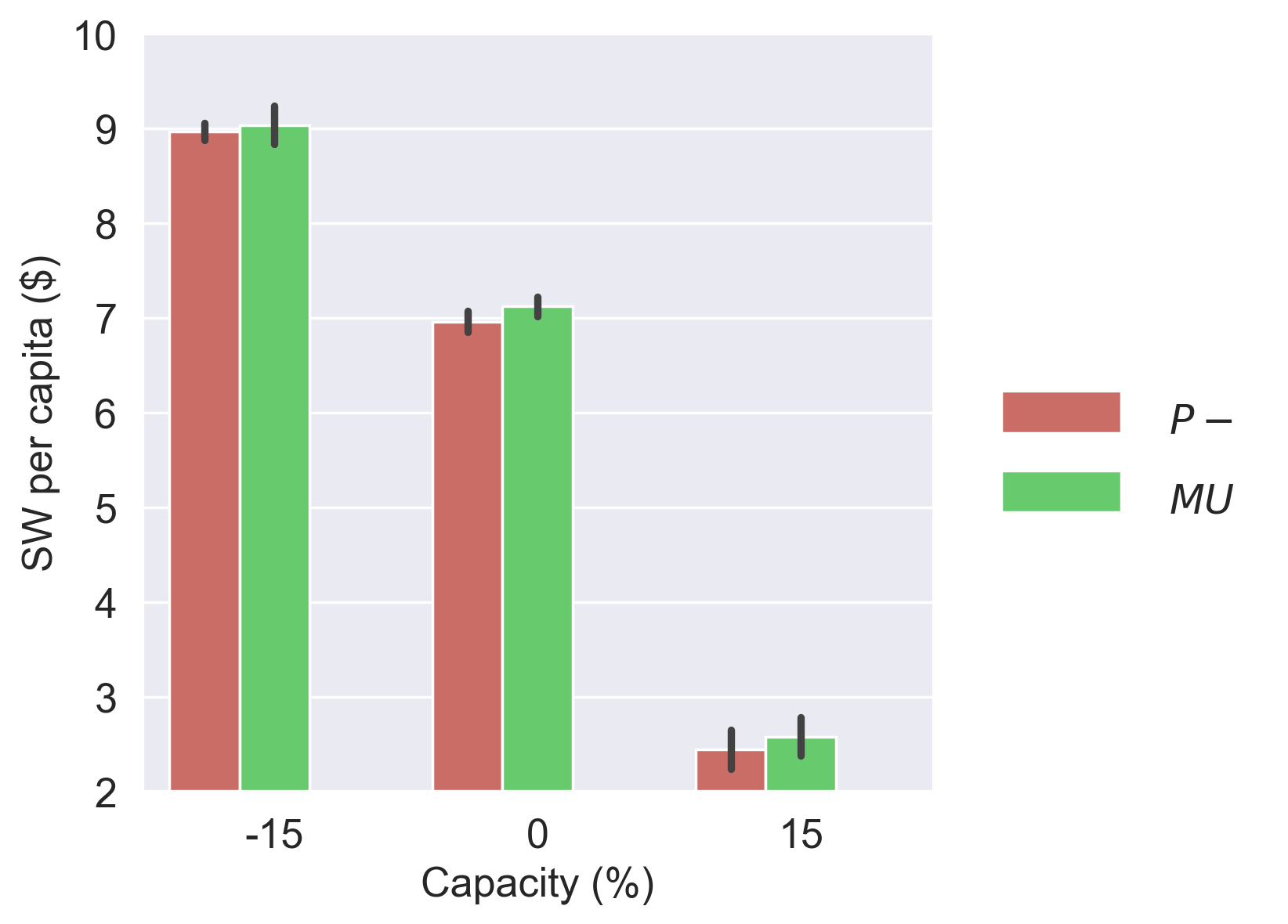}
    \caption[]{\small{Social welfare}}
    \label{fig:capSW}

  \end{subfigure}
    \begin{subfigure}[b]{0.55\textwidth}
    \centering\includegraphics[width=\textwidth]{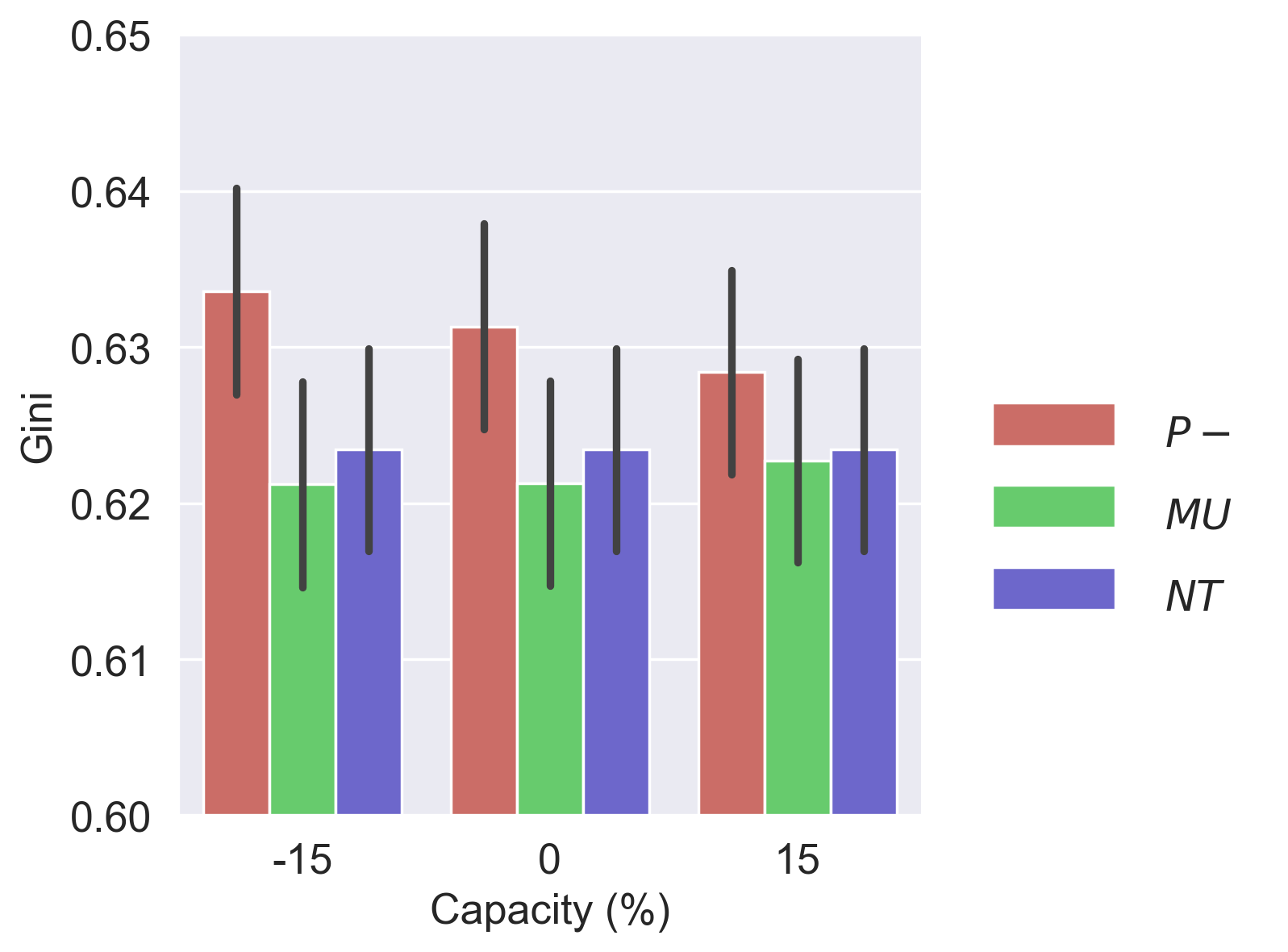}
    \caption[]{\small{Gini coefficient}}
    \label{fig:CapGini}

  \end{subfigure}
  \begin{subfigure}[b]{0.55\textwidth}
    \centering\includegraphics[width=\textwidth]{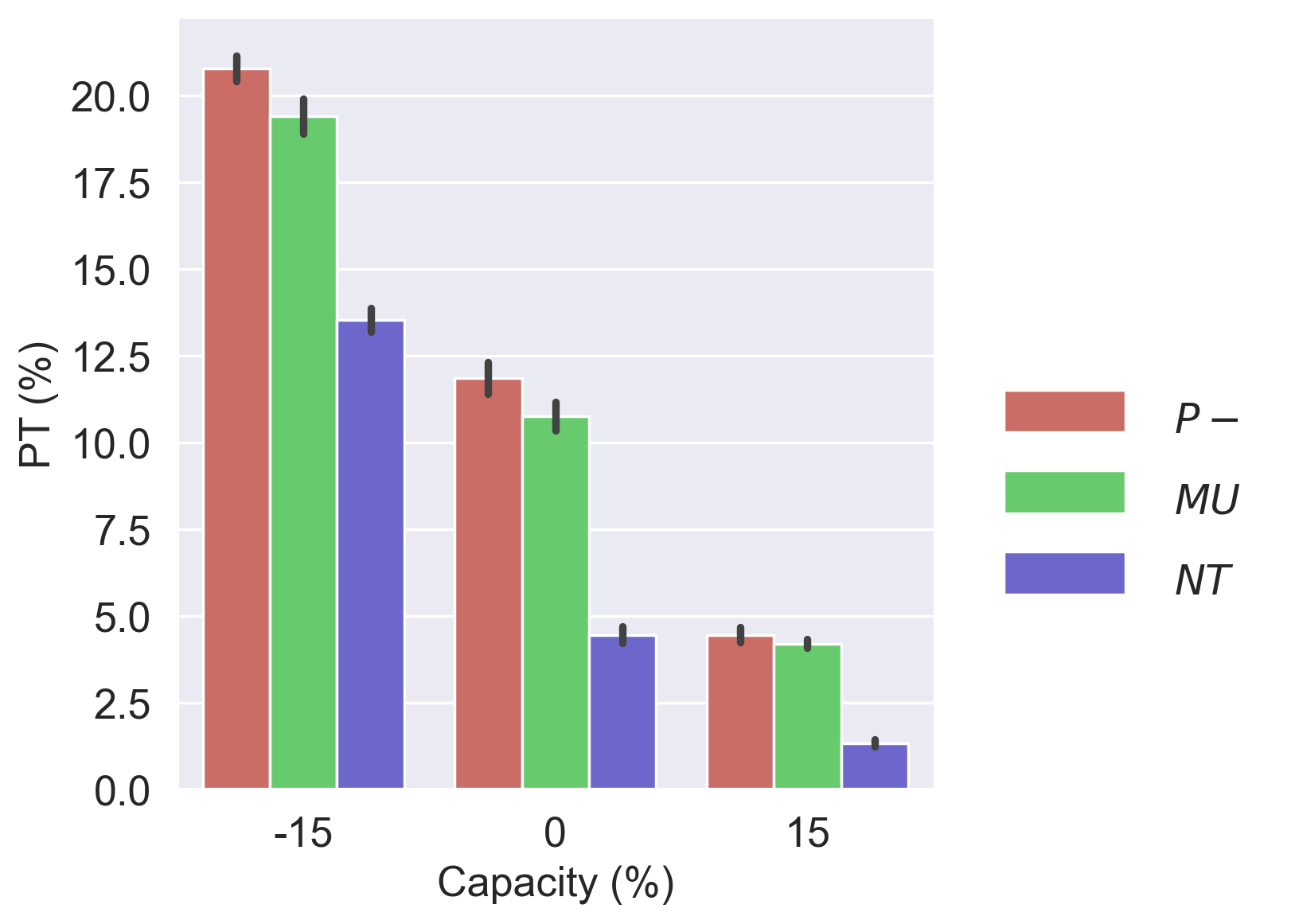}
    \caption[]{\small{PT share}}
    \label{fig:CapPT}

  \end{subfigure}
   \begin{subfigure}[b]{0.55\textwidth}
    \centering\includegraphics[width=\textwidth]{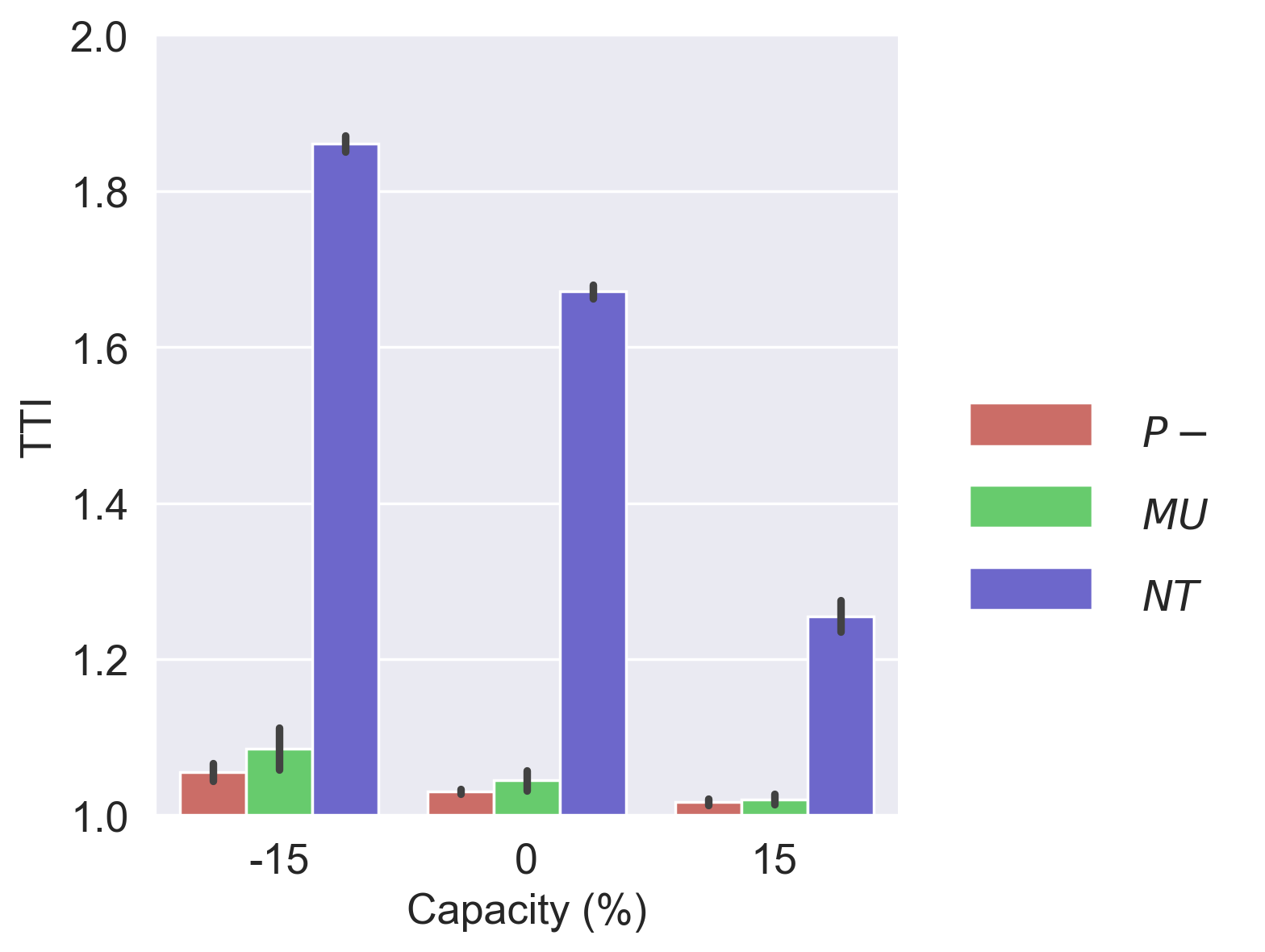}
    \caption[]{\small{Travel Time Index }}
    \label{fig:CapTTI}

  \end{subfigure}

  \caption{Variation of performance measures with capacity}
  \label{fig:cap}
\end{figure}
%Then we compare Trinity to CP with various conditions and under two different toll profile specifications. For Gaussian toll profile, we choose to use 3 Gaussian mixtures. In Table \ref{tab:res3Inc}, we present results of three different levels of income effect through varying income effect parameter $\lambda$ in behavioral model. We run 5 realizations for each case and report average social welfare (top) and Gini coefficient (bottom) with their standard deviations in parenthesis. As we can see, Gaussian toll always yields higher SW than step toll as it is continuously changeable. But in terms of equity, Gaussian toll and step toll have statistically the same Gini coefficient. Trinity also performs the same as CP but always has lower Gini coefficient.

\subsubsection{Capacity}\label{sec:Res_Cap}
The comparative performance of the various instruments under varying levels of capacity in terms of social welfare (relative to the NT scenario), Gini coefficient, PT share and travel time index (TTI) are shown in Figure \ref{fig:cap}. First, we observe that the overall welfare of pricing and the TMC scheme are similar at all levels of capacity, and in fact, marginally higher for the TMC scheme despite the small fixed transaction fees (due to the income effect). This is in line with the general finding that both pricing and tradable credits are equivalent in terms of efficiency under deterministic demand/supply and in the absence of transaction costs and income effects (\cite{Yang2011,de2018congestion}). As expected, overall welfare gains decrease as the capacity increases and congestion effects are less severe. 

\begin{figure}[h!]
  \begin{subfigure}[b]{0.48\textwidth}
    \centering\includegraphics[width=\textwidth]{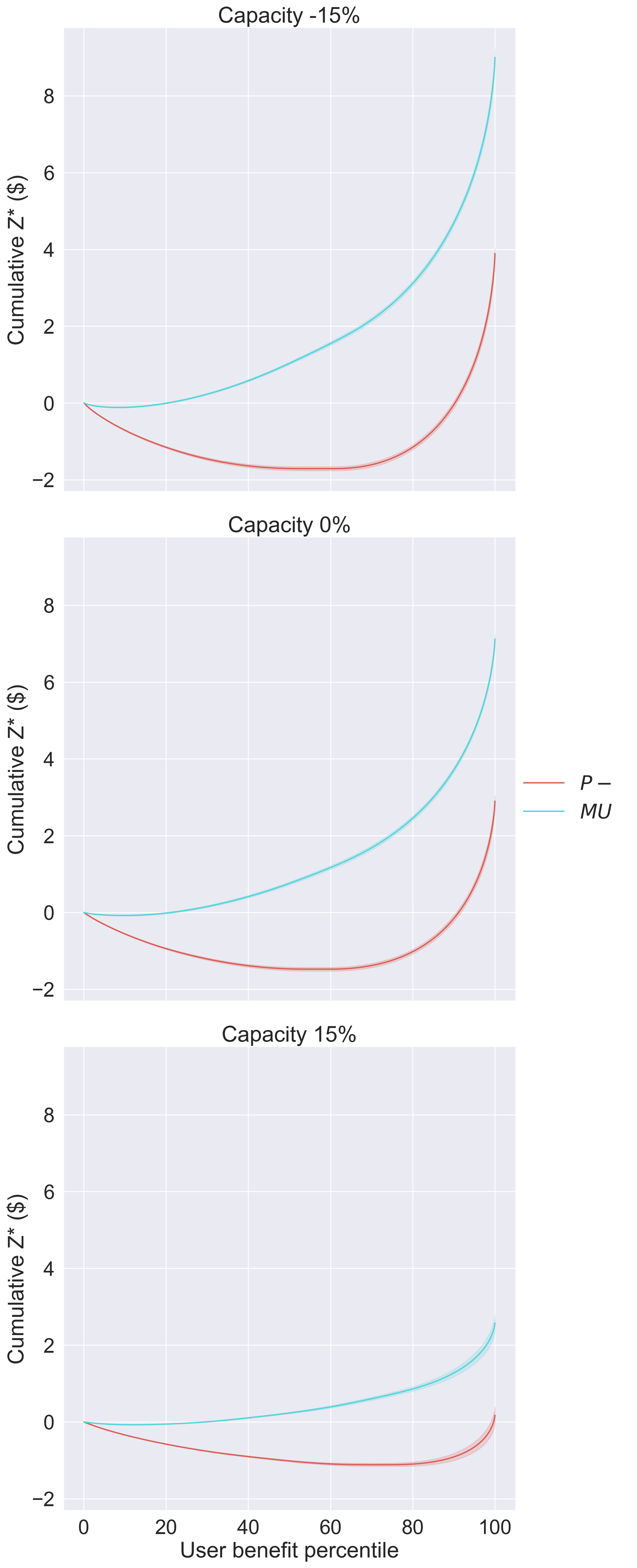}
    \caption[]{\small{Ordered by benefit}}
    \label{fig:CapUB}
  \end{subfigure}
    \begin{subfigure}[b]{0.48\textwidth}
    \centering\includegraphics[width=\textwidth]{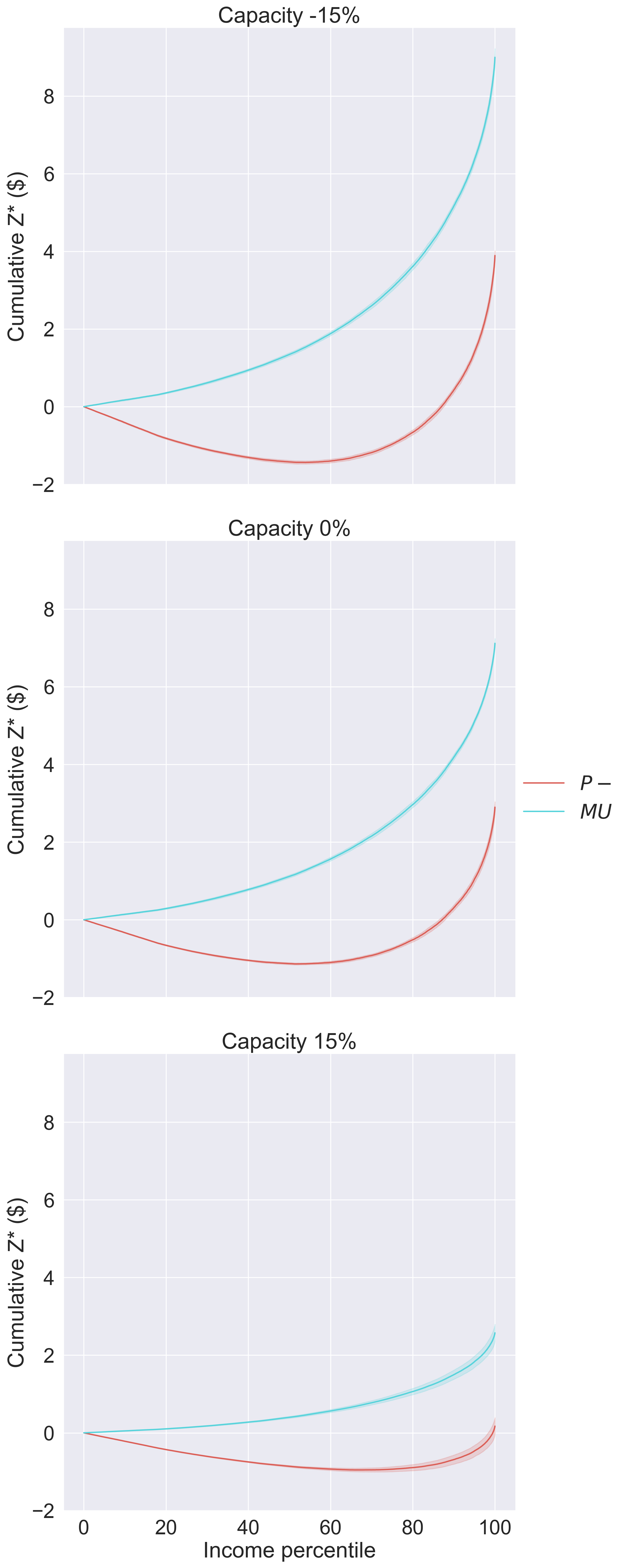}
    \caption[]{\small{Ordered by income}}
    \label{fig:CapUBI}
  \end{subfigure}

  \caption{Variation of cumulative user benefits with capacity}
  \label{fig:CapDist}
\end{figure} 

Next, Figure \ref{fig:CapGini} shows that when toll revenues are not redistributed, the congestion pricing scheme is regressive, as seen by the increase in the Gini coefficient or GC (computed based on individual disposable income $I_n$ plus user benefit $z_n$, see Equation \ref{eq:zj}) relative to the NT case. Observe also that the GC ($P-$) increases as capacity level decreases, which implies that $P-$ becomes less equitable because as capacity decreases, the tolls increase to deal with the increasing congestion leading to the greater losses of low income users. 

In contrast, the TMC scheme improves the GC relative to the NT case, due to the free uniform allocation of tokens to all travelers (a uniform allocation increases the proportion of cumulative benefits obtained by the travelers with lower values of time). The regressive nature of the congestion pricing scheme can also be observed in Figure \ref{fig:CapUB} which plots cumulative user benefits (normalized by population size) as a function of the user benefit percentile and Figure \ref{fig:CapUBI} which plots cumulative user benefits (normalized by population size) as a function of income percentile. Clearly,  at all capacity levels, one can observe that a large proportion of users are worse off  from pricing (negative user benefits) whereas in the case of the TMC scheme the proportion of `losers' is significantly smaller. Although not clearly visible in the plots, there are still some `losers' with small negative benefits in the TMC scheme. Thus, under a uniform allocation of tokens, a tradable credit scheme does not necessarily guarantee Pareto improvement. This conclusion accords with the finding in \cite{arnott1994welfare} that under pricing, even  with a uniform revenue rebate, some users may still be worse off. \cite{fan2022managing} discuss the conditions under which Pareto Improvement is guaranteed for a tradable credit scheme using the standard bottleneck model with homogeneous users. 

It should be pointed out that that the above discussion on the regressiveness of pricing is premised on the assumption that value of time is correlated with income and that there is a one-to-one relationship between VOT and income. Empirical studies have identified other correlates of income and hence, \cite{verhoef2004product} have cautioned against viewing the VOT distribution as simply representing the income distribution (see also \cite{lehe2020winners} on this). Moreover, as noted in \cite{eliasson2006equity}, ultimately, the distributional outcomes of congestion pricing depend largely on how the toll revenues --which can be significantly larger than net user benefits-- are used. The ratio between tolls revenues and user benefits under pricing in our experiments are in the range of the empirical values reported in \cite{eliasson2006equity}.

% From Elaisson: We conclude that the two most important factors for the net impact of congestion pricing are the initial travel patterns and how revenues are used. Differences in these respects dwarf differences in other factors such as values of time. This is accentuated by the fact that the total collected charges are more than three times as large as the net benefits.

\begin{figure}[h!]
  \begin{subfigure}[b]{0.48\textwidth}
    \centering\includegraphics[width=\textwidth]{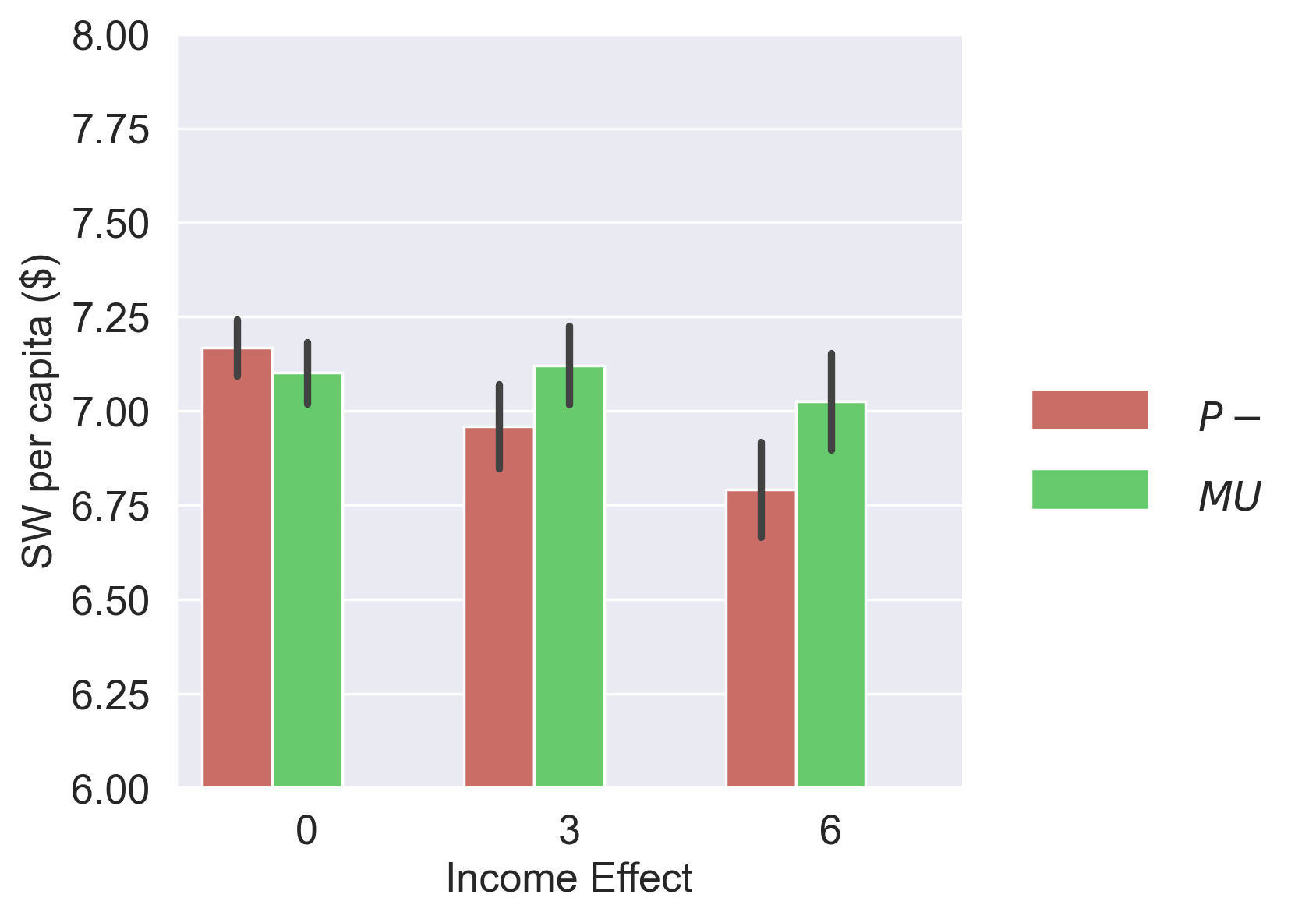}
    \caption[]{\small{Social welfare}}
    \label{fig:IncSW}
  \end{subfigure}
    \begin{subfigure}[b]{0.48\textwidth}
    \centering\includegraphics[width=\textwidth]{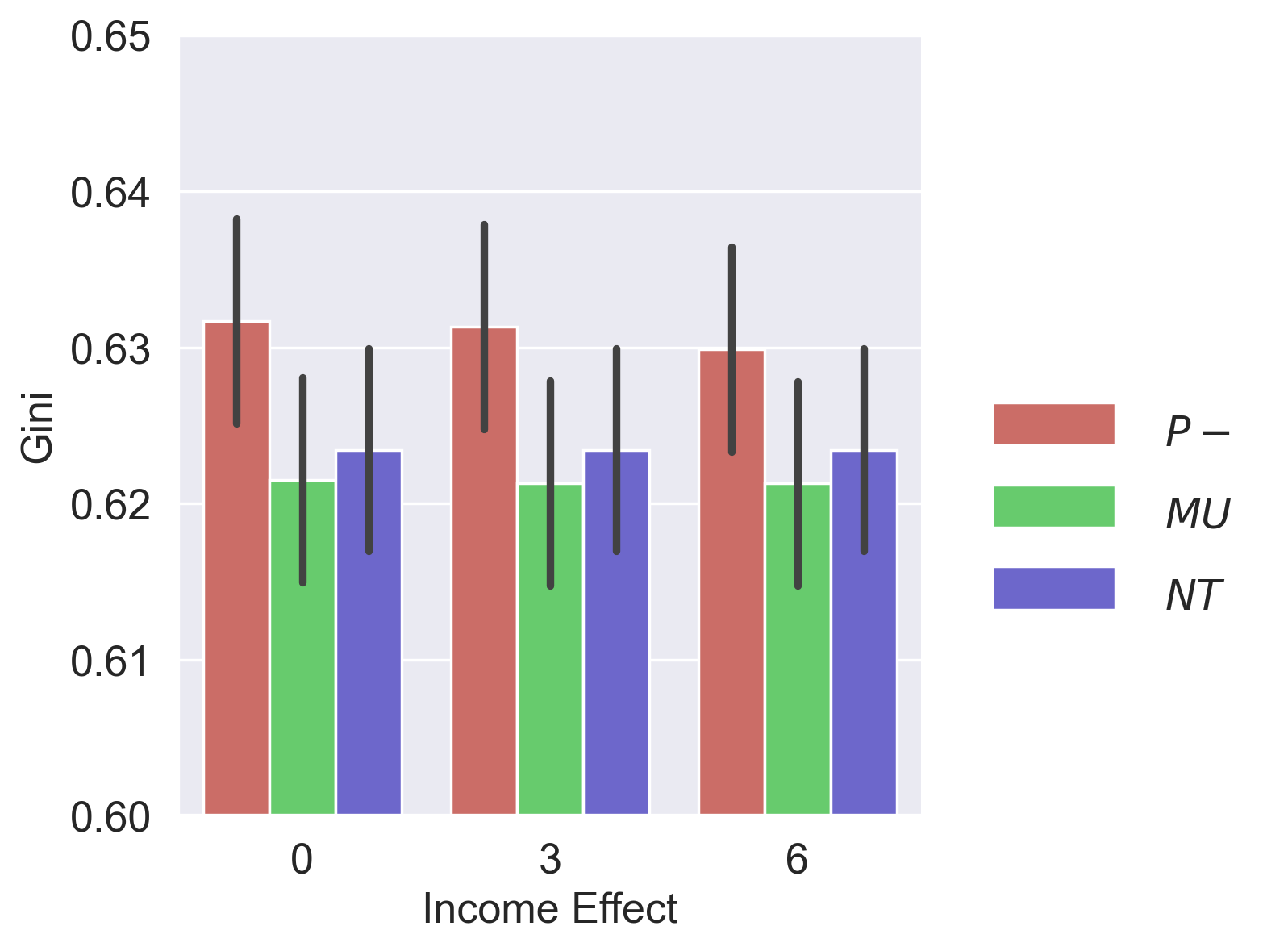}
    \caption[]{\small{Gini coefficient}}
    \label{fig:IncGini}
  \end{subfigure}
  \begin{subfigure}[b]{0.48\textwidth}
    \centering\includegraphics[width=\textwidth]{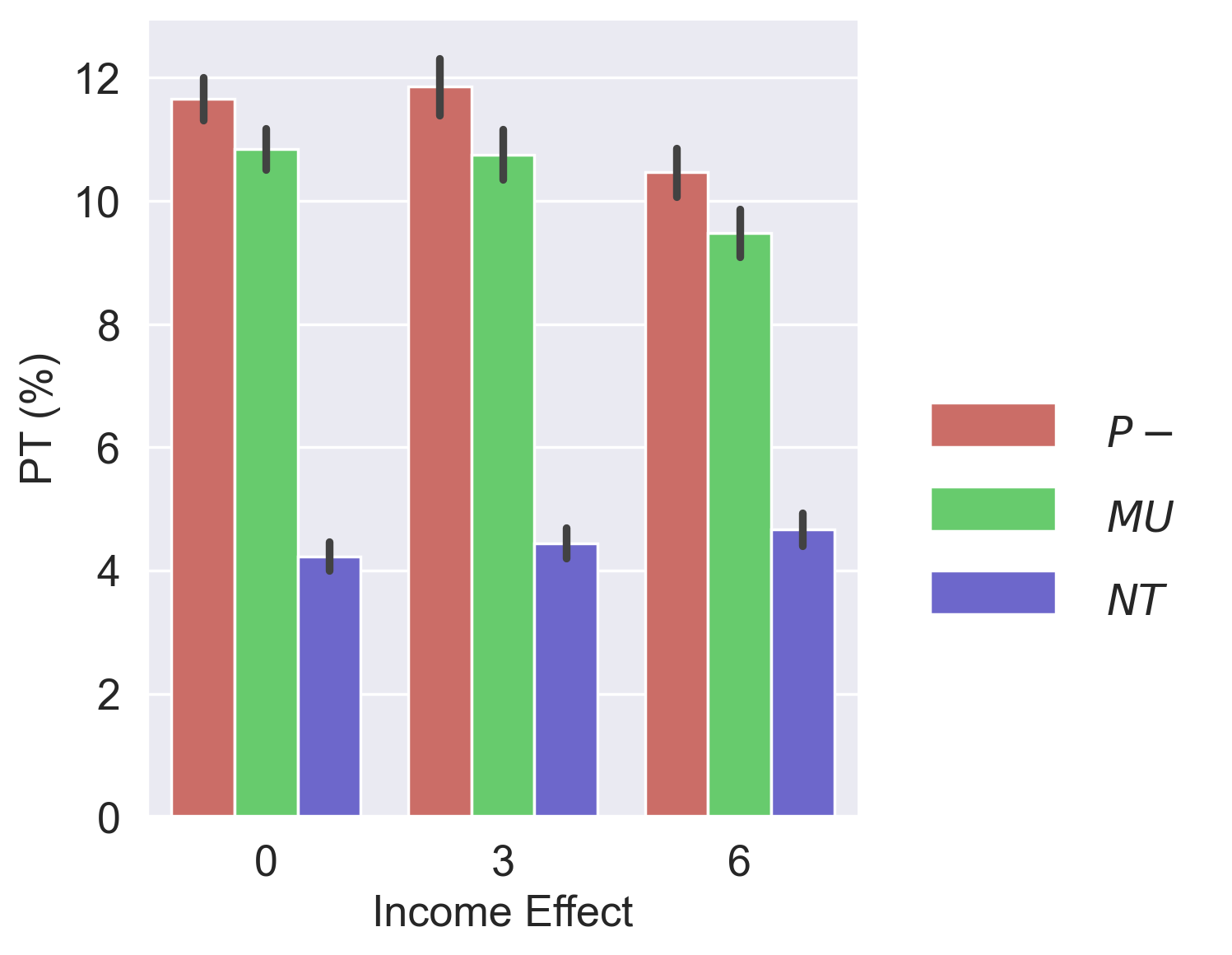}
    \caption[]{\small{PT share}}
    \label{fig:IncPT}
  \end{subfigure}
   \begin{subfigure}[b]{0.48\textwidth}
    \centering\includegraphics[width=\textwidth]{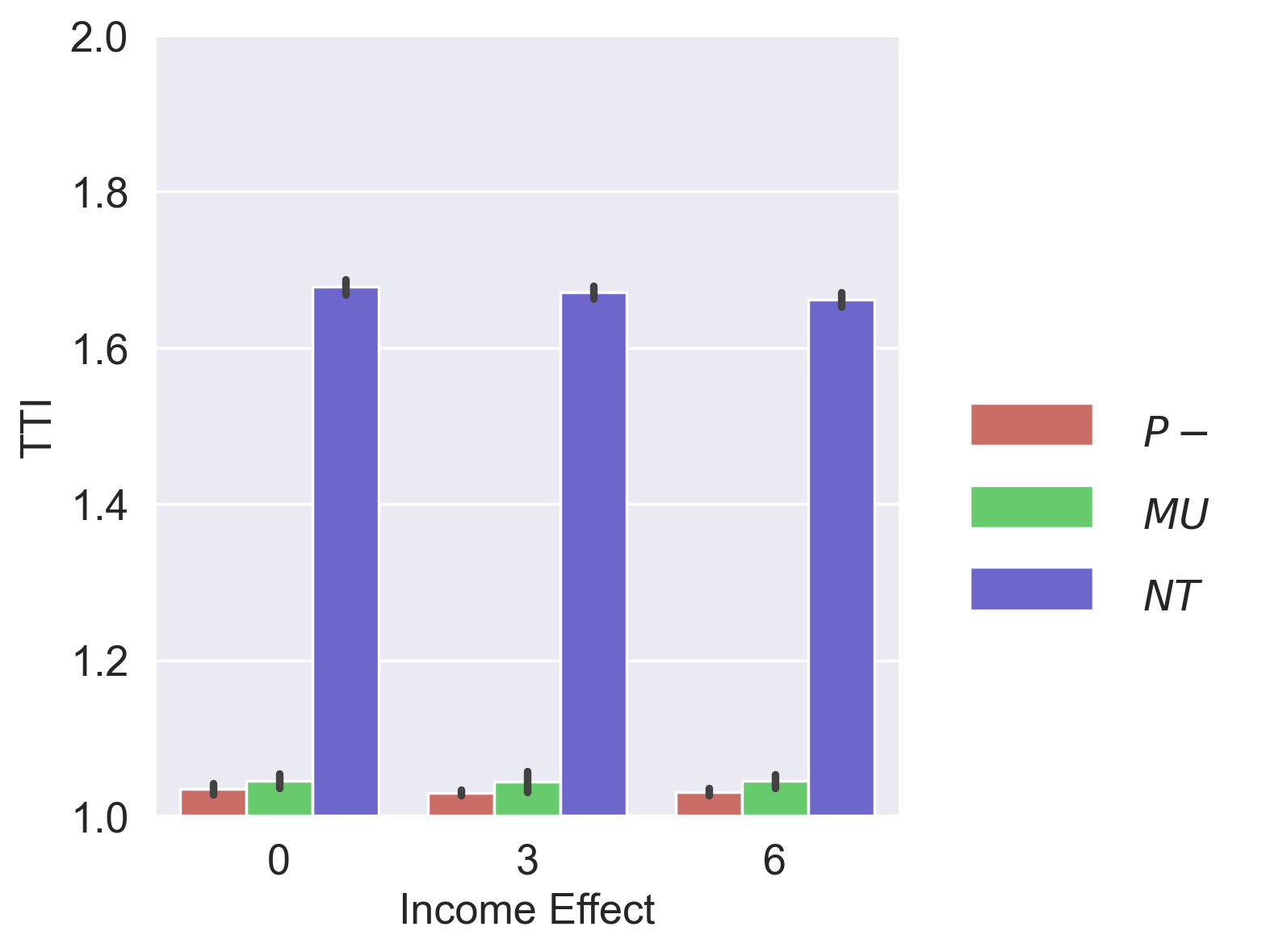}
    \caption[]{\small{Travel Time Index}}
    \label{fig:IncTTI}
  \end{subfigure}
  \caption{Variation of performance measures with income effect level}
  \label{fig:Inc}
\end{figure}

In terms of network performance, significant improvements in the TTI and larger transit shares are observed under both instruments (relative to NT) at all capacity levels (Figures \ref{fig:CapPT} and \ref{fig:CapTTI} ).  
%A similar plot of the ratios of these four metrics to the baseline values under varying levels of capacity can be found in Figure \ref{fig:Capper} in Appendix \ref{appa}. 
%The social welfare is computed relative to the NT and consists of the user benefit and regulator revenue. As it is revenue neutral for all instruments with distribution ($\delta = 1$), their social welfare are also their user benefits. For the pricing without distribution $P-$, since regulator revenue is not neutral under $P-$, it is expected that the user benefit of $P-$ is less than other instruments. This is confirmed by the following plots on distributional impacts. 

%Among all instruments, $PI_S$ achieves the highest social welfare as its distribution rule is to maximize social welfare directly. The pricing with hybrid distribution rule $PI_H$ has social welfare less than that of $PI_S$ as its revenue is distributed to compensate all users' losses (not only low-income users) to ensure Pareto improvement. The pricing with uniform distribution $PU$ has social welfare less than that of $PI_H$ as its revenue is distributed uniformly to all users including those who do not have losses.

%We can also observe than TMC with uniform token allocation $MU$ performs the same as $PU$ and TMC with personalized token allocation $MI$ performs the same as $PI$ given the effect of transaction fess are minimal. This is because the market value of token allocation is roughly equal to the dollar value of the corresponding refund, which causes similar behavior changes as the income effects are similar.

\subsubsection{Income Effect}\label{sec:Res_Income}
Next we examine the impact of income effects, measured by the parameter $\lambda$. As noted previously, non-linear income effects do impact behavior and the relative efficiency of pricing and TMCs. This impact on behavior along with the fact that the valuation of the token allocation is higher for lower income users (who have a higher marginal utility of income - recall that our measure of user benefit is directly the money metric utility) results in increasing welfare differences as $\lambda$ increases (this can be seen in Figure 
\ref{fig:IncSW}). However, we caution that in terms of magnitudes, these differences are still small. As expected, when income effects are absent, the welfare of the TMC is marginally lower than P- due to transaction fees. Observe also that the social welfare of $P-$ decreases as the income effect increases since users are more sensitive to the tolls.

The impact on behavior can also be seen in the lower PT shares with increasing $\lambda$ in Figure \ref{fig:IncPT}. In terms of distributional impacts, we observe similar trends in the GC and patterns in user benefits as described in Section \ref{sec:Res_Cap} (which show the regressiveness and larger proportion of `losers' under pricing; see the base capacity case in Figures \ref{fig:CapUB} and \ref{fig:CapUBI}). There are small variations across $\lambda$ and hence, for brevity, we omit plots of user benefits. 

%noting that in the absence of the income effects, all instruments are expected to have almost identical social welfare given that the losses due to the TMC transaction fees are minimal. This is because the distribution does not change the user behavior as utility differences are the same. The distribution also does not bring additional benefit to low-income users because their marginal utilities of income are 1. This is demonstrated in the following experiments on varying levels of the income effects. Considering the nonlinear income effects, the results indicate that the behavior changes and the marginal utility benefit of income due to the refunding and the credit allocation lead to small gains in welfare. This is promising given the other benefits, especially the distributional benefits, shown in next.

%\begin{figure}[!h]
%  \begin{subfigure}[b]{0.48\textwidth}
%    \centering\includegraphics[width=\textwidth]{Plots/MD_comparison/IncUB.png}
%    \caption[]{\small{Lorenz curve of user benefits}}
%    \label{fig:IncUB}
%  \end{subfigure}
%    \begin{subfigure}[b]{0.48\textwidth}
%    \centering\includegraphics[width=\textwidth]{Plots/MD_comparison/IncUBI.png}
%    \caption[]{\small{Distribution of user benefits by income}}
 %   \label{fig:IncUBI}
 % \end{subfigure}

 % \caption{Distributional impacts of various instruments by income effect levels}
 % \label{fig:IncDist}
%\end{figure}

\begin{figure}[h!]
  \begin{subfigure}[b]{0.48\textwidth}
    \centering\includegraphics[width=\textwidth]{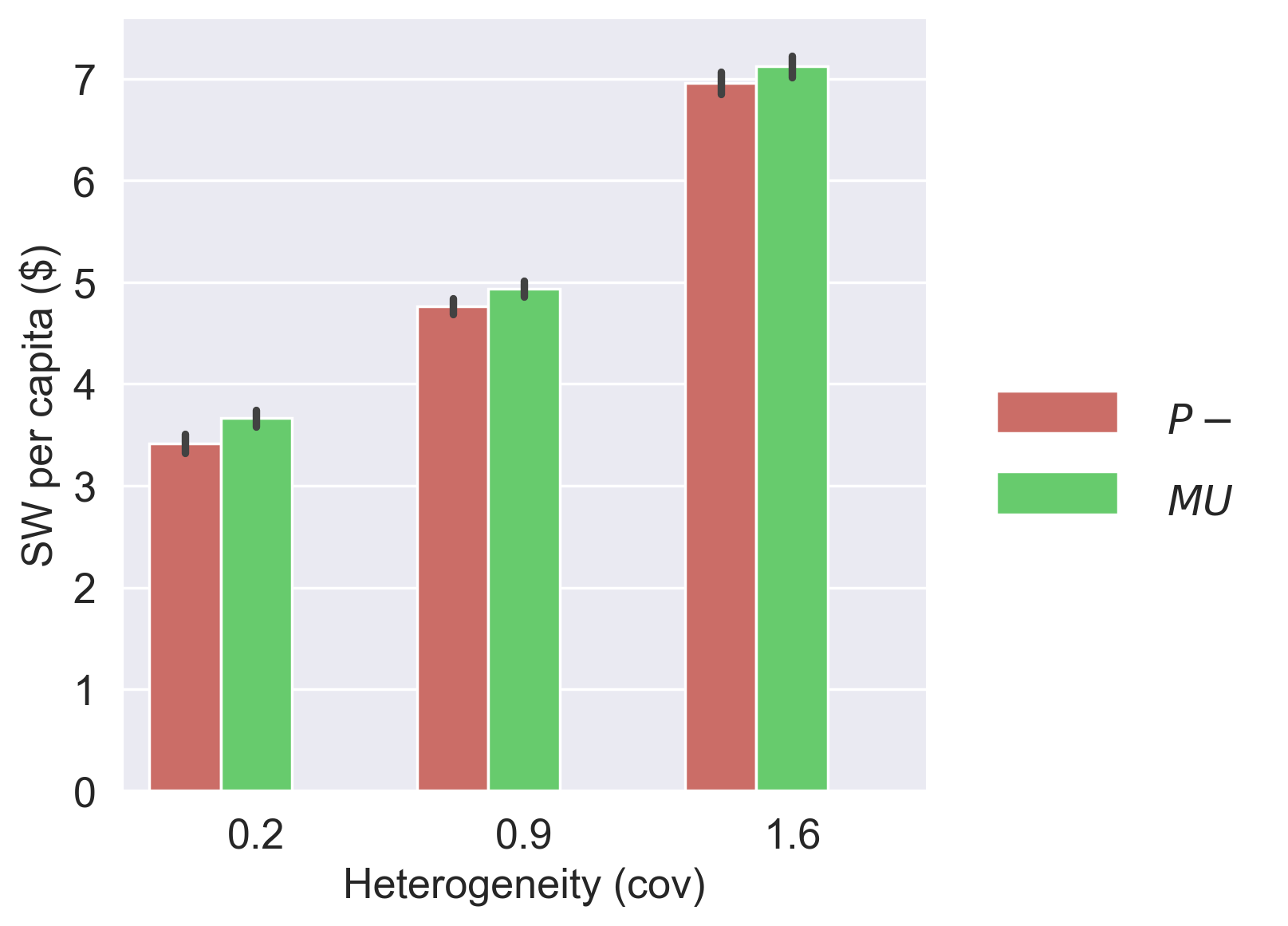}
    \caption[]{\small{Social welfare }}
    \label{fig:HeteroSW}

  \end{subfigure}
    \begin{subfigure}[b]{0.48\textwidth}
    \centering\includegraphics[width=\textwidth]{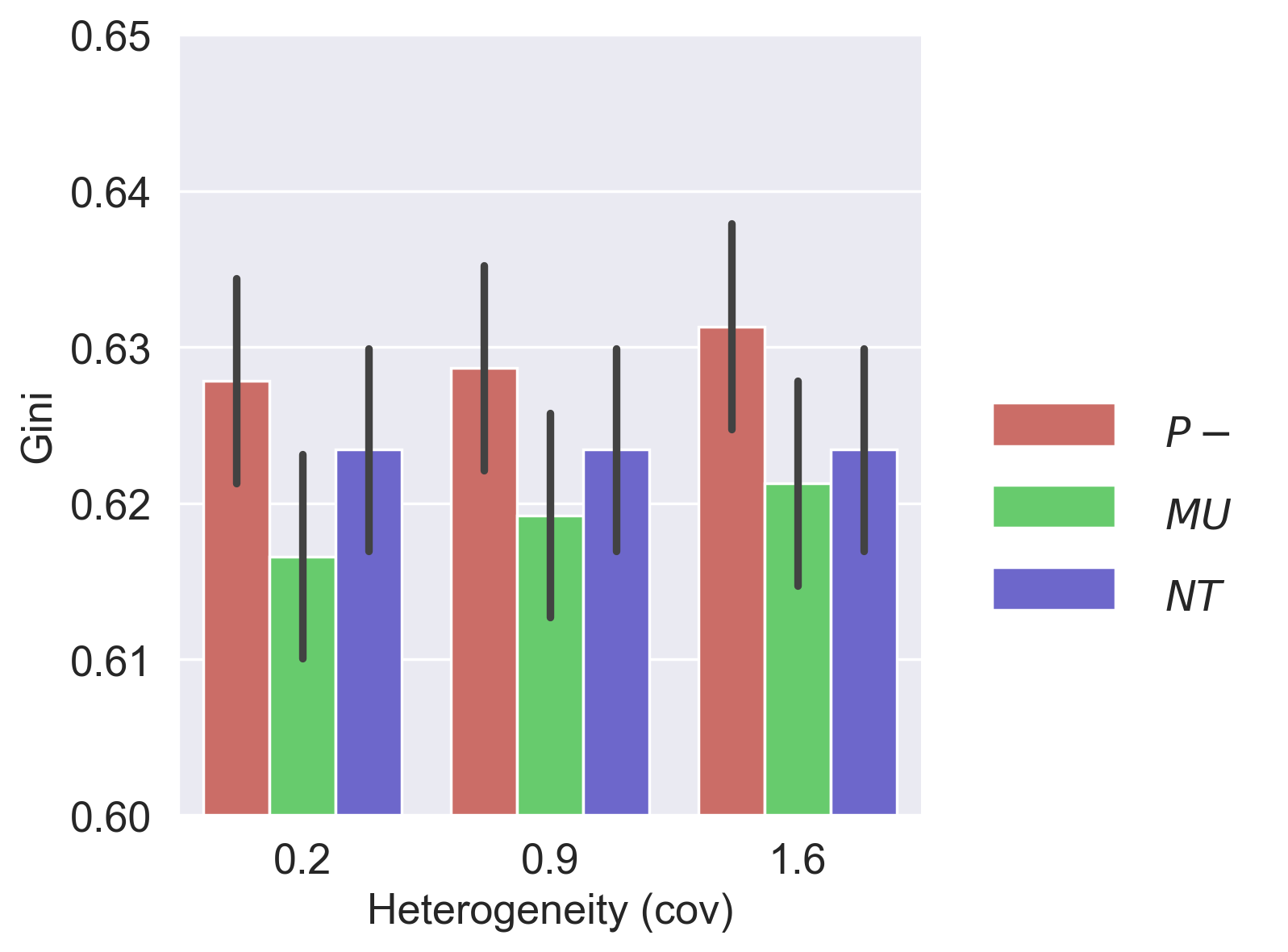}
    \caption[]{\small{Gini coefficients }}
    \label{fig:HeteroGini}

  \end{subfigure}
  \begin{subfigure}[b]{0.48\textwidth}
    \centering\includegraphics[width=\textwidth]{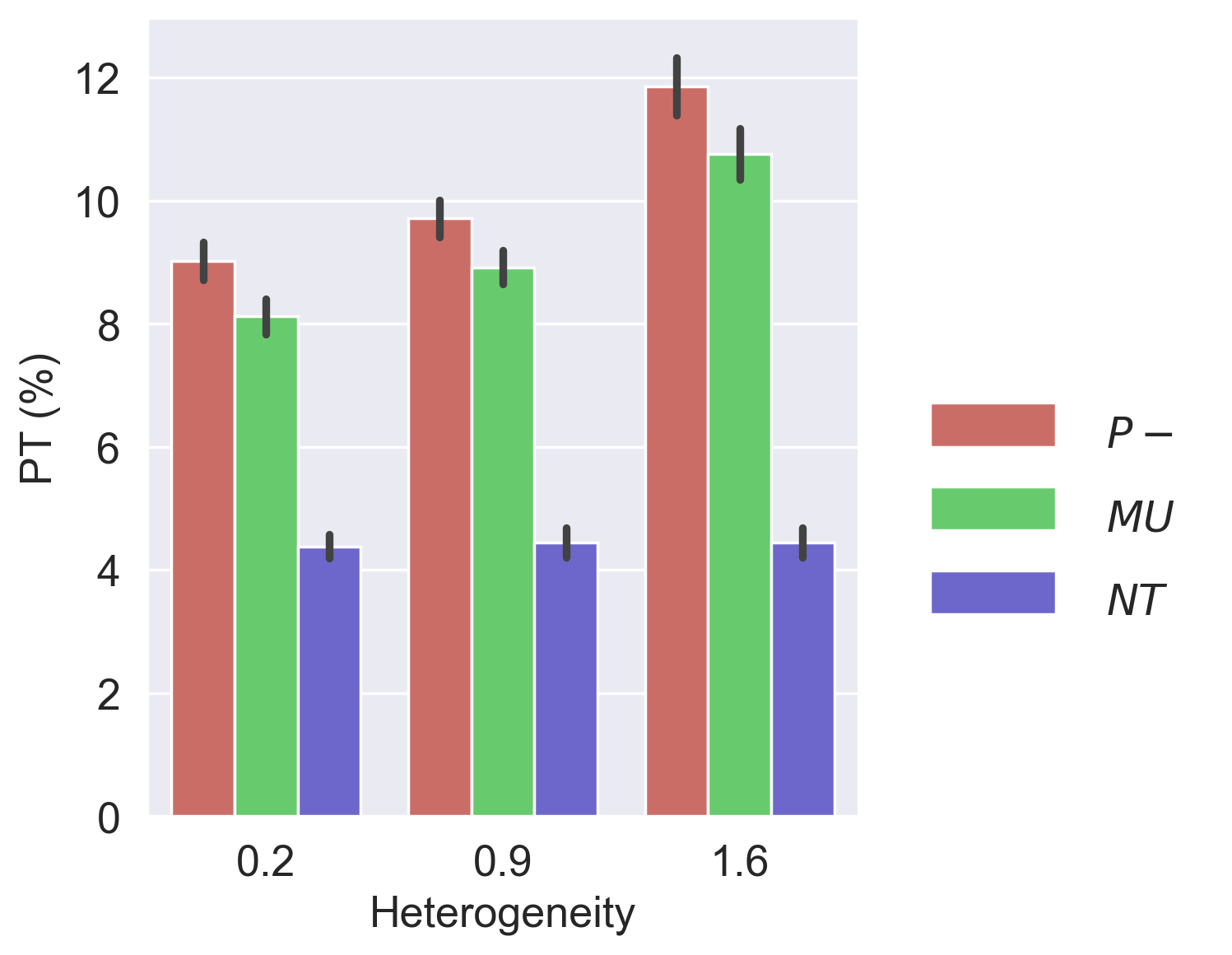}
    \caption[]{\small{PT share }}
    \label{fig:HeteroPT}

  \end{subfigure}
   \begin{subfigure}[b]{0.48\textwidth}
    \centering\includegraphics[width=\textwidth]{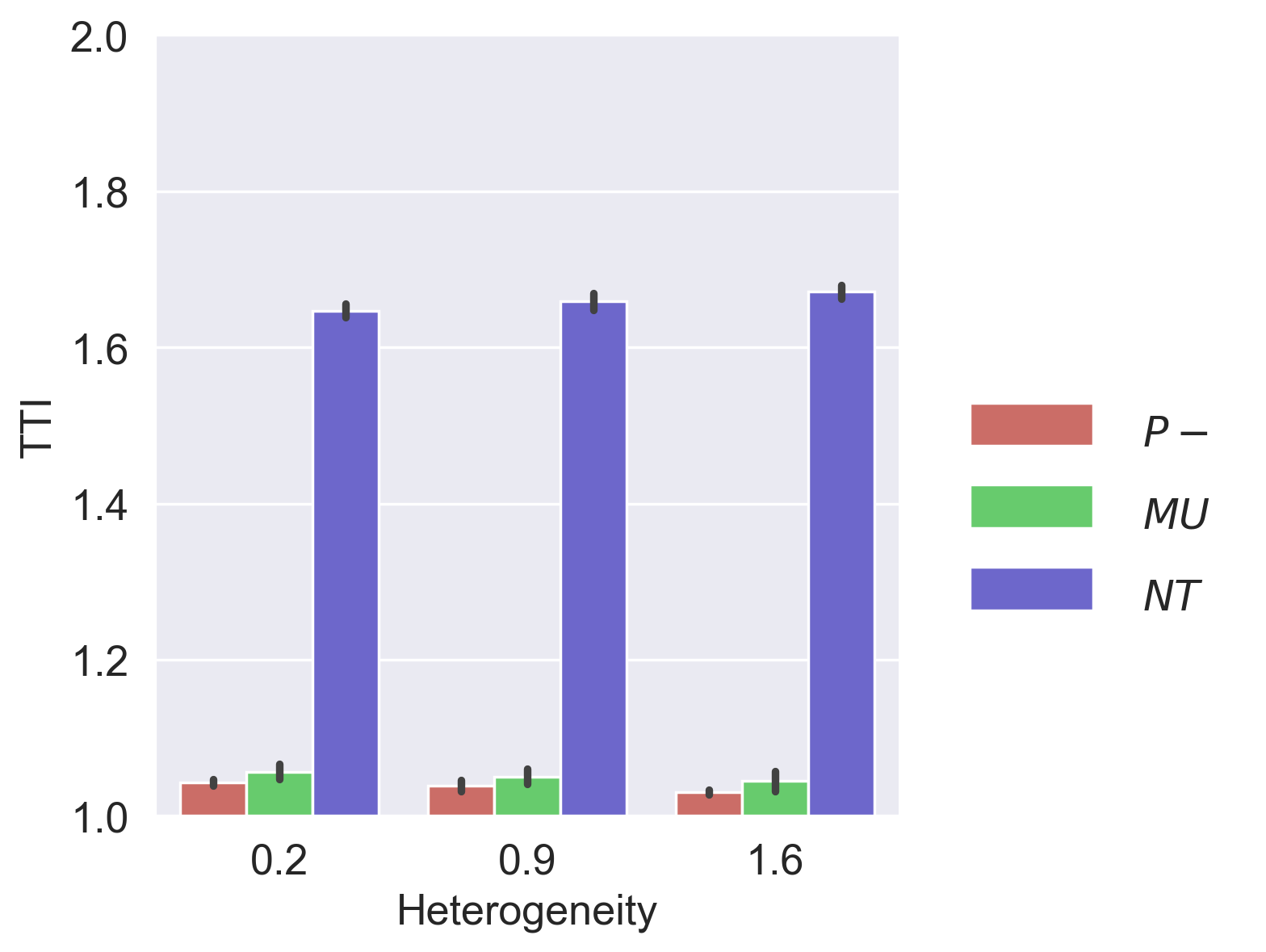}
    \caption[]{\small{TTI }}
    \label{fig:HeteroTTI}

  \end{subfigure}

  \caption{Variation of performance measures with heterogeneity levels }
  \label{fig:Hetero}
\end{figure}

\begin{figure}[h!]
  \begin{subfigure}[b]{0.48\textwidth}
    \centering\includegraphics[width=\textwidth]{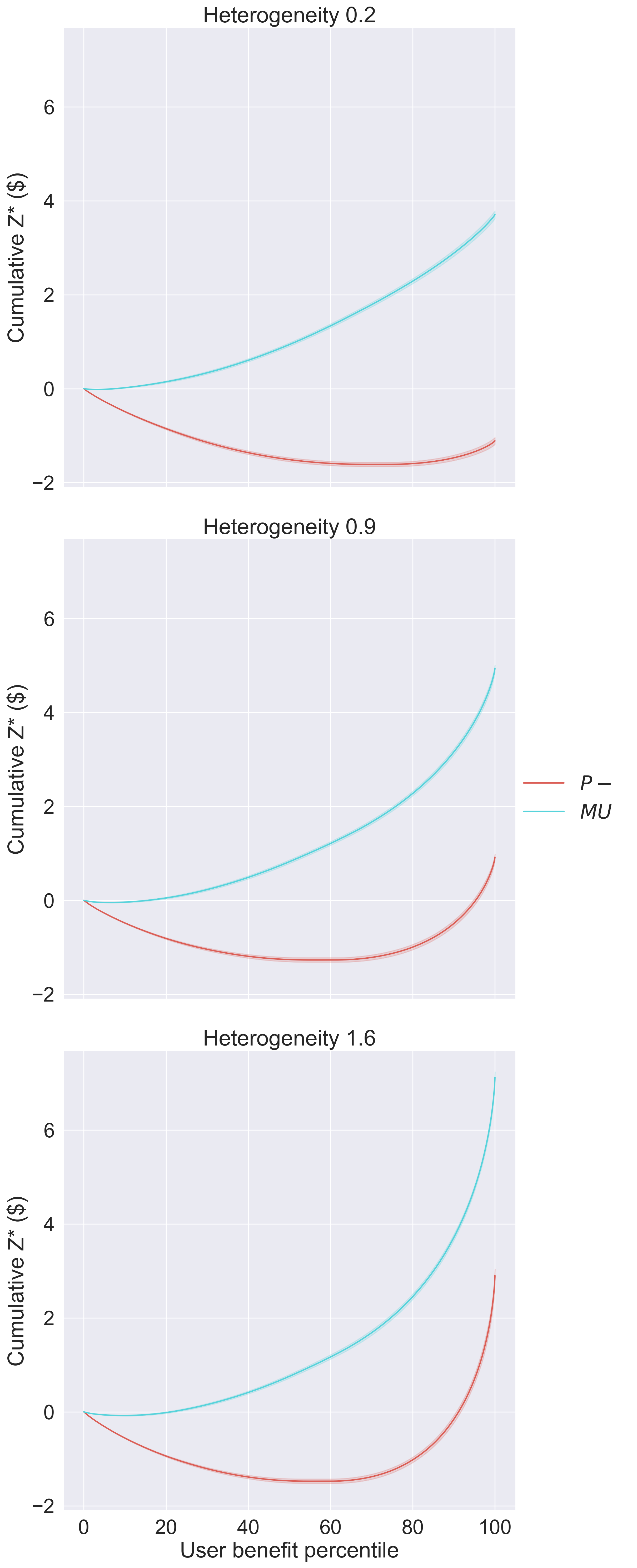}
    \caption[]{\small{Lorenz curve of user benefits}}
    \label{fig:HeteroUB}
  \end{subfigure}
    \begin{subfigure}[b]{0.48\textwidth}
    \centering\includegraphics[width=\textwidth]{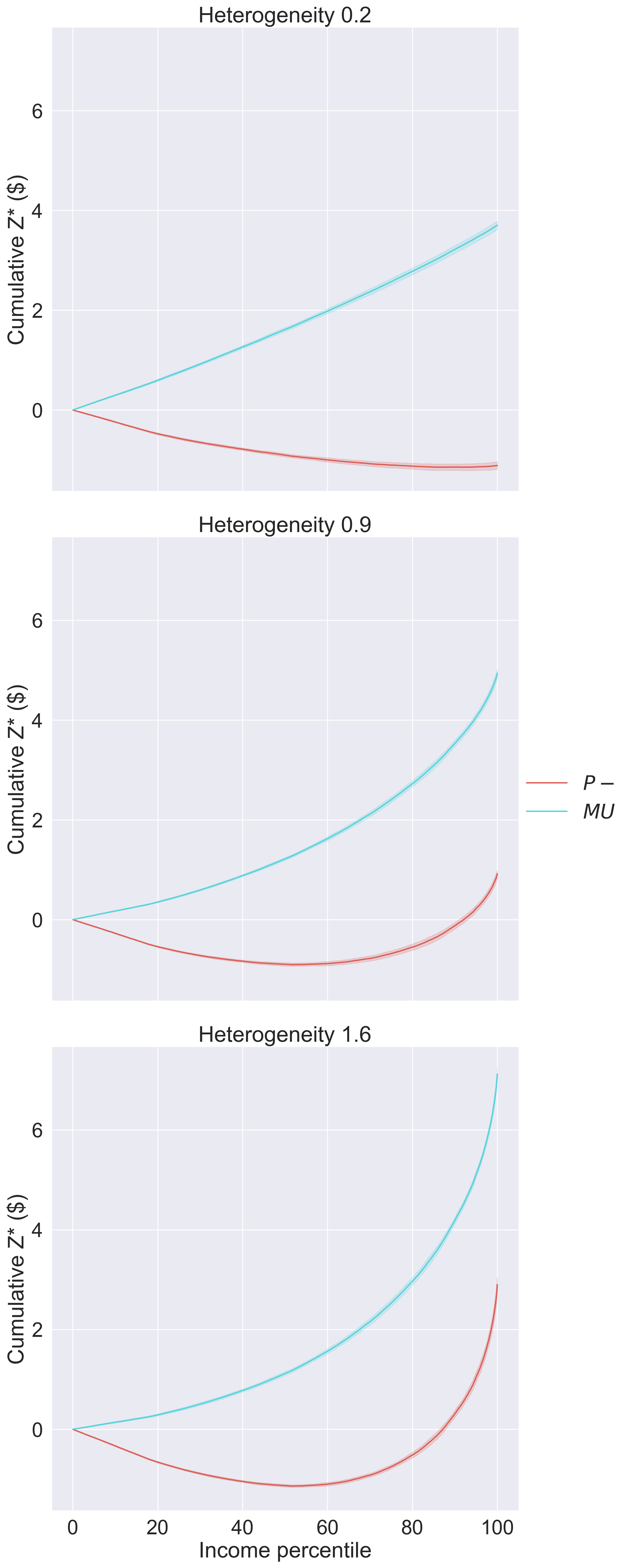}
    \caption[]{\small{Distribution of user benefits by income}}
    \label{fig:HeteroUBI}
  \end{subfigure}

  \caption{Distributional impacts of various instruments by heterogeneity levels}
  \label{fig:HeteroDist}
\end{figure}

\subsubsection{Heterogeneity}
As can be seen in Figure \ref{fig:HeteroSW}, as the extent of heterogeneity increases, the social welfare of $P-$ increases, which is consistent with findings in the literature (e.g. \cite{van2011winning}). Neglecting heterogeneity underestimates the benefits of pricing \citep{verhoef2004product}).  The relative performance of the two instruments does not change appreciably with the extent of heterogeneity, with the TMC scheme having marginally higher welfare and significantly better GC at all levels of heterogeneity. In terms of distributional impacts, at low levels of heterogeneity (COV of 0.2), it can be seen in Figure \ref{fig:HeteroUB} that all users are worse off under P$-$ whereas most of them are better off under the credit system.

\subsection{Robustness}\label{sec:robustness}

In practice, toll profiles may often be sub-optimal because of changing conditions, forecast errors and uncertainty. Practically, it is difficult to update these toll profiles (especially at the network level) regularly in practice. For example, Singapore updates the ERP scheme once every three months. In contrast, some market elements of the TMC scheme (e.g. allocation rate) are easier to adapt and have the potential to influence travelers' behavior (through the market price) to recover efficiency losses. In this section, two scenarios of a sub-optimal toll profile are investigated, including forecast error and non-recurrent events.

\begin{figure}[!]
    \centering
    \includegraphics[width=0.6\linewidth]{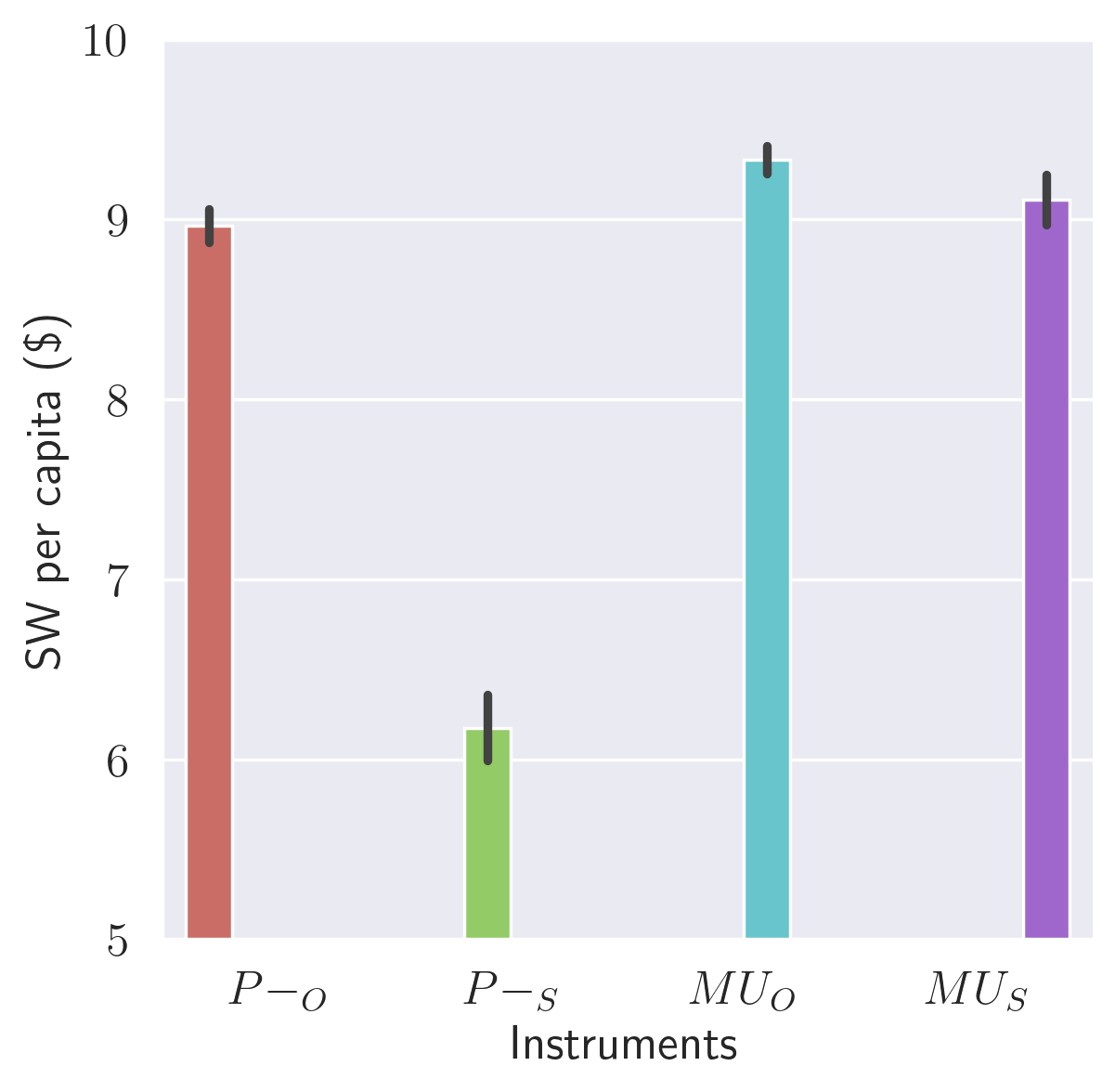}
    \caption{The social welfare of pricing and TMC with sub-optimal and optimal toll profiles}
\label{fig:FE}
\end{figure}

The first scenario is forecast error wherein actual road capacity is assumed to be 15\% less than the anticipated road capacity used to optimize the toll profile. The social welfare of pricing and TMC with this sub-optimal toll profile (based on anticipated road capacity) are plotted in Figure \ref{fig:FE} and denoted by $P-_S$ and $MU_S$. The social welfare of pricing and TMC with the associated optimal toll profiles are also plotted and denoted as $P-_O$ and $MU_O$. As we can see, $MU_S$ (TMC with sub-optimal tolls) is able to recover efficiency losses through a reduction in the allocation rate, which reduces token supply and increases token price. The optimal allocation rate is determined using a grid search and is found to be 15\% lower than the original allocation rate.

\begin{figure}[!]
    \centering
    \includegraphics[width=0.6\linewidth]{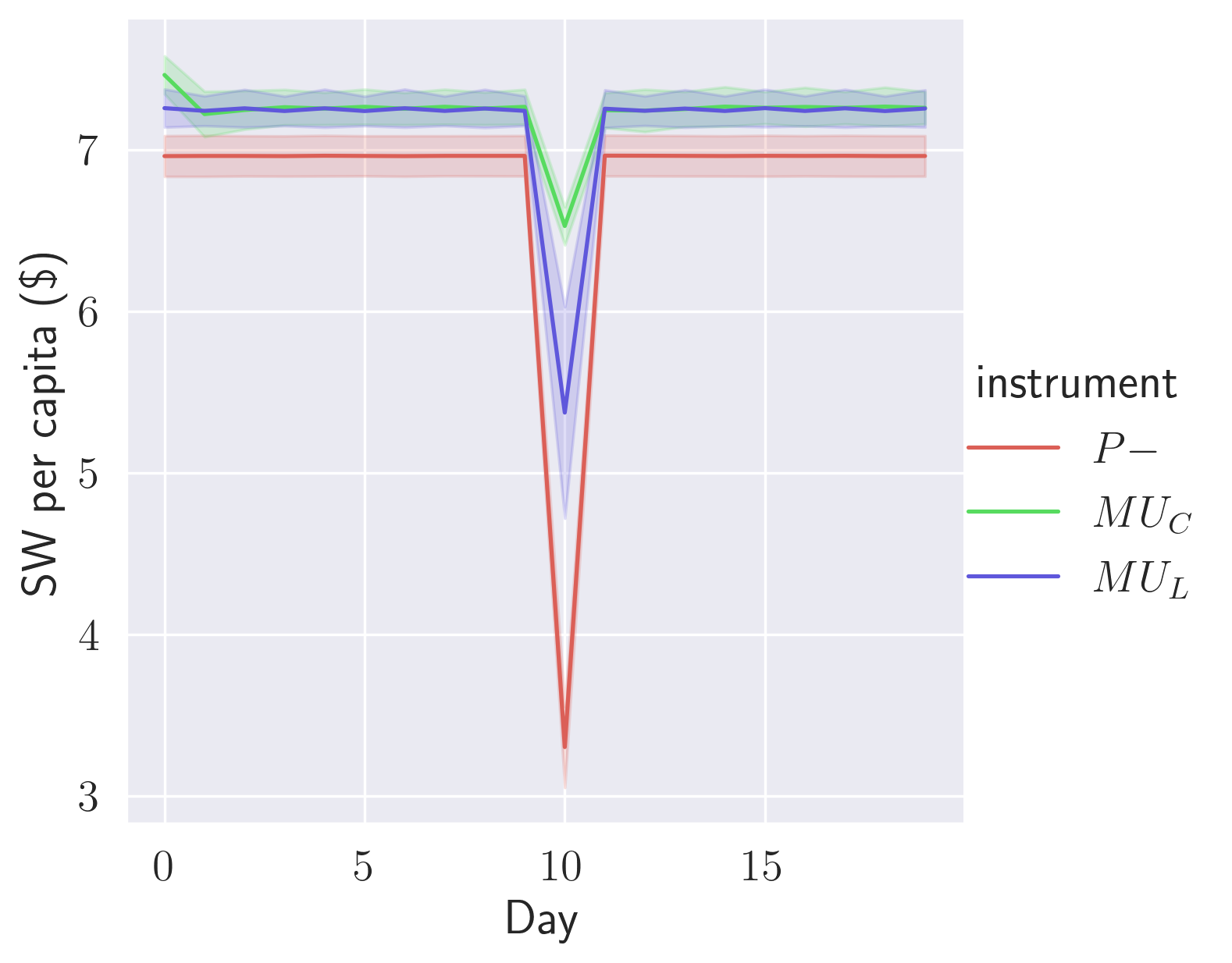}
    \caption{Social welfare of pricing and TMC with sub-optimal and optimal toll profiles}
\label{fig:UE}
\end{figure}

The second scenario is a non-recurrent event. Specifically, it is assumed that there is a sudden within-day capacity drop by 15\% (e.g., due to an accident or incident) from 7AM to 8:30AM on the 10th day after the system has reached an equilibrium. The social welfare across days for the three instruments are plotted in Figure \ref{fig:UE}. The first instrument is pricing without distribution and denoted as $P-$. The second one is TMC with a lump-sum allocation and denoted by $MU_L$. The third instrument is TMC with a continuous allocation and denoted as $MU_C$. 

% in whichMC with lump-sum allocation is denoted as $MU_L$ and TMC with continuous allocation is denoted a Ts $MU_C$.

Under $MU_L$, travelers receive the entire day's token allocation at the beginning of the day in the form of a `lump-sum' allocation. This form of token allocation is the standard design of TMC schemes in the literature (e.g. \cite{Yang2011,brands2020tradable}). Regarding trading, they can buy additional tokens at the time of traveling for immediate use and redeem unused tokens at the end of the day. Since trading is automated, there is no transaction fee considered under the lump-sum allocation. The regulator has three market parameters to control including within-day token price, regulation starting time and ending time. It cannot control allocation rate as all the tokens have already been allocated at the beginning of the day. Users who have not traveled yet before the new token price takes effect can update their plans according to the new information. We assume the same behavioral model used in the pre-day decision applies (with updated token prices and transaction fees). Using the DE algorithm, we determine that it is optimal for the regulator to increase the token price to $\$1.8$ between 6:55AM and 9:15AM. As shown in Figure \ref{fig:UE}, it performs better than $P-$.

For $MU_C$, the regulator can control not only token price, regulation starting and ending time, but also the  allocation rate (which affects the forecasted account balance of users over the rest of the day) and transaction fees. We optimize the fixed transaction fees of buying and selling together. Through optimization, between 7:05AM and 8:50AM, the regulator should set token price equal to $\$1.25$, allocation rate $r$ equal to 0 and fixed transaction fee equal to $\$0.5$. It performs better than the lump-sum allocation $MU_L$ as shown in Figure \ref{fig:UE}. This is intuitive because the travel behavior is impacted by both the allocation rate and transaction fees, which provide the regulator more degrees of freedom to intervene. This demonstrates the advantages of a continuous allocation of tokens over a lump-sum allocation of tokens. 

\section{Conclusions}\label{sec:Sec7}
This paper proposes and analyzes alternative market models for a tradable mobility credit system focusing on allocation/expiration of tokens, rules governing trading, transaction fees, and regulator intervention. We develop a methodology to explicitly model the dis-aggregate behavior of individuals within the market. Extensive simulation experiments are conducted within a combined mode and departure time context for the morning commute problem using a day-to-day assignment framework wherein transportation demand is modeled using a logit-mixture model with income effects and supply is modeled using a standard bottleneck model. 

The results show that small fixed transaction fees can effectively mitigate undesirable behavior in the market without a significant loss in efficiency whereas proportional transaction fees are less effective both in terms of efficiency and in avoiding undesirable market behavior. The market design we adopt is shown to yield stable and convergent market prices and is revenue neutral. 
We also show that an allocation of tokens in continuous time can be beneficial in dealing with non-recurrent events. One undesirable property of the system we do observe is that the selling behavior of individuals at equilibrium is dependent on the initial account states (although optimal welfare, flows and market prices are unique), and this deserves further investigation. With regard to the comparative performance relative to congestion pricing, in the presence of income effects, the TMC system yields a marginally higher social welfare. Finally, the TMC scheme is more equitable (when revenues from congestion pricing are not redistributed) although it is not guaranteed to be Pareto-improving when tokens are distributed equally. 
  
There are several promising directions for future research. First, given that a uniform allocation of tokens does not guarantee Pareto-improvement, it is clear that personalization is necessary to ensure there are no `losers'. The design of personalized token allocation  schemes is an important direction of future research. Second, more systematic experiments on day-to-day variability and its impact on the robustness of the TMC scheme and market are warranted. Finally, large scale  simulations on real-world networks with disaggregate agent based models is a natural next step in studying market design and other aspects of the tradable credit schemes towards real-world deployments.

\section*{Acknowledgements}
This research was funded by the U.S. National Science Foundation (Award CMMI-1917891) and the NEMESYS project funded by the DTU (Technical University of Denmark)-NTU (Nanyang Technical University) Alliance.

\bibliography{main}

\appendix

\section{Existence and Uniqueness of Equilibrium}\label{sec:App_uniqueness}
\begin{figure}[!h]
  \begin{subfigure}[b]{0.48\textwidth}
    \centering\includegraphics[width=\textwidth]{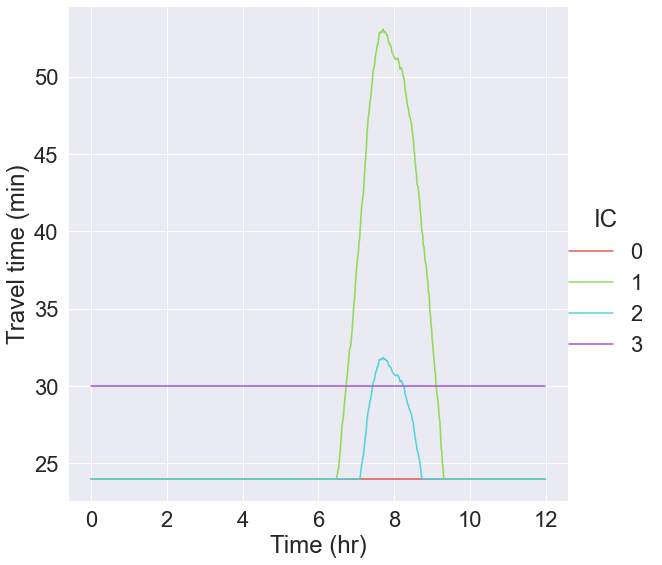}
    \caption[]{\small{Initial travel times}}
  \end{subfigure}
  \begin{subfigure}[b]{0.48\textwidth}
    \centering\includegraphics[width=\textwidth]{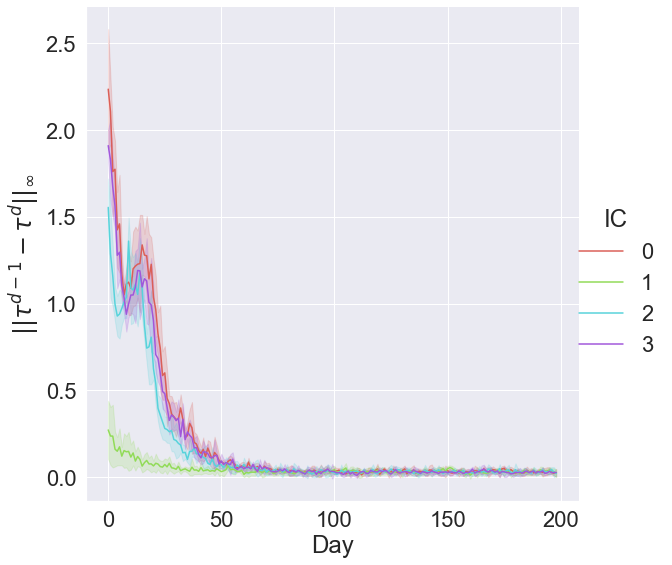}
    \caption[]{\small{Convergence of travel time of driving}}
  \end{subfigure}
  
  \caption{Initial travel time information of driving and corresponding convergence}
  \label{fig:NTIC}
\end{figure}

In this section, we examine convergence of the day-to-day model (Section \ref{sec:Sec5}). Using data and parameters discussed in the Section \ref{sec:Data_parameters}, a No Toll (NT) scenario with different sets of initial travel time information for the driving alternatives is simulated. Five independent replications or draws of the error terms in the utility (Equation \ref{eqn:utility}) are performed.   
%The population size $N$ is 7500. Because the simulation is stochastic, five replications with different initial seeds for random number generation are performed for each initial travel time of driving condition. 
The four different sets of initial travel time information of driving are plotted in Figure \ref{fig:NTIC}. Note that the initial travel time information serves as the basis for the departure time decisions in day 0 of the day-to-day simulation. The initial travel time set 0 represents free flow travel times across the entire day; the initial travel time set 1 consists of equilibrium travel times of driving obtained from the simulation using the initial travel time set 0 for a particular random seed; the initial travel time set 2 is 0.6 times the travel time set 1; finally, the initial travel time set 3 represents a constant 30 minutes travel time across the entire day.

\begin{figure}[!]
  \begin{subfigure}[b]{0.48\textwidth}
    \centering\includegraphics[width=\textwidth]{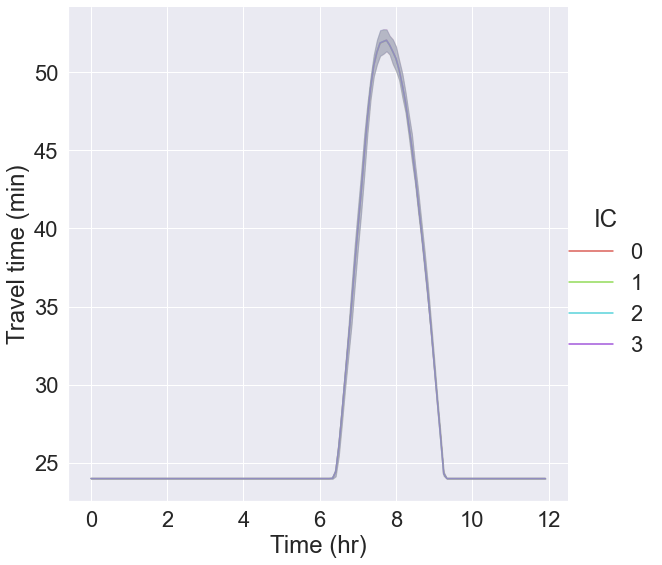}
    \caption[]{\small{Travel time of driving at equilibrium}}
  \end{subfigure}
  \begin{subfigure}[b]{0.48\textwidth}
    \centering\includegraphics[width=\textwidth]{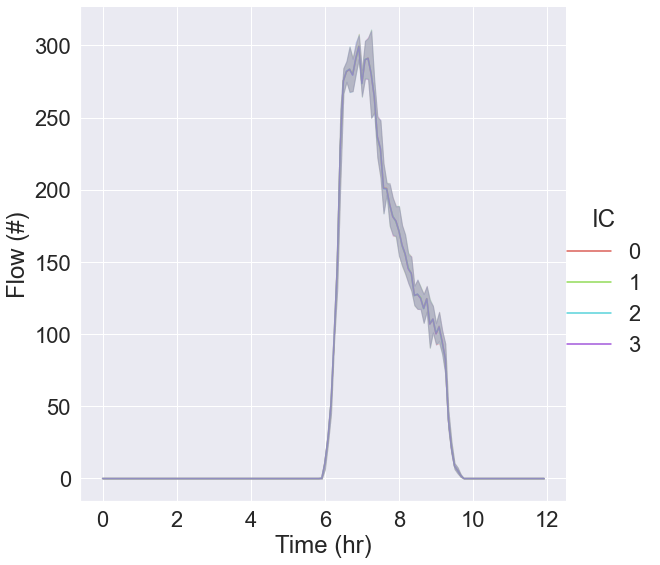}
    \caption[]{\small{Flow rate of driving at equilibrium}}
  \end{subfigure}
  
  \caption{Travel time and flow of driving at equilibrium with different sets of initial travel time information}
  \label{fig:NTUNI}
\end{figure}

The corresponding convergence of travel times of driving is plotted in Figure \ref{fig:NTIC}. The lines and bands in the plot represent averages and standard deviations, respectively, across the five replications. As we can see from the plot, the infinity norm (Equation \ref{eqn:Infinity_norm}) converges to within a threshold of 0.01 by about 50 days, indicating acceptable convergence of the day-to-day model.
%which implies the point wise maximum difference between the travel time vector of day $d-1$ and travel time vector of day $d$ converges according to the definition of infinity norm. 
%This suggests that simulations with different initial travel time sets have all converged in terms of travel times. 
%The convergence of green line does not start from 0 because the initial travel time set 1 of driving is an equilibrium travel time of a particular random seed.
%However, this still does not establish whether the equilibrium is unique or not. 
In order to examine uniqueness, the travel time vectors and flow vectors of driving at equilibrium are plotted in Figure \ref{fig:NTUNI}. The results indicate that simulations with different initial travel times of driving converge to the same travel time and departure flow rate patterns at equilibrium. As only two mode choices are considered and the travel time and headway of PT are constant, once the travel time of driving converges, the resulting departure flow of PT also converges.

\begin{figure}[!ht]
  \begin{subfigure}[b]{0.48\textwidth}
    \centering\includegraphics[width=\textwidth]{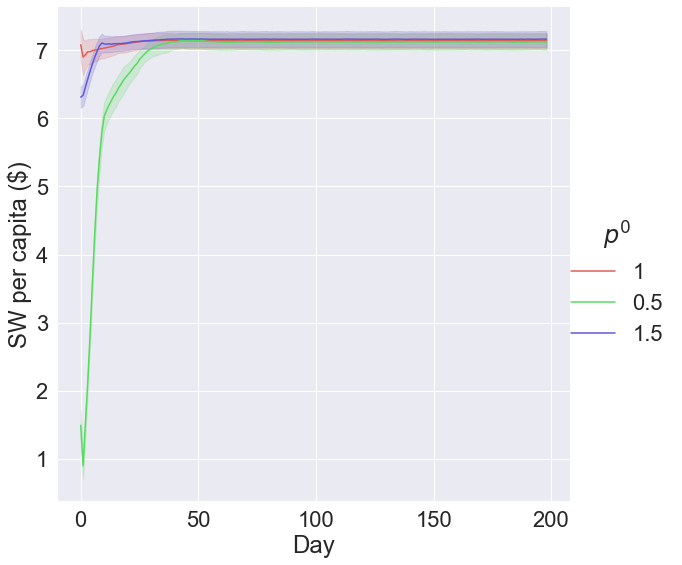}
    \caption[]{\small{ Social Welfare }}
  \end{subfigure}
  \begin{subfigure}[b]{0.48\textwidth}
    \centering\includegraphics[width=\textwidth]{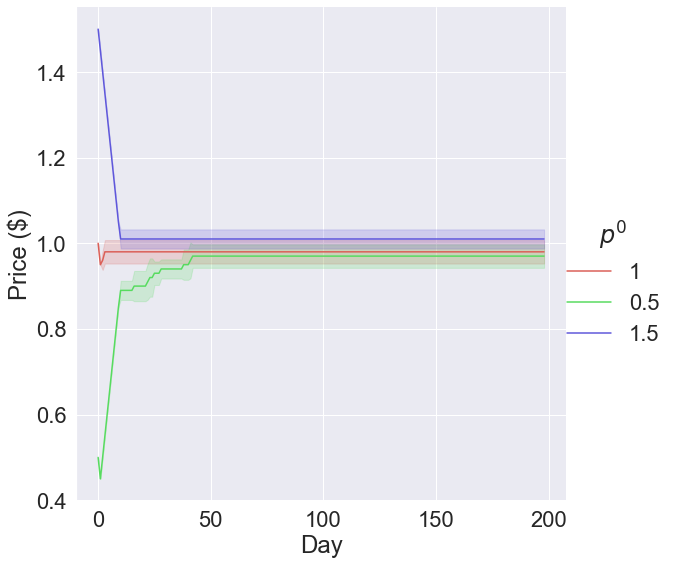}
    \caption[]{\small{Token market price}}
  \end{subfigure}
    \begin{subfigure}[b]{0.48\textwidth}
    \centering\includegraphics[width=\textwidth]{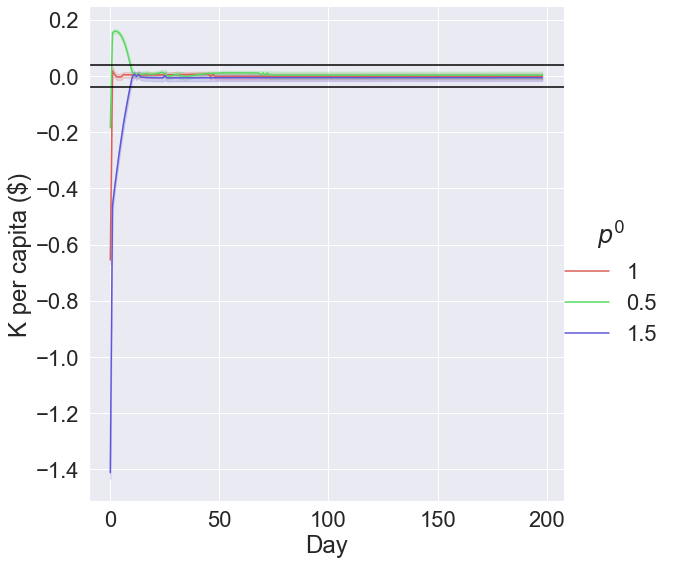}
    \caption[]{\small{Regulator Revenue}}
    \label{fig:PICRR}
  \end{subfigure}
    \begin{subfigure}[b]{0.48\textwidth}
    \centering\includegraphics[width=\textwidth]{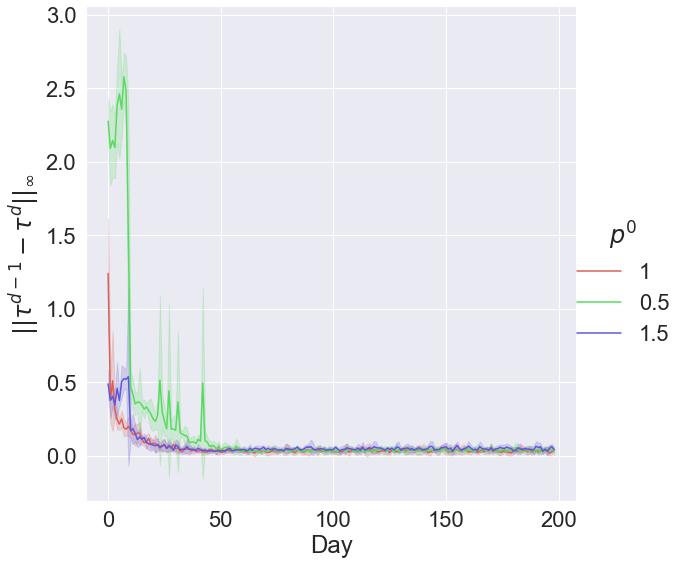}
    \caption[]{\small{Travel time of driving }}
    \label{fig:PICTT}
  \end{subfigure}
  \caption{Convergence of various metrics with different initial prices $p^0$ }
  \label{fig:PIC}
\end{figure}

\section{Convergence of Market Prices}\label{sec:App_convg_market_price}
The effect of various initial market prices $p^0$ on convergence of token market price and social welfare are examined in Figure \ref{fig:PIC}. The price and social welfare converge to values that are not statistically significantly different at a significance level of 0.05 regardless of the initial price. The regulator revenues under the three initial prices converge to be within the regulator revenue threshold band (the black lines) as shown in Figure \ref{fig:PICRR}. Travel times under the three initial prices converge too as shown in Figure \ref{fig:PICTT}. Similar experiments under a range of scenarios showed that the price adjustment scheme converges (in terms of travel times, flows and token prices) to the same equilibrium for different initial market prices.

%\section{Decaying selling price}
%In order to account time value of tokens, selling price of tokens decay linearly as tokens expire. Let $t_a$ denote acquisition time of the oldest token, $t_a^i$ denote acquisition time of token $i$, and $t$ denote current time. The selling price function of token $i$, denoted as $\dot{p}(t,t_a^i)$, can be written as 

%\begin{align}
 %   \dot{p}(t,t_a^i) = p(1-c_s)(1-\lambda\frac{t-t_a^i}{24}) = p_s(1-\lambda\frac{t-t_a^i}{24})
%\end{align}

%where $p$ represents token market price and $c_s$ represents proportional part of selling transaction cost. $\lambda$ represents a factor of decaying. If $\lambda=0$, selling price of all tokens is equal to market price $p$; if $\lambda=1$, selling price of token $i$ at age of one day ($t-t_a^i=24$) is equal to 0.

%The selling revenue at time $t$ of a user $n$ with account balance equal to $y$ can be written as

%\begin{align}
%S(y,t) =& \int_{t_a}^{t} \dot{p}(t,z)rdz -TC_s \\ \notag = & \int_{t_a}^{t} p_s(1-\lambda\frac{t-z}{24}) rdz -TC_s \\ \notag = &\int_{t_a}^{t} p_s rdz -\int_{t_a}^{t}\lambda\frac{t-z}{24} rdz -TC_s \\ \notag = & yp_s-\lambda \frac{1}{2}\frac{p_sr}{24}(t-t_a)^2 -TC_s \\ \notag = & yp_s-\lambda \frac{1}{2}\frac{p_sy^2}{24r} -TC_s 
%\end{align}

%where $TC_s $ represents fixed selling transaction cost.

\end{document}